\documentclass[a4paper,fleqn,usenatbib]{mnras}

\usepackage{newtxtext,newtxmath}

\usepackage[T1]{fontenc}
\usepackage{ae,aecompl}

\usepackage{threeparttable}
\usepackage{verbatim} 
\usepackage[usenames,dvipsnames]{color}
\usepackage{amssymb, amsmath}
\usepackage{multicol}
\usepackage{longtable}
\usepackage{rotating}
\usepackage{pdflscape}
\usepackage{url}
\usepackage{times}
\usepackage{multirow}
\usepackage{lscape}

\usepackage{graphicx}
\usepackage{epstopdf}
\usepackage[labelfont=bf,labelsep=space]{caption}

\usepackage{enumitem}
\usepackage[justification=centering]{caption}
\usepackage{subfig, float}

\usepackage{natbib}
\bibpunct{(}{)}{;}{a}{}{,}

\usepackage{array}
\newcolumntype{x}[1]{>{\centering\arraybackslash\hspace{0pt}}m{#1}}

\makeatletter
\def\url@leostyle{%
  \@ifundefined{selectfont}{\def\UrlFont{\sf}}{\def\UrlFont{\small\ttfamily}}}
\makeatother
\usepackage{newtxtext,newtxmath}

\DeclareRobustCommand{\ion}[2]{%
\relax\ifmmode
\ifx\testbx\f@series
{\mathbf{#1\,\mathsc{#2}}}\else
{\mathrm{#1\,\mathsc{#2}}}\fi
\else\textup{#1\,{\mdseries\textsc{#2}}}%
\fi}

\title[Self-consistent modelling of starbursts]{\color{black}Stellar populations and physical properties of starbursts in the Antennae galaxy from self-consistent modelling of MUSE spectra  }

\color{black} 
\author[Gunawardhana \it{et.\,al}]{M.\,L.\,P.\,Gunawardhana$^{1}$\thanks{E-mail: gunawardhana@strw.leidenuniv.nl}, {J.\,Brinchmann$^{2,\,1}$}, {P.\,M.\,Weilbacher$^3$}, {P.\,Norberg$^{4,\,5,\,6}$},  
\newauthor
{A.\,Monreal-Ibero$^{7,\,8}$}, {T.\,Nanayakkara$^1$}, {M.\,den Brok$^3$}, {L.\,Boogaard$^1$}, {W.\,Kollatschny$^{9}$}\\
\\
$^{1}$Leiden Observatory, Leiden University, PO Box 9513, 2300 RA, Leiden, The Netherlands\\
$^{2}$Instituto de Astrof\'isica e Ci\^{e}ncias do Espa\c{c}o, Universidade do Porto, CAUP, Rua das Estrelas, PT4150-762 Porto, Portugal\\
$^{3}$Leibniz-Institut f\"{u}r Astrophysik Potsdam (AIP), An der Sternwarte 16, D-14482 Potsdam, Germany\\
$^{4}$Institute for Computational Cosmology (ICC), Department of Physics, Durham University, South Road, Durham, DH1 3LE, UK\\
$^{5}$Centre for Extragalactic Astronomy (CEA), Department of Physics, Durham University, South Road, Durham, DH1 3LE, UK\\
$^{6}$Institute for Data Science, Durham University, South Road, Durham, DH1 3LE, UK\\
$^{7}$Instituto de Astrof\'{\i}sica de Canarias (IAC), E-38205 La Laguna, Tenerife, Spain\\
$^{8}$Universidad de La Laguna, Dpto.\ Astrof\'{\i}sica, E-38206 La Laguna, Tenerife, Spain\\
$^{9}$Institut f\"ur Astrophysik, Universit\"at G\"ottingen, Friedrich-Hund Platz 1, D-37077 G\"ottingen, Germany\\
}

\begin{document}
\color{black}
\date{Accepted date. Received date; in original form date}

\pagerange{\pageref{firstpage}--\pageref{lastpage}} \pubyear{2002}

\maketitle

\label{firstpage}

\begin{abstract}
We have modelled the stellar and nebular continua and emission-line intensity ratios of massive stellar populations in the Antennae galaxy using high resolution and self-consistent libraries of model HII regions around central clusters of aging stars. The model libraries are constructed using the stellar population synthesis code, \textsc{Starburst99}, and photoionisation model, \textsc{Cloudy}. The Geneva and PARSEC stellar evolutionary models are plugged into \textsc{Starburst99} to allow comparison between the two models. Using a spectrum-fitting methodology that allows the spectral features in the stellar and nebular continua (e.g.\, Wolf-Rayet features, Paschen jump), and emission-line diagnostics to constrain the models, we apply the libraries to the high-resolution MUSE spectra of the starbursting regions in the Antennae galaxy. Through this approach, we were able to model the continuum emission from Wolf-Rayet stars and extract stellar and gas metallicities, ages, electron temperatures and densities of starbursts by exploiting the full spectrum. From the application to the Antennae galaxy, we find that (1) the starbursts in the Antennae galaxy are characterised by stellar and gas metallicities of around solar, (2) the star-forming gas in starbursts in the Western loop of NGC 4038 appear to be more enriched, albeit slightly, than the rest of galaxy, (3) the youngest starbursts are found across the overlap region and over parts of the western-loop, though in comparison, the regions in the western-loop appear to be at a slightly later stage in star-formation than the overlap region, and (4) the results obtained from fitting the Geneva and Parsec models are largely consistent.  
\end{abstract}

\begin{keywords}
galaxies: starburst -- galaxies: individual: NGC 4038, NGC 4039 -- galaxies: ISM -- ISM: structure -- ISM: HII regions	
\end{keywords}

\section{Introduction}
The non-uniform expanding bubbles of ionised gas surrounding young massive stellar clusters are ideal laboratories to study the star formation processes at early stages of stellar evolution. Encoded in the photons escaping these regions is a wealth of information on their star formation histories, physical, chemical and dynamical conditions, dust and gas reservoirs; in the form of absorption and emission peaks in the stellar continuum, intensity variations and jumps in the nebular continuum, and Balmer, Paschen, Helium, and forbidden transitions in the emission spectrum. Despite the abundant information in spectra, the modelling of starbursting HII regions tend to rely on a limited amount of information to describe such a region.  

Understandably, describing what can necessarily be considered a galactic ecosystem in its entirety can be highly non-trivial. At a given instance, propelled by the mechanical energy output of its massive stars, an HII region is in an initial state of expansion, eventually stalling as the surrounding pressure becomes sufficiently high. During an expansion phase, the outer layers of an HII region are expanding more rapidly, are larger and have lower internal pressures and densities than the inner regions, thus generating temperature and pressure zones. Moreover, the properties of their young, massive stellar populations vary systematically with chemical abundance. Single massive stars with higher chemical abundances lose more mass through stronger stellar winds, rapidly evolve out of the main sequence to become supergiants, and have lower effective temperatures. In later stages of their lives, the massive stars may become a short-lived Wolf-Rayet star, briefly resurging in their ionising photon flux and mechanical energy production. The higher the chemical abundance, the easier it becomes for stars to achieve the temperature and chemical conditions required to evolve into a Wolf-Rayet star, which can lead to more low mass stars entering the Wolf-Rayet phase. Some of the chemical conditions required can, however, be modified by binary evolution \citep[e.g.][]{Eldridge2017}.

There are several approaches to characterising HII regions with varying levels of sophistication. One of the most common is utilising the optical emission line ratio diagnostics to measure key physical properties. The gas-phase metallicity, for example, is frequently derived using single line ratios, e.g.\,$R_{23}\equiv$([\ion{O}{ii}]$\lambda\lambda$3727,29 + [\ion{O}{iii}]$\lambda\lambda$4959, 5007)/H$\beta$ based on the two strongest lines of the strongest coolant of HII regions \citep{Pagel1979}, $S_{23}$:([\ion{S}{ii}]$\lambda\lambda$6717,31 + [\ion{S}{iii}]$\lambda\lambda$9069, 9532)/H$\beta$ \citep{Oey2002}, [\ion{Ar}{iii}]$\lambda$7135/[\ion{O}{iii}]$\lambda$5007 and [\ion{S}{iii}]$\lambda$9069/[\ion{O}{iii}]$\lambda$5007 \citep{Stasinska2006}. Likewise, the electron temperatures can be predicted from the intensity ratios of N$^{+}$, S$^{++}$ and Ne$^{++}$, and electron densities from the ratios of S$^{+}$, Cl$^{++}$ and O$^{+}$ \citep{Osterbrock, Peimbert2017}. The emission line diagnostics allow large datasets or data with limited spectral coverages to be analysed efficiently. Generally, the studies that employ these techniques tend to use several optical line ratio diagnostics together to overcome any biases that may arise from relying one diagnostics, for example, measuring electron temperatures and therefore, abundances \citep[e.g.][]{Bian2018}. 

A more sophisticated approach to interpreting the spectra of HII regions involves evolutionary population synthesis. Indeed, to understand the age and the evolutionary stage of HII regions requires self-consistent models for the ionising stars and the ionised gas. In this respect, the stellar population synthesis (SPS) codes and photoionisation models have proven to be powerful tools in synthesising model spectra for HII regions. 

The SPS codes allow the modelling of the light emitted by a coeval stellar population of a given metallicity and the reprocessing of that light by dust. The three main ingredients in any SPS model are stellar spectral libraries, the instructions on how a star evolves in the form of isochrones, and a stellar initial mass function (IMF). Most SPS codes, such as \textsc{Starburst99} \citep{Leitherer99}, \textsc{FSPS} \citep{Conroy2009, Byler2017}, \textsc{Galev} \citep{Kotulla2009}, \textsc{P\'egase} \citep{Fioc1999}, \textsc{PopStar} \citep{Molla2009}, SLUG \citep{daSilva2012}, BC03 \citep{BC2003}, and the models of \citet{Goerdt1998}, assume single stellar evolutionary paths for the stellar populations, while a few, such as BPASS \citep{Eldridge2017} and the models of \citet{Belkus2003}, take the full effect of binary stellar evolution into account. Coupled with SPS codes, photoionisation models, \textsc{Cloudy} \citep{Ferland2013} and \textsc{Mappings-iii} \citep{Sutherland1993, Dopita2013}, simulate the physical conditions within the interstellar medium, predicting the nebular emission as a function of the properties of stars and gas. 

The large girds of simulated emission-line spectra of \ion{H}{ii} regions generated using evolutionary population synthesis techniques have been routinely used in the development of optical and infrared diagnostics for abundance determinations \citep[e.g.][]{Kewley2002, Pereira-Santaella2017}, for the classification of starburst galaxies in optical line ratio diagnostic diagrams \citep[e.g.][]{Dopita2000, Kewley2001, Levesque2010}, for the determination of ages of the exciting stars from the positions of the observations on the theoretical HII region isochrone \citep[e.g.][]{Dopita2006}, for the derivations of gas-phase metallicities \citep[e.g.][]{Tremonti2004} and star formation rates \citep[e.g.][]{Brinchmann2004}, and for estimating gas masses \citep[e.g.][]{Brinchmann2013}.

The next stage of sophistication in modelling star-forming regions involves the self-consistent approaches to identifying star formation histories that best reproduce the stellar and nebular properties of star-forming regions.  In this respect, the publicly available FADO code \citep[Fitting Analysis using Differential evolution Optimization; ][]{Gomes2017} presents a mechanism to enable consistency between the observed nebular properties and the best-fitting star formation history of a star-forming region. FADO is, however, not set up to describe the stellar and nebular properties of massive, young stellar populations in starburst environments. 

In this paper, we introduce the development of a self-consistent and high-resolution spectral library for modelling both stellar and nebular properties of young, massive stellar populations, and a model-fitting strategy to fully exploit the spectra of starbursts to extract physical properties simultaneously. 

Based on the understanding that five physical parameters can effectively define a star-forming region, we construct self-consistent and time-dependent spherical models for HII regions. These parameters are the temperature and age of the ionising stars, chemical abundance, ionisation parameter and density (or pressure). The underlying physical processes, like photoionisation, collisional excitation, and radiative cooling, create the observed connections between these parameters \citep{Dopita2006, Kashino2019}. The central stellar clusters are modelled as simple stellar populations using the \textsc{Starburst99} code \citep{Leitherer99} updated to include the latest information on spectral libraries and isochrones. The stellar models of \textsc{Starburst99} are, then, coupled with the \textsc{Cloudy} \citep{Ferland2013} photoionisation model to self-consistently predict the nebular line and continuum emission as a function of age and different physical properties. To explore the capabilities of the developed library of models in extending our understanding of massive star-formation, we apply the models to the high resolution and high signal-to-noise MUSE \citep[Multi-Unit Spectroscopic Explorer;][]{Bacon2012} observations of the Antennae galaxy \citep{Weilbacher2018}.  
 
The Antennae galaxy (NGC 4038/39, Arp 244), at a distance of $22\pm3$ Mpc \citep{Schweizer2008}, is the closest major merger of two gas-rich spirals. Due to their proximity, the stellar populations and interstellar medium of the two galaxies are highly accessible, and the Hubble Space Telescope imaging has revealed the presence of thousands of compact young, massive star clusters \citep{Whitmore2010}, making it an ideal laboratory to study the physical processes and conditions prevalent in the environments of extreme star formation. 

The emission from Wolf-Rayet (WR) stars, a crucial evolutionary stage of massive stars, are prominent in the spectra of the HII regions in the Antennae galaxy. The appearance and the exact evolutionary timeframe of WR stars is largely chemical abundance-dependent, although, the first WR stars typically appear about 2 Myr after a star formation episode and disappear within some 5 Myr \citep{Crowther2007, Brinchmann2008}. Therefore, WR stars are high-resolution temporal tracers of the recent star formation history and powerful probes of the high mass slope of the stellar IMF. The effects of WR stars progressively enhance with increasing metallicity, and at solar-like metallicities, the WR features appear more prominent and diverse. So the Antennae starbursts with solar-like abundances \citep[e.g.][]{Bastian2009, Lardo2015} provide an ideal opportunity to exploit the WR features to explore massive star formation.   

The structure of the paper is two-fold. The focus of the first part of the paper (\S\,\ref{high_res_models} through to \S\,\ref{sec:method}) is the construction of a comprehensive stellar + nebular model library using an updated \textsc{Starbust99} code (\S\,\ref{subsec:stellar_lib} to \S\,\ref{subsec:props_wr}) coupled to \textsc{Cloudy} (\S\,\ref{subsec:nebular_model}) for characterising starbursts, and the development of the fitting routines on top of the existing \textsc{Platefit} code (\S\,\ref{sec:method}). In the second part of the paper (from \S\,\ref{sec:data} onwards), we present and discuss the Antennae dataset used for this study (\S\,\ref{sec:data}), the distribution of the physical properties (e.g.\,stellar and gas metallicities, light-weighted ages of the ionising stars, electron density and temperatures) derived simultaneously from applying the model library to the MUSE spectra of starbursting HII regions (\S\,\ref{sec:props}), as well as compare our results with properties derived using other methodologies and with the literature. 

The assumed cosmological parameters are H$_0$ = 70 km\,s$^{-1}$\,Mpc$^{-1}$, $\Omega_M$ = 0.3, and $\Omega_{\Lambda}$= 0.7. A \citet{Kroupa2001} stellar initial mass function is assumed throughout. 

\section{A high resolution stellar \& nebular template library for modelling starbursts}\label{high_res_models}
A key goal of our study is to constrain the physical properties, including relative ages of the stellar populations in the HII regions. To do so, we need to leverage the spectral resolution and range of the MUSE data and in particular to accurately model the Wolf-Rayet features we see in many of our spectra. So in this section, we discuss the implementation of a theoretical spectral library aimed at modelling the stellar and nebular spectral features of the young, massive stellar populations observed in the starbursting regions in the Antennae galaxy. 

To generate the stellar and nebular templates self-consistently, we use the \textsc{starburst99} code \citep{Leitherer99} coupled with the photoionisation model \textsc{Cloudy} \citep{Ferland2013}. We update and modify several key constituents of \textsc{starburst99} code in order to increase the wavelength resolution and improve the accuracy of the modelling of massive stellar populations, which we discuss in the subsequent sections. 

\subsection{The spectral libraries}\label{subsec:stellar_lib}
One of the main ingredients in SSP codes is the spectral library. The spectral resolution and wavelength range of a spectral library directly define the inherent spectral resolution and the wavelength coverage of the models. Moreover, the range of stellar types included in a spectral library determine how well the models reproduce a particular stellar population.      

As our goal is to create a model library with high spectral resolution over a broad range in wavelengths that captures the effects of massive stars with high temperatures, we update the spectral libraries - the stellar and the Wolf-Rayet libraries - incorporated into \textsc{Starburst99} in order to create high-resolution stellar templates, as discussed below.   

\subsubsection{The stellar spectral library}
We incorporate the synthetic stellar library of \citet{Munari2005} into \textsc{starburst99}. The \cite{Munari2005} library spans $2500-10\,500\,\AA$ in wavelength, and maps the full HR diagram, exploring $51\,288$ combinations of atmospheric parameters. The library was computed using the \textsc{synthe} code of \cite{Kurucz1993} with the input model atmospheres of \cite{Castelli2003}, which incorporate newer opacity distribution functions, at a resolution of 500\,000 and subsequently degraded to lower resolutions of 2500, 8500 and 20\,000. Additionally, the solar-scaled abundances of \cite{Grevesse1998} were adopted, along with the atomic and molecular line lists by \cite{Kurucz1992} in the computation of synthetic spectra. 

Lines for several molecules, such as C2, CN, CO, CH, NH, SiH, SiO, MgH and OH, were included for all the stars, and TiO molecular lines were included for stars cooler than T$_{\rm{eff}}$ of 5000 K. The atomic lines with predicted energy levels (``predicted lines") are, however, excluded from the calculation of the synthetic spectra as the wavelengths and intensities of predicted lines can be significantly uncertain, resulting in polluting high-resolution model spectra with false absorption features.  The polluting effects become progressively worse with increasing metallicity \citep{Munari2005}. On the other hand, predicted lines are essential to the total line blanketing in model atmospheres and for the computation of spectral energy distributions. The effect of the lack of predicted lines is to underestimate the blanketing, mostly affecting the predictions of broadband colours, and it becomes increasingly more significant towards bluer wavelengths, cooler temperatures (typically T$_{\rm{eff}}$<7000 K) and higher stellar metallicities \citep[][]{Munari2005, Coelho2007}. The high-resolution spectral libraries, therefore, need to be flux calibrated if they are to also be used to obtain good predictions for broadband colours \citep{Coelho2014}. For this study, however, we have not flux calibrated the \citet{Munari2005} library for the effects of line blanketing. As our goal is to create a model library that captures the effects of massive stars with high temperatures over which the blanketing due to predicted lines is less important, we expect the uncertainties arising from the lack of predicted lines to be relatively low. 

As described before, synthetic stellar libraries like the \citet{Munari2005} library offer both a more extensive mapping of the HR diagram and better coverage in wavelength than empirical libraries. The impact of utilising synthetic versus empirical stellar libraries, and libraries with limited versus full HR coverage on predicted properties, such as colours and magnitudes of SSPs, and age, metallicity and reddening measures extracted from spectral fits, are investigated by \cite{Coelho2019}. In the case of synthetic vs empirical, they find a null effect on the mean light-weighted ages, but find metallicity to be underestimated by $\sim$0.13, whereas in the case of limited vs full wavelength coverage, the ages are found to be underestimated while metallicity shows a little impact.

In Table\,\ref{tab:munari_vs_us}, we present our selection of spectra from the extensive \cite{Munari2005} stellar library.  From the R=8500 library, we select the spectra spanning the full range in effective temperatures, surface gravities and stellar metallicities, and assume a microturbulent velocity of 2 km\,s$^{-1}$ and no $\alpha$-to-Fe enhancement. Furthermore, following \cite{deJager1980} and \cite{Martins2005}, we adopt an effective temperature dependent initial stellar rotational velocity prescription (see Table\,\ref{tab:munari_vs_us} notes for T$_{\rm{eff}}$ dependent V$_{\rm{rot}}$ relation) for the selection of spectra. Our assumption of [$\alpha$/Fe] of 0.0 is motivated by the study of \citet{Lardo2015} that find that a solar-scaled composition for the Antennae HII regions to be reasonable. Although, according to \citet{Lardo2015}, the errors deriving from the assumption of a solar-scaled composition instead of an alpha enhanced one has minimal impact on their metallicity estimates. 
\begin{table*}. 
\caption{The range in various parameters available in the \citet{Munari2005} stellar spectral library and our selection of spectra for the construction of the high-resolution templates}
\begin{threeparttable}[t]
\centering
\begin{tabular}{l|c|c}
\hline
 Parameter			& \cite{Munari2005} library 	&  our selection \\
\hline
Wavelength 								& $2500-10\,500$ \AA  				& $3000-10\,000$ \AA \\
Resolution (R) 								& 2000, 8500, 20\,000        			& 8500      \\				
Effective temperature (T$_{\rm{eff}}$)   			& $3500 - 47\,500$ K  	 			& Full range \\
Surface gravity (log\,$g$)  					&  $0.0 - 5.0$ [0.5 dex sampling]  		& Full range  \\
Metallicity ([M/H]) 							& -2.5 to 0.5     	 					& Full range \\
$\alpha$ abundance ([$\alpha$/Fe]) 				& $0.0$, $+0.4$ 					& $0.0$ \\
Microturbulent velocity ($\xi$)    				& 1, 2, 4 km\,s$^{-1}$   				& 2 km\,s$^{-1}$ \\
Rotational velocity (V$_{\rm{rot}}$)    			& $0.0$ to $500$  	   				& T$_{\rm{eff}}$ dependent V$_{\rm{rot}}$\tnote{1} \\
\hline
\end{tabular}
\begin{tablenotes}
     \item[1] {Following \cite{deJager1980} and \cite{Martins2005} we assume, \\ V$_{\rm{rot}} \approx 100$ [km\,s$^{-1}$] for T$_{\rm{eff}}$ [K] $\geqslant7000$ \\ V$_{\rm{rot}}\approx50$ [km\,s$^{-1}$] for $6000\leq$T$_{\rm{eff}}$ [K]$\leq7000$  \\  V$_{\rm{rot}}\approx10$ [km\,s$^{-1}$] for T$_{\rm{eff}}$ [K] $<6000$}  
\end{tablenotes}
    \end{threeparttable}%
\label{tab:munari_vs_us}
\end{table*}%

\subsubsection{The Wolf-Rayet spectral library}\label{subsubsec:wr_lib}
We replace \textsc{starburst99}'s low resolution CMFGEN library \citep{Hillier1998} with the higher resolution Potsdam grids\footnote{\url{http://www.astro.physik.uni-potsdam.de/~wrh/PoWR/powrgrid1.php}} (Galactic, LMC and SMC) of model atmospheres for Wolf-Rayet (WR) stars \citep{Hamann2004, Sander2012, Todt2015}. The Potsdam Wolf-Rayet (PoWR) library provides comprehensive grids of expanding, non-local thermodynamic equilibrium, iron group line-blanketed\footnote{The large number of iron line transitions (Fe\,\textsc{iv}, Fe\,\textsc{v}, Fe\,\textsc{vi}) form a pseudo continuum in the ultra-violet dominating the spectral energy distribution. The iron group line-blanketing is, therefore, important for reproducing the observed spectra of WR stars, particularly, WC subtypes \citep{Grafener2002}} atmospheres of WR subtypes; WR stars with strong Helium and Nitrogen lines (WN stars) and WR stars with strong Helium and Carbon lines (WC stars). 

The WR spectra are parameterised by luminosity (L), effective temperature (T$_{\rm{eff}}$) and wind density, which is higher than the wind density of O-stars and is thought to be caused by the high luminosity-to-mass ratios of WR stars \citep{Leitherer2014}. The wind density plays a significant role in determining the types of features in a given WR spectrum.  
The PoWR library provides grids of WC and WN subtypes as a function of wind density, parameterised as `transformed radius', and T$_{\rm{eff}}$. 
Therefore, in order to combine the PoWR library with \textsc{Starburst99}, we adopt the `transformed' radius definition of \cite{Schmutz1989} as given in \cite{Leitherer2014}, 
\begin{equation}
R_t = R_* \Big( \frac{v_{\infty}}{2500} \Big/ \frac{\sqrt{D}\dot{M}}{1\times10^{-4}}\Big)^{2/3},
\label{eq:rt_calculation}
\end{equation}
where $R_*$ is the effective stellar radius in the unit of m, $v_{\infty}$ is the terminal wind velocity in the unit of km\,s$^{-1}$,  and $\dot{{M}}$ is the mass loss rate in M$_{\odot}$ yr$^{-1}$. The winds of WR stars are thought to be `clumped' based on the observations of line profile variabilities \citep{Moffat1988} and the over-prediction of the strengths of the theoretical line profiles \citep{Hillier1991}. The wind clumping factor, $D$, therefore, has the effect of adjusting $\dot{{M}}$. The typical values adopted for $D$ in the literature \citep[e.g.][]{Dessart2000, Crowther2002, Smith2002, Crowther2007, Leitherer2014}, which we also adopt in this study, range from $4-10$ for WC to approximately 4 for WN subtypes \citep{Crowther2007}. The exact $D$ values needed for WC and WN subtypes, though, is unclear \citep{Smith2002}.  

At a given time, the parameters T$_{\rm{eff}}$ and R$_*$ in Eq.\,\ref{eq:rt_calculation} guide the selection of a WC or WN PoWR spectrum closest to a given T$_{\rm{eff}}$ and R$_t$, with the WC and WN classification determined based on the surface abundances of Hydrogen, Carbon, Nitrogen and Oxygen. All of this information is drawn from the isochrones (i.e.\,evolutionary stage of stars of different initial masses in the HR diagram at a fixed age) incorporated into \textsc{Starburst99}. We introduce and discuss the different isochrones used in this study in \S\,\ref{subsec:stellar_isoc}. 

Furthermore, as the PoWR grids distinguish between WN-early and late types of WR stars, we also incorporate those into \textsc{starburst99} following the prescription of \cite{Chen2015}, which uses the surface abundance of Hydrogen for the WN-early/late classification. Finally, as noted above, the PoWR grids are only available for Galactic, LMC and SMC metallicities. As such, in our incorporation of the PoWR library to \textsc{starburst99}, we assume that the contribution from the WR stars to be small for metallicities lower than SMC, and for metallicities higher than the Galactic, we use the Galactic WR grid. While this is not ideal, as the present study is focussed on investigating metal-rich starbursts, the uncertainties arising from our assumption is likely minimal.   

\subsection{The stellar evolutionary tracks}\label{subsec:stellar_isoc}
Regardless of temperature, massive stars develop strong stellar winds that lead to a significant decrease in stellar mass over their lifetimes. As such, the mass-loss rates are of critical importance in estimating massive stars' contribution to enriching the ISM \citep{Maeder1994}. All stellar evolutionary models incorporate mass-loss rates for massive stars at varying levels. Some of the most widely used single-star stellar evolutionary models for massive stars in stellar population synthesis include the models of the Geneva \citep{Schaller1992, Charbonnel1993, Meynet1994} and the Padova \citep{Bertelli1994, Girardi2000, Girardi2010} groups. Relatively more recently, stellar evolutionary tracks incorporating stellar rotational effects on massive stars \citep[e.g.\,Geneva and MIST models;][]{Meynet1994, Paxton2013}, and binary evolutionary effects \citep[e.g.\,BPASS;][]{Eldridge2009, Eldridge2017, Gotberg2019} have also been published. 

For the present work, we use the latest single stellar evolutionary models of the Geneva and the Padova groups. The Geneva isochrones are already part of the \textsc{Starburst99} code that is publicly available, and we incorporate the Parsec models (from the Padova group) into \textsc{Starburst99}. We outline some of the main features of the two sets of isochrones, and the changes made to the Parsec models below.

\subsubsection{The Geneva stellar evolutionary tracks}
The Geneva `high mass loss' isochrones \citep{Meynet1994} are generally preferred in the modelling of starbursts \citep[e.g.][]{Brinchmann2008, Levesque2010, Byler2017} as they are meant to accurately model the observations of the crucial WR phase, particularly the low-luminosity WR stars. The enhanced mass loss rates \citep[$\approx$ 2x the rates of the `standard' grid from the][]{deJager1988} provide a reasonable approximation of the high mass losses experienced by massive stars entering the WR phase. The mass-loss rates during the WR phase (i.e.\,early-type WN, WC and WO) were left unchanged, except for late-type WN WR. Likewise, the mass loss rates are uncorrected for any initial metallicity effects \citep{Meynet1994}. 

For the present analysis, we adopt the full range of metallicities provided with code, i.e.\,$Z=0.001, 0.004, 0.008, 0.02, 0.04$. All isochrones extend up to an upper initial mass of 120M$_{\odot}$, but terminate at different lower initial masses of 25, 20, 15, 15, 12 M$_{\odot}$, respectively, from low-to-high metallicity. Therefore, to extend the isochrones down to a stellar mass of approximately 0.1 M$_{\odot}$,  the `high mass loss' tracks are combined with the `standard mass loss' tracks. In total, the evolutionary information for around 22 stellar mass types over 51-time intervals are provided in the Geneva isochrones.  

\subsubsection{The Padova stellar evolutionary tracks}

We use the PARSECv1.2S library\footnote{The PARSECv1.2S library is downloaded from the CMD3.1 web interphase (\url{http://stev.oapd.inaf.it/cgi-bin/cmd})} of isochrones \citep{Bressan2012} from the Padova group that is constructed using an extensively revised and extended version of the Padova code \citep{Bertelli1994, Girardi2000, Girardi2010}. This release also incorporates updates to a better treatment of boundary conditions in low-mass stars \citep[M\,$\lesssim$\,0.6\,M$_{\odot}$;][]{Chen2014}, and envelope overshooting and latest mass-loss rates for massive stars \citep[14\,$\,\lesssim$\,M/M$_{\odot}$$\lesssim$\,350;][]{Tang2014, Chen2015}.        

The Parsec grid includes tracks for 15 initial metallicities ($Z_i=0.001$ to $0.06$), with the initial Helium content relating to the initial metallicities through $Y_i = 1.78\times Z_i + 0.2485$ \citep{Komatsu2011, Bressan2012}. For each metallicity, $\sim120$ mass values between 0.1 and 350 M$_{\odot}$ are also included in the grid. For this analysis, we select isochrones covering the same range in metallicities as available for the Geneva models and mass in the range 0.1 to 200 M$_{\odot}$.  

From the PARSECv1.2S library, we draw evolutionary tracks for 50 stellar mass types with masses in the range $0.1-350$ M$_{\odot}$ to construct the isochrones that are incorporated into \textsc{Starburst99}. Each isochrone is sampled finely in time (400-time steps) to minimise the uncertainties arising from the interpolations performed within \textsc{Starburst99}. The metallicities are chosen to match the range of metallicities available with the Geneva isochrones. 

One of the caveats of using the Parsec (or Padova) isochrones for modelling massive stars is that they lack the information on the effective temperature correction needed to model the crucial WR evolutionary phase of massive stars. In the absence of this information, \textsc{Starburst99}, for example, defaults to using the effective stellar temperature instead, which can lead to several orders of magnitude difference in far-ultraviolet spectra computed using Parsec and Geneva isochrones \citep{D'Agostino2019}.  Therefore, we calculate an effective temperature correction for WR stars following the prescriptions of \citet{Leitherer2014} and \citet{Smith2002} to incorporate into our implementation of Parsec isochrones.  

\subsection{On the properties of the generated stellar models}\label{subsec:props_wr}
\begin{figure*}
\begin{center}
\includegraphics[scale=0.55, trim={0.1cm 9.9cm 16.5cm 1.0cm},clip]{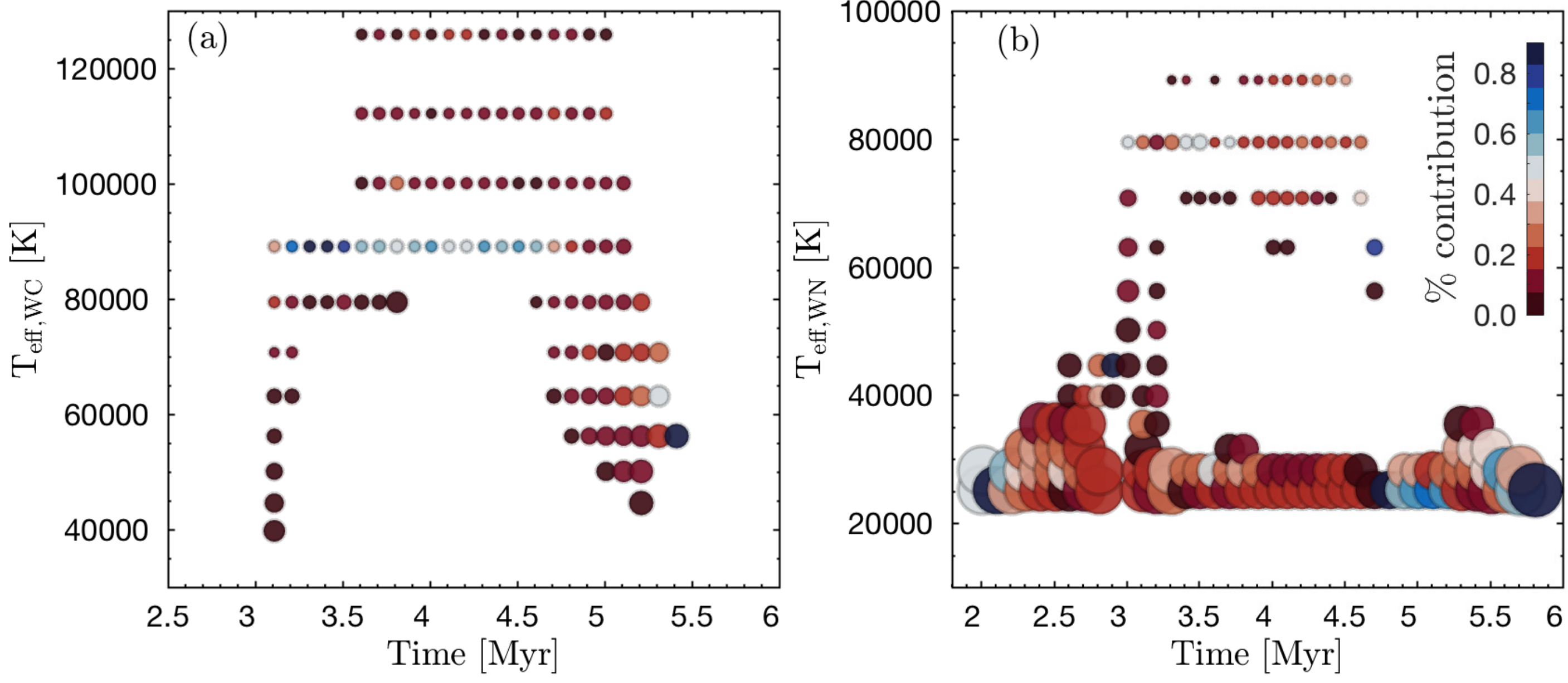}
\includegraphics[scale=0.55, trim={0.1cm 9.9cm 16.5cm 1.0cm},clip]{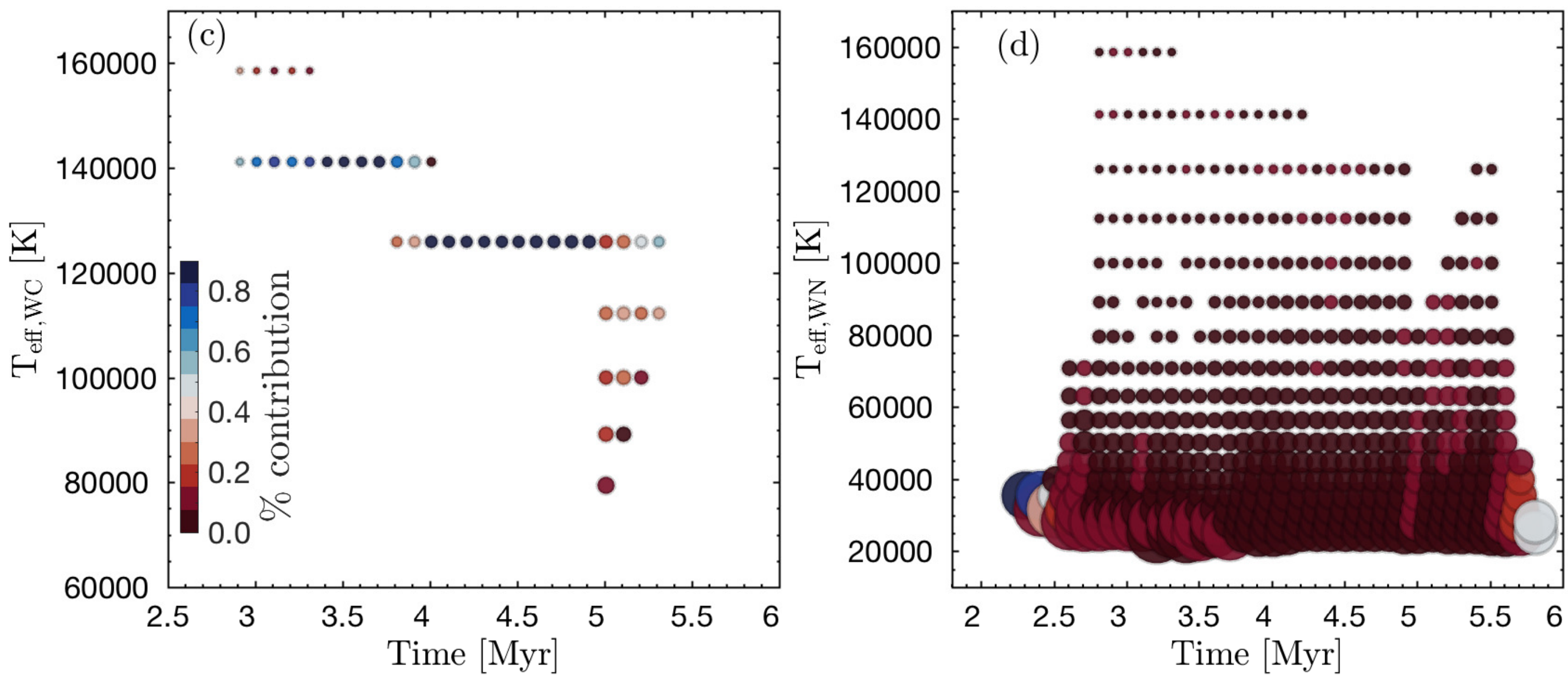}
\caption{The selection of WR spectra as a function of time for Geneva high mass-loss (a-b) and Parsec (c-d) stellar evolutionary models assuming an IMF with a high mass cut off of 120 M$_{\odot}$. At a given time, T$_{\rm{eff}}$ and surface abundances of Hydrogen, Carbon, Nitrogen and Oxygen of the isochrones are used to determine the WR subtype, then T$_{\rm{eff}}$ and R$_t$ are used to choose the spectrum from the PoWR grids. \textit{Left}: the selection of the WC spectra as a function of time and T$_{\rm{eff}}$ (of the WR star), colour coded to show the percentage of WC spectra of a given T$_{\rm{eff}}$ (WR) which are chosen at a given time. \textit{Right}: the same as left, but for WN. The marker size denote the relative R$_t$, demonstrating that the spectra of WC subtypes chosen have R$_t$ smaller than that of WN. Note that the spectral features between spectra of a given WR subtype can vary significantly depending on the R$_t$ chosen. }
\label{fig:evolution_WR_fractions}
\end{center}
\end{figure*}

\subsubsection{The improvements to the spectral features}
\begin{figure*}
\begin{center}
\includegraphics[scale=0.38, trim={2.2cm 1.4cm 22.5cm 0.1cm},clip]{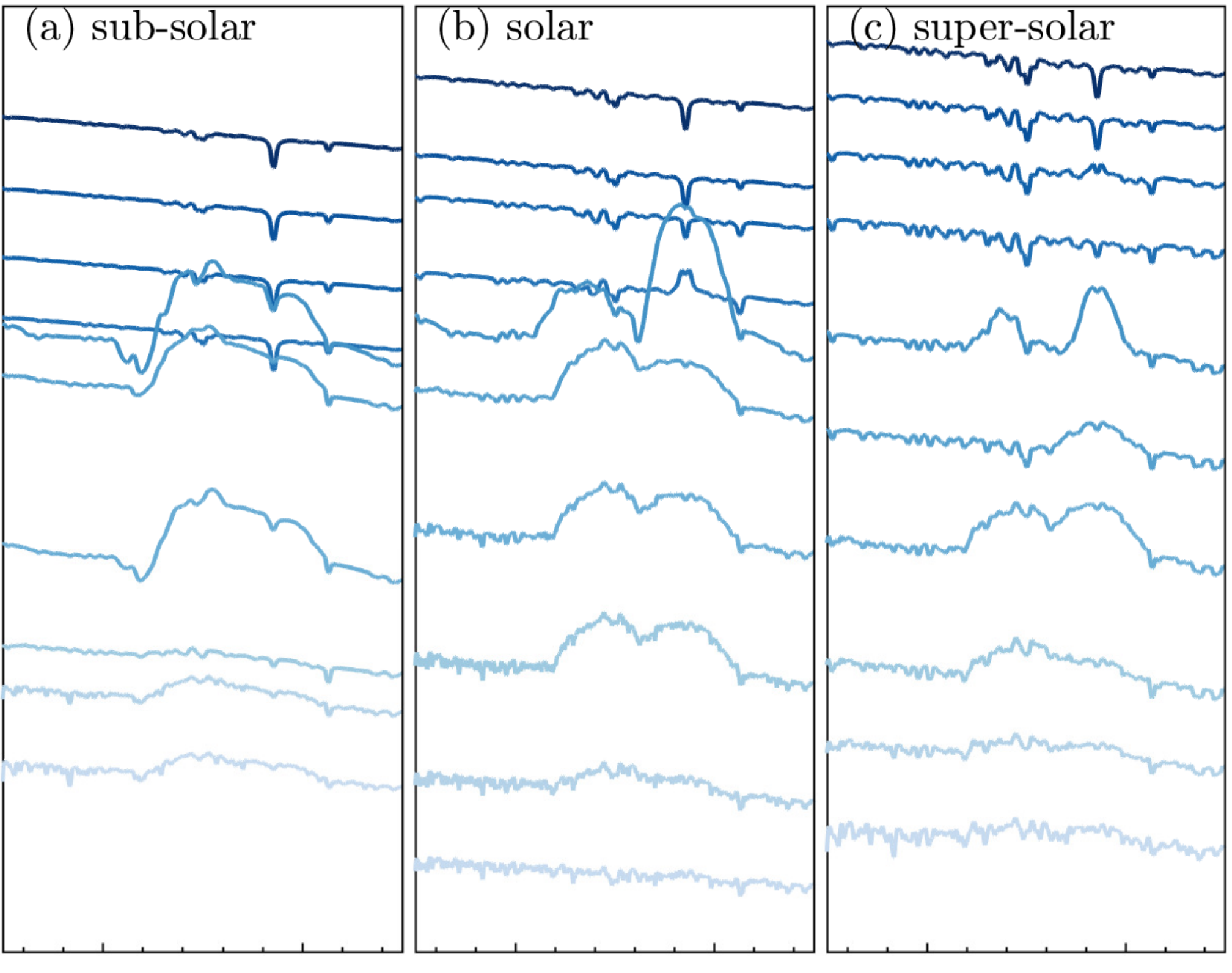}
\includegraphics[scale=0.38, trim={2.2cm 1.4cm 22.5cm 0.1cm},clip]{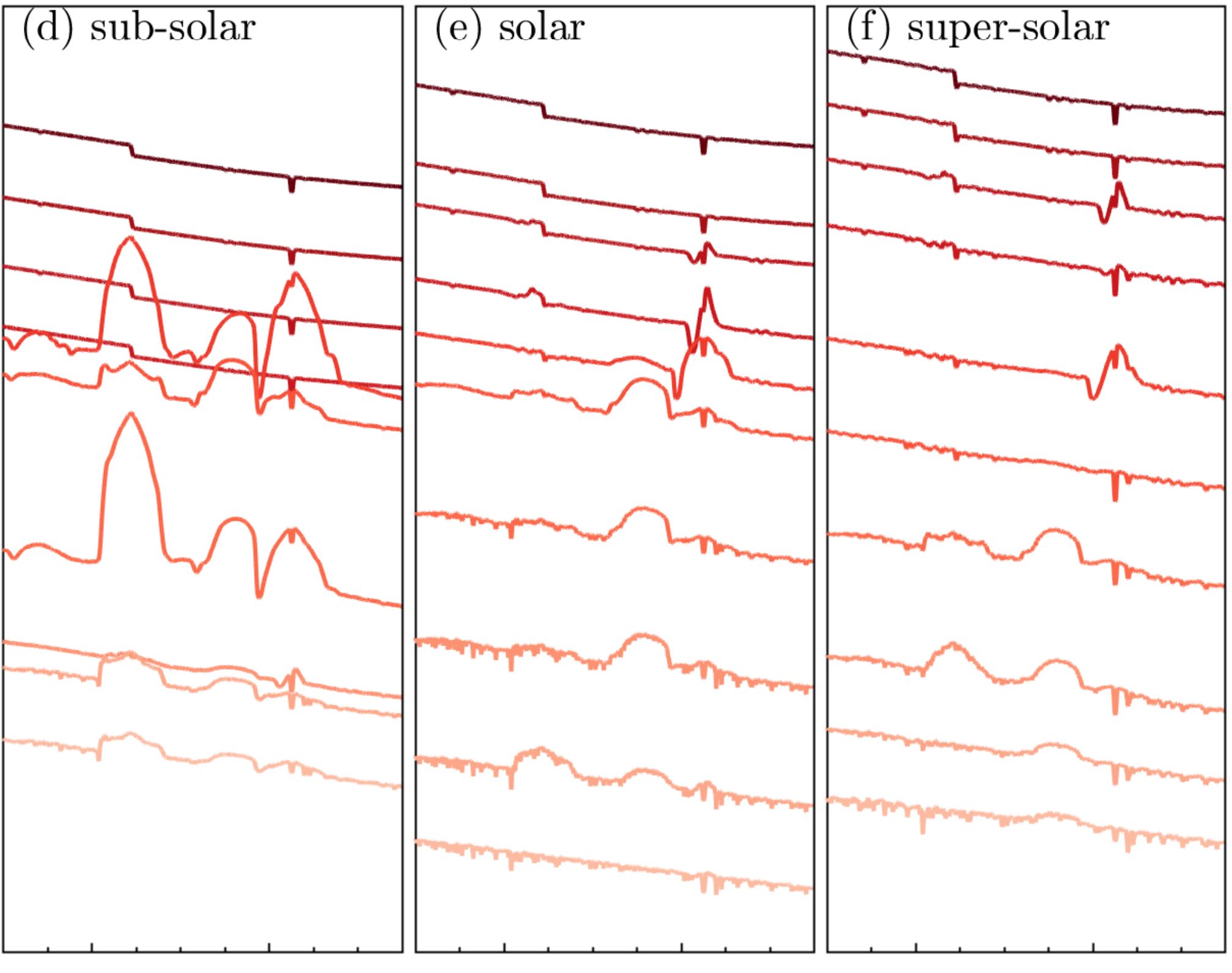}
\includegraphics[scale=0.38, trim={2.2cm 0.cm 22.5cm 0.1cm},clip]{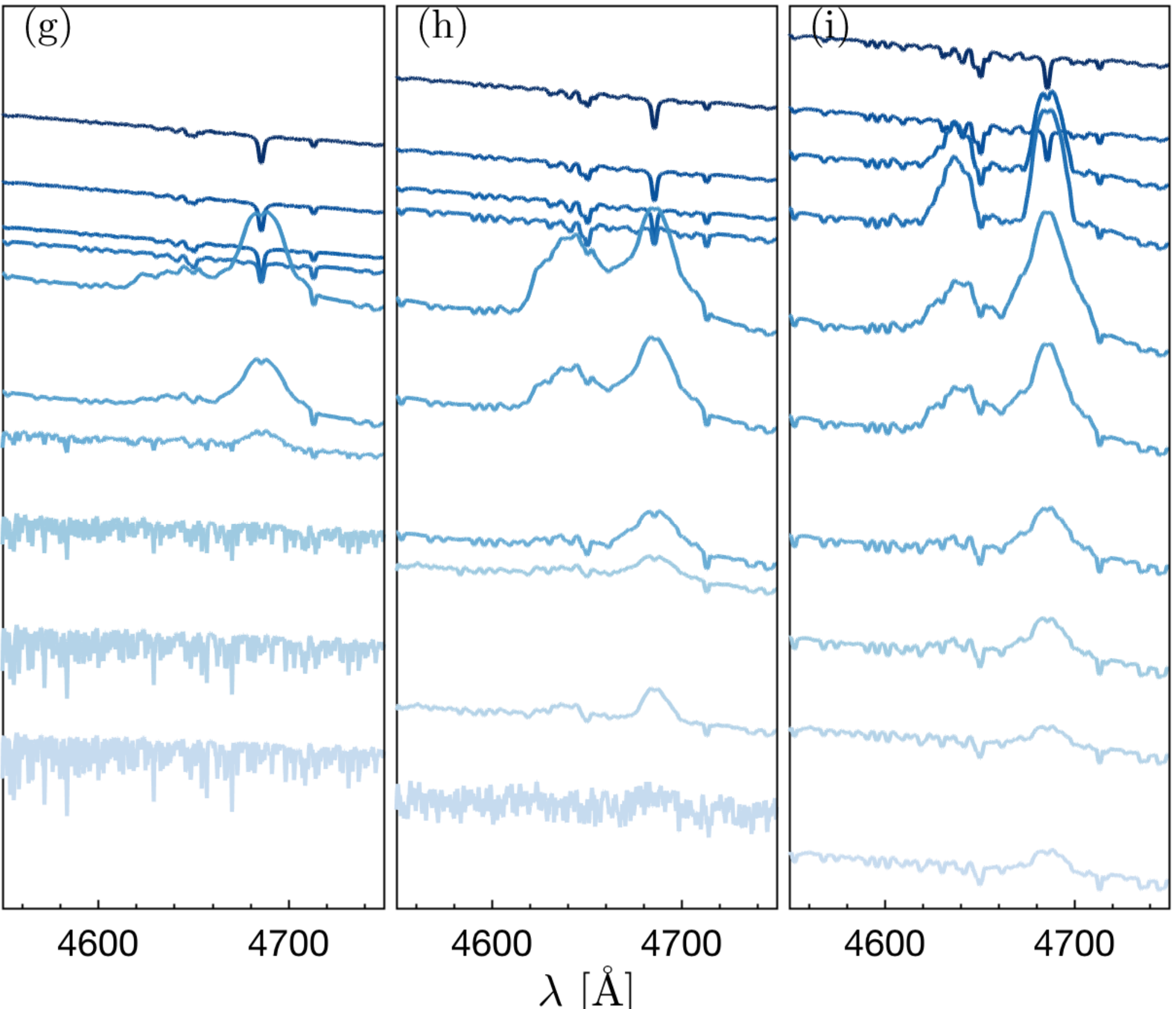}
\includegraphics[scale=0.38, trim={2.2cm 0.cm 22.5cm 0.1cm},clip]{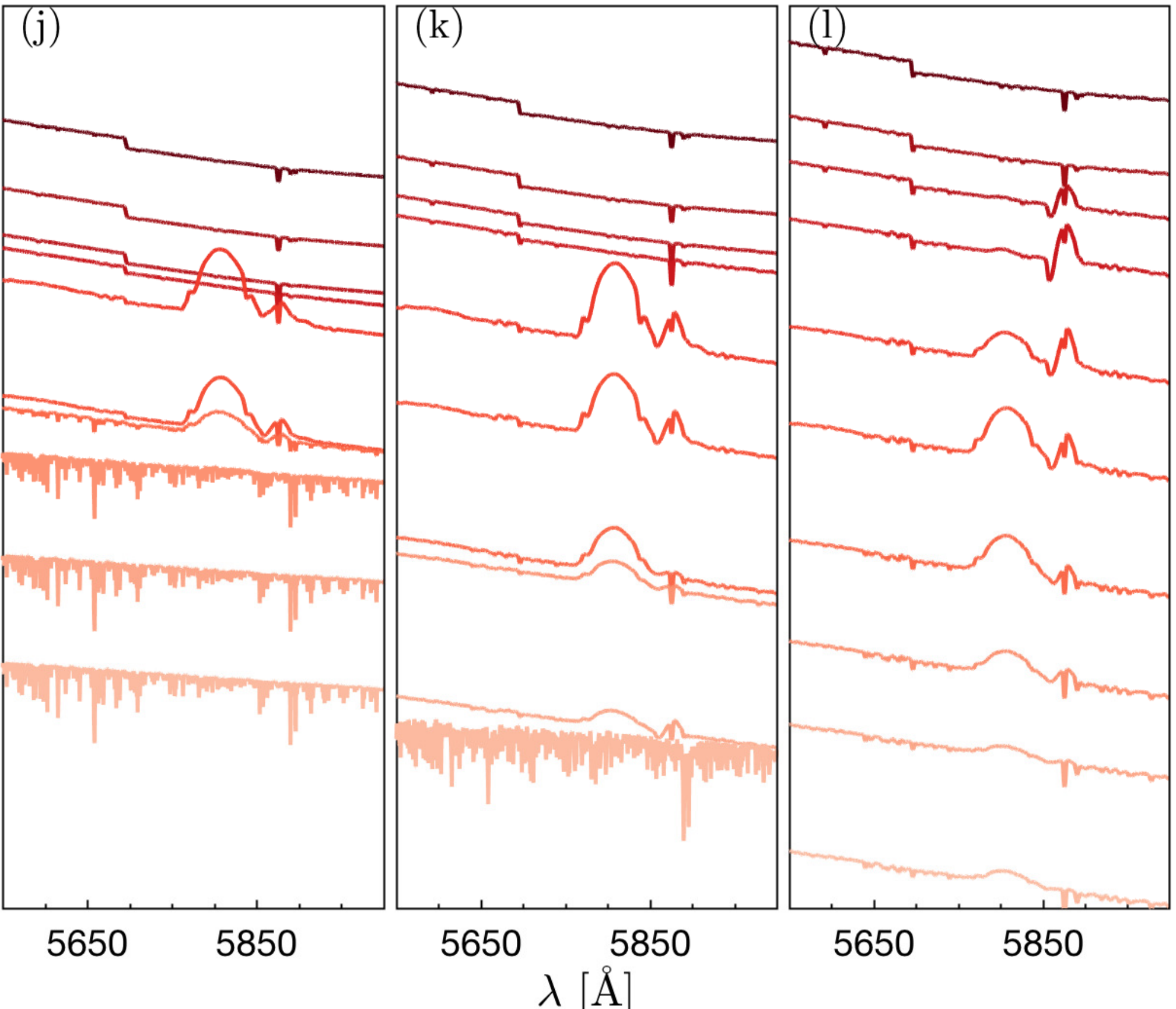}
\caption{The evolution of the blue and red Wolf-Rayet features as a function of stellar metallicity, $\log$ Z$_{\rm{s}}$/Z$_{\odot}$=-0.5 (sub-solar), 0.0 (solar), +0.4 (super-solar), for Geneva high mass loss (a--f) and Parsec (g--l) isochrones. The time increases from the top-to-bottom from 1 to 5.5 Myr, in intervals of 0.5 Myr. The range in y axes is the same for all the blue (and red) sets, and the assumed IMF upper mass cut is 120 M$_{\odot}$. }
\label{fig:evolution_WR_diffZs}
\end{center}
\end{figure*}
\begin{figure*}
\begin{center}
\includegraphics[scale=0.4, trim={2.2cm 0.8cm 4.5cm 0.1cm},clip]{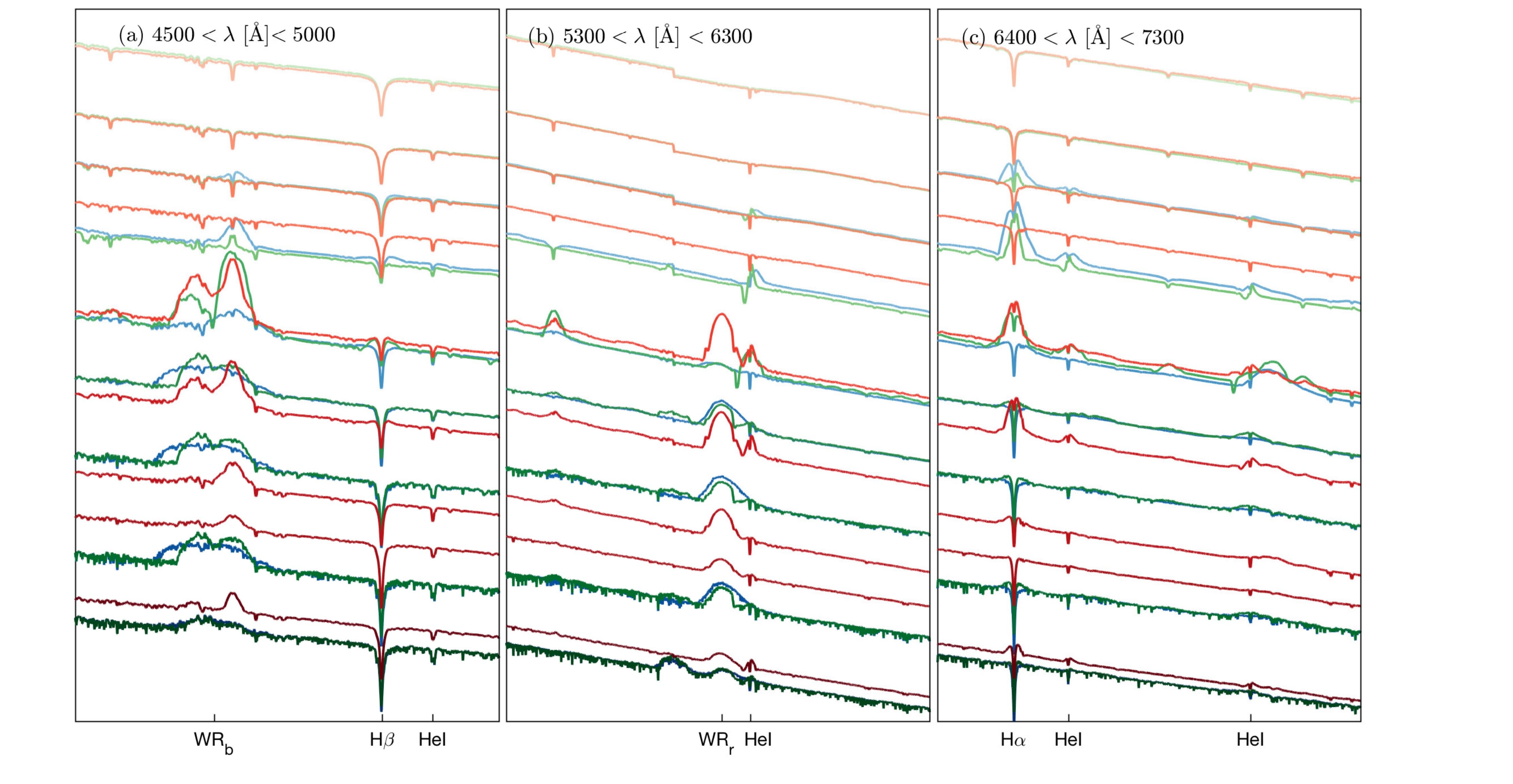}
\caption{The evolution of simple stellar populations of solar stellar metallicity as predicted by Munari stellar library/Geneva isochrones/CMFGEN WR library (in blue), Munari stellar library/Geneva isochrones/PoWR WR library (in green), and Munari stellar library/Parsec isochrones/PoWR WR library combinations. The time runs from 1--5 Myr (top-to-bottom) in intervals of 0.5 Myr. The rest-frame wavelengths of HeI from left-to-right are 4921.9\AA, 5875.5\AA,  6678.3\AA, and 7065.7\AA, and the assumed IMF upper mass cut is 120 M$_{\odot}$. }
\label{fig:evolution_diff_isochrones}
\end{center}
\end{figure*}

The minimum initial stellar mass capable of becoming a WR star (M$_{\rm{min, WR}}$), T$_{\rm{eff}}$, and the surface abundances of H, C, N and O from the isochrones determine whether a star enters into a given WR phase. The M$_{\rm{min, WR}}$ is metallicity dependent. According to single stellar evolutionary theories, at solar metallicity, a star with M$_{\rm{min, WR}}$$\gtrsim$ 20 M$_{\odot}$ may become a late-type WN star if its surface H abundance drops below $0.4$ while retaining a T$_{\rm{eff}}$$>25\,000$ K. At the same metallicity, a M$_{\rm{min, WR}}$ higher than 20 M$_{\odot}$ is required for a star to enter early-type WN or WC phases. These M$_{\rm{min, WR}}$ thresholds progressively increase with decreasing stellar metallicities \citep{Georgy2012, Georgy2015}. For the present analysis, we have assumed a fixed M$_{\rm{min, WR}}$ for a given stellar metallicity following the convention of \textsc{starburst99} \citep{Leitherer99}.  

In Figure\,\ref{fig:evolution_WR_fractions}, we show the T$_{\rm{eff}}$ and R$_t$ based selection of spectra from the PoWR library guided by Geneva high-mass loss (top panels) and Parsec (bottom panels) isochrones of solar metallicity. The size of the symbols indicate relative R$_t$, and the colour code denotes the number of times (as a percentage) a WR spectrum of a given T$_{\rm{eff}}$ and R$_t$ is selected at each time step. At a given T$_{\rm{eff}}$ and age, the WC stars have relatively lower R$_t$ than WN subtypes. On average, the Geneva high-mass loss isochrones lead to the selection of WC- (top-left) and WN- (top-right) subtypes with lower T$_{\rm{eff}}$ than Parsec isochrones. Under Parsec isochrones, the stars enter WC phase at earlier times than Geneva, and in turn, are characterised by higher T$_{\rm{eff}}$ and generally lower R$_{t}$. The WN stars, on the other hand, appear at somewhat earlier times under Geneva than Parsec. Moreover, the Geneva isochrones allow a relatively more (less) varied selection of WC (WN) WR subtypes in the T$_{\rm{eff}}$, R$_t$ and age grid than the Parsec. 

The overall impact of the Geneva versus Parsec selection is illustrated in Figure\,\ref{fig:evolution_WR_diffZs}, which shows the evolutionary predictions for two prominent WR features; the blue ($4550\lesssim\lambda$ [\AA]$\lesssim4750$, left-panels) and red ($5550\lesssim\lambda$ [\AA]$\lesssim6000$, right panels) WR bumps, shown by the same colours in the figure. In panels $a$-$c$ and $g$-$i$, we show the evolution of the blue WR feature as predicted by the combinations of the Geneva high-mass loss isochrones and PoWR, and Parsec and PoWR libraries, respectively, for sub-solar, solar and super-solar stellar metallicities. The variations in the shape of the blue WR feature reflect the differences in WR spectra selected from the PoWR library based on the Geneva and Parsec evolutionary models of the different stellar metallicities considered.   

As evident in Figure\,\ref{fig:evolution_WR_diffZs}, the blue WR feature tend to appear either as two emission peaks or as a single bump. If there are two peaks at approximately $4650$\AA\,and $4686$\AA, then the bluer of the two is typically attributed to the presence WC subtypes, though there can be some contribution from the late-type WN stars. A narrow blue peak indicates the presence of late-type WC stars, while early-type WC stars produce a broad emission peak, which may even extend and dominate over the redder component of the blue WR feature. The redder emission peak of the blue WR feature is attributed to WN subtypes.  A broad redder component signals the presence of early-type WN stars, while a single or two narrow peaks indicates the presence of late-type WN stars \citep{Crowther2007, Sander2012, Todt2015}.   

Similarly, in panels $d$-$f$ and $g$-$l$ of Figure\,\ref{fig:evolution_WR_diffZs}, we show the evolution of the red WR feature, which is a signature of WC stars. The red WR feature consists of three peaks (clearly evident in panel $d$). The bluer of the three is a strong indicator of the presence of late-type WC stars, while the second is associated with early-type WC stars. The third, and the relatively weaker of the three, only appear as a separate peak if the star-forming region has a high number of late-type WC stars. Otherwise, it is usually blended with the second peak \citep{Crowther2007}. So the presence of the red WR feature as three peaks in specific Geneva SSPs and lack thereof altogether in Parsec SSPs points to Geneva isochrones predicting a higher number of WC, particularly late-type WC, stars than Parsec. Note also that Figure\,\ref{fig:evolution_WR_fractions} (a) shows that the Geneva isochrones sample the T$_{\rm{eff}}$,  R$_t$ and age grid of WC stars more broadly than the Parsec isochrones. 

The appearance and duration of the WR features are stellar metallicity dependent. With increasing stellar metallicity, the WR phase progressively extends in duration, the blue and red WR features start to appear at earlier times, and the features become more prominent. The Parsec SSPs, in particular, show this behaviour clearly. The slight decrease in strength of the WR features apparent in the super-solar metallicity SSPs, especially the blue component of the blue WR bump,  is likely a result of the decrease in effective stellar temperatures with increasing metallicity, limiting the number of stars achieving the T$_{\rm{eff}}$ threshold needed to become a WC star. In Geneva SSPs, however, these trends are evident to a lesser degree, likely as a result of the high-mass loss rates of the Geneva isochrones leading to the selection of a large number of late-type WC spectra. Nevertheless, the WR features (in both Geneva and Parsec SSPs) show an overall behaviour with stellar metallicity that qualitatively agrees with the spectroscopic observations of star-forming regions in nearby galaxies \citep[][]{Crowther2007, Neugent2019}. 

In Figure\,\ref{fig:evolution_diff_isochrones}, we compare the SSPs produced using the Geneva HML+CMFGEN, Geneva HML+PoWR, and Parsec+PoWR combinations over three different wavelength windows centred around prominent spectral features. As a result of the low-resolution of CMFGEN spectra \citep{Hillier1998}, the Geneva HML+CMFGEN combination is unable to distinguish different peaks associated with the blue and red WR features, whereas both Geneva HML+PoWR and Parsec+PoWR allow the peaks to be resolved. It is also worth noting the nature of the evolution of the Balmer stellar absorption features -- there is a WR feature centred around the H$\alpha$ wavelength over some of the young ages considered in the figure, effectively decreasing the absorption equivalent width of H$\alpha$ relative to H$\beta$. 

Finally, we have made some modifications to the PoWR library in order to produce WR features that are comparable with observations to-date.  We detail these modifications and their impact on the generated stellar templates in Appendix\,\ref{appA}. Briefly, we remove four spectra, two each, from the PoWR Galactic and LMC WC libraries, respectively, in order to reduce the significant enhancement of the blue component of the red WR bump evident in Figure\,\ref{fig:wr_removal}. This is a feature that has not been observed in the spectra of starbursting regions.  

\subsubsection{Chemical abundance dependance of the ionising continuum}
\begin{figure}
\begin{center}
\includegraphics[scale=0.45, trim={0.cm 0.cm 0.cm 0.cm},clip]{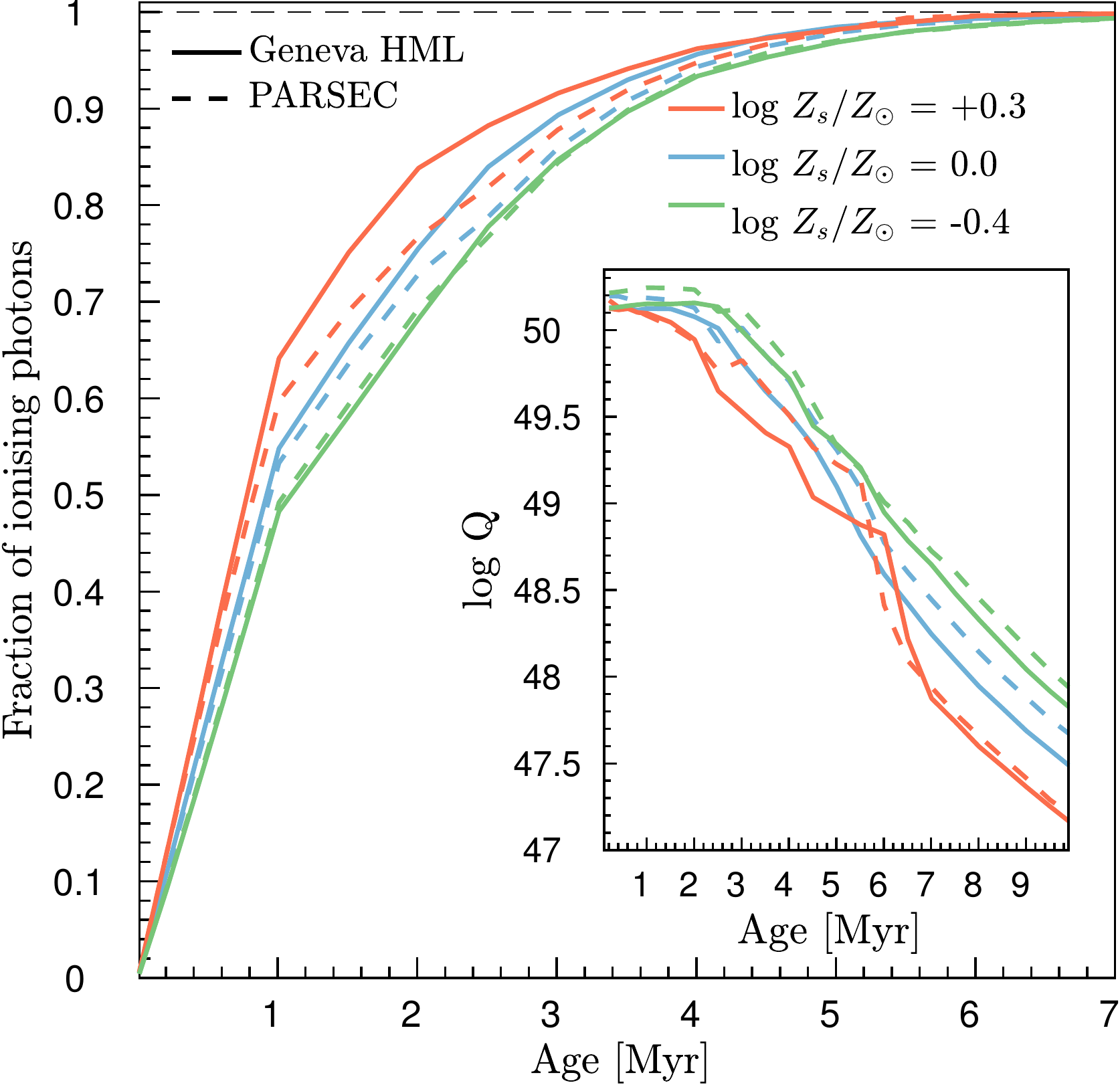}
\caption{Chemical abundance dependence of the ionising continua predicted by Geneva HML versus Parsec isochrones. The cumulative fraction of ionising photons as a function of time and stellar metallicity (main panel), demonstrating that all the ionising photons have essentially been emitted by approximately 6 Myr. The evolution of the ionising photons ($\log Q$) as a function of time for the same three metallicity for a starburst of 3000 M$_{\odot}$ is shown within the inset. The zig-zag pattern, most notable in the super-solar models, highlight the short period over which the effects of the WR stars are prominent. }
\label{fig:evolution_logQ_logU}
\end{center}
\end{figure}

The chemical abundance dependence of the ionising continuum predicted by the Geneva and Parsec isochrones is shown in Figure\,\ref{fig:evolution_logQ_logU}. 
The number of ionising photons emitted is a strong function of time, and within approximately 6 Myr virtually all ionising photons have been emitted. At a fixed age, the cumulative fraction of ionising photons emitted also show a clear dependence on stellar metallicity; the fraction of ionising photons produced increases with increasing metallicity. Between Geneva HML and Parsec isochrones, the Geneva models lead to the production of a higher fraction of ionising photons at a fixed age and metallicity. Within the inset of Figure\,\ref{fig:evolution_logQ_logU}, we show the evolution of $\log Q$ as a function of metallicity.  On average, the higher metallicity models lead to a lower number of ionising photons than lower metallicity counterparts, and between the Geneva HML and Parsec, the Parsec models produce relatively more photons than the Geneva HML. The zig-zag patterns evident in the $\log Q$ evolution, most notable in the super-solar metallicity models, correspond to the ages where the effects of the WR stars are heightened, wherein Parsec models occur at an earlier age.  

\subsection{The Nebular Model}\label{subsec:nebular_model}

We construct the nebular models self-consistently with \textsc{Cloudy} \citep{Ferland2013} using our
modified \textsc{Starburst99} models, with different ages and stellar metallicities, as input. For the chemical composition of the gas, we adopt the solar-scaled abundances, except for Nitrogen, and H/He relation provided in \citet{Dopita2000}. Nitrogen is assumed to be a secondary nucleosynthesis element above metallicities of 0.23 solar. Therefore, to scale Nitrogen with metallicity, we adopt the piecewise empirical relation, again, from \citet{Dopita2000}. The assumption of a solar-scaled composition for the Antennae HII regions is found to be reasonable by \citet{Lardo2015}. Note also that the abundance of the stellar library incorporated into \textsc{Starburst99} is [$\alpha$/Fe]=0.0 (Table\,\ref{tab:munari_vs_us}). So we have consistently used the same abundance pattern for the construction of the stellar and nebular components for the models of the HII regions.

For the construction of the models for HII regions, we assume a simple spherical approximation. While this is not strictly true, for regions dominated young, massive stars and their birth clouds, \citet{Efstathiou2000} and \citet{Siebenmorgen2007}, for example, find a simple spherical approximation to be reasonable. Moreover, in the application of model library to the observed spectra (described in \S\,\ref{sec:method}), we rely on emission line intensity ratios rather absolute luminosities to minimise the effects of the assumption of spherical geometry.   

\begin{figure*}
\begin{center}
\includegraphics[width=0.33\textwidth, trim={1.0cm 1.cm 1.8cm 1.2cm},clip]{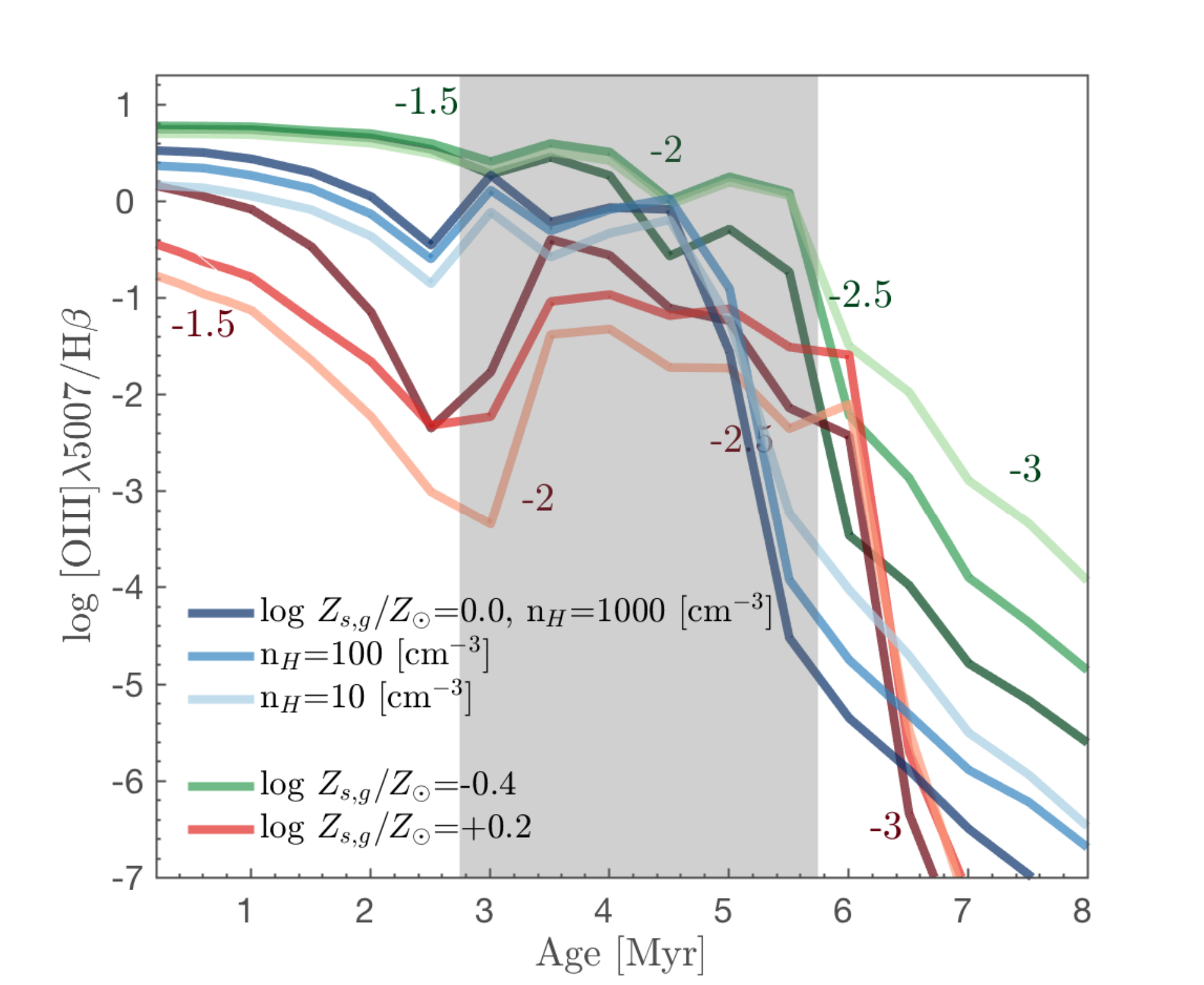}
\includegraphics[width=0.33\textwidth, trim={0.05cm 0.7cm 0.05cm 0.05cm},clip]{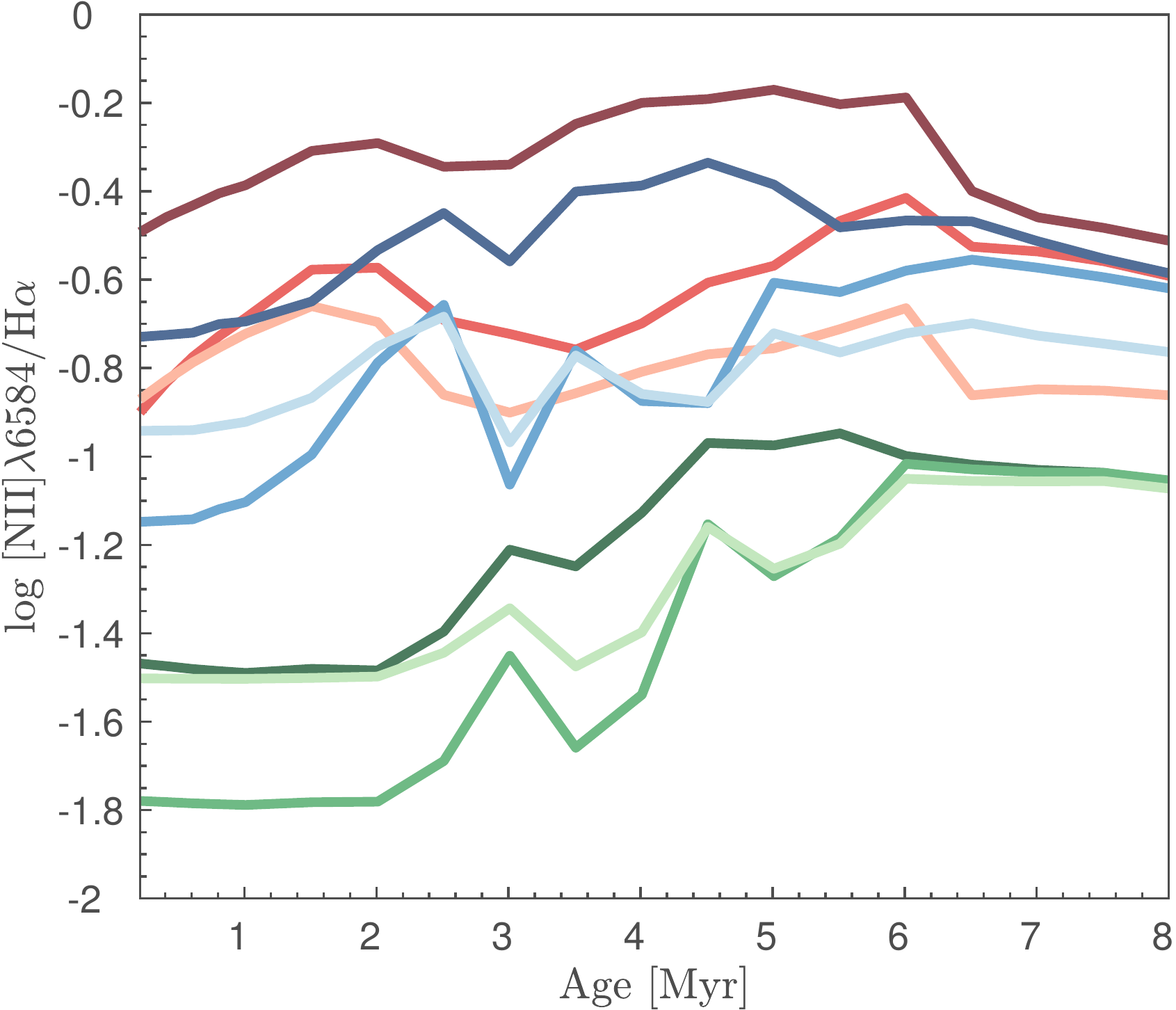}
\includegraphics[width=0.33\textwidth, trim={0.05cm 0.7cm 0.05cm 0.03cm},clip]{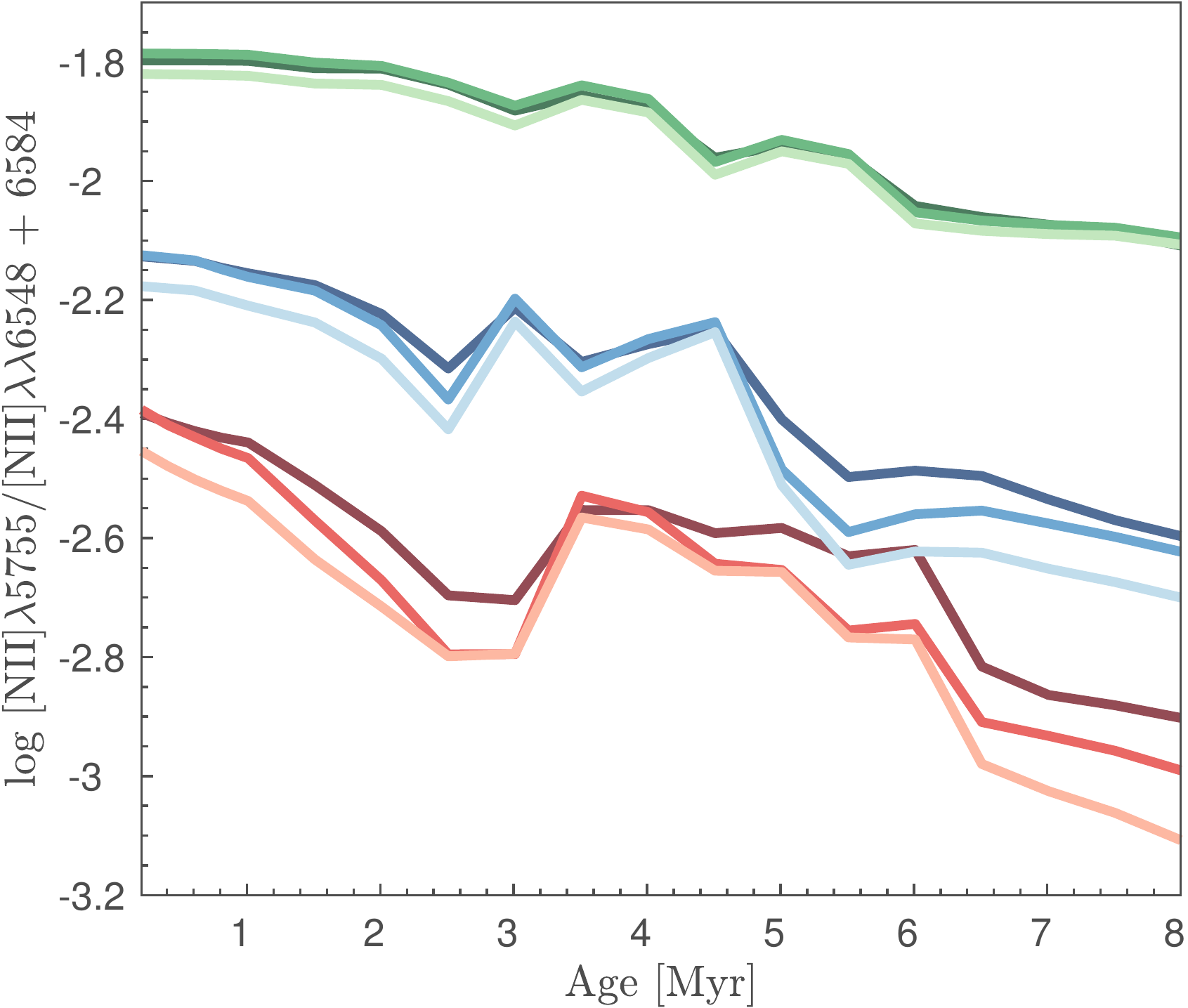}\\
\includegraphics[width=0.33\textwidth, trim={0.05cm 0.05cm 0.05cm 0.03cm},clip]{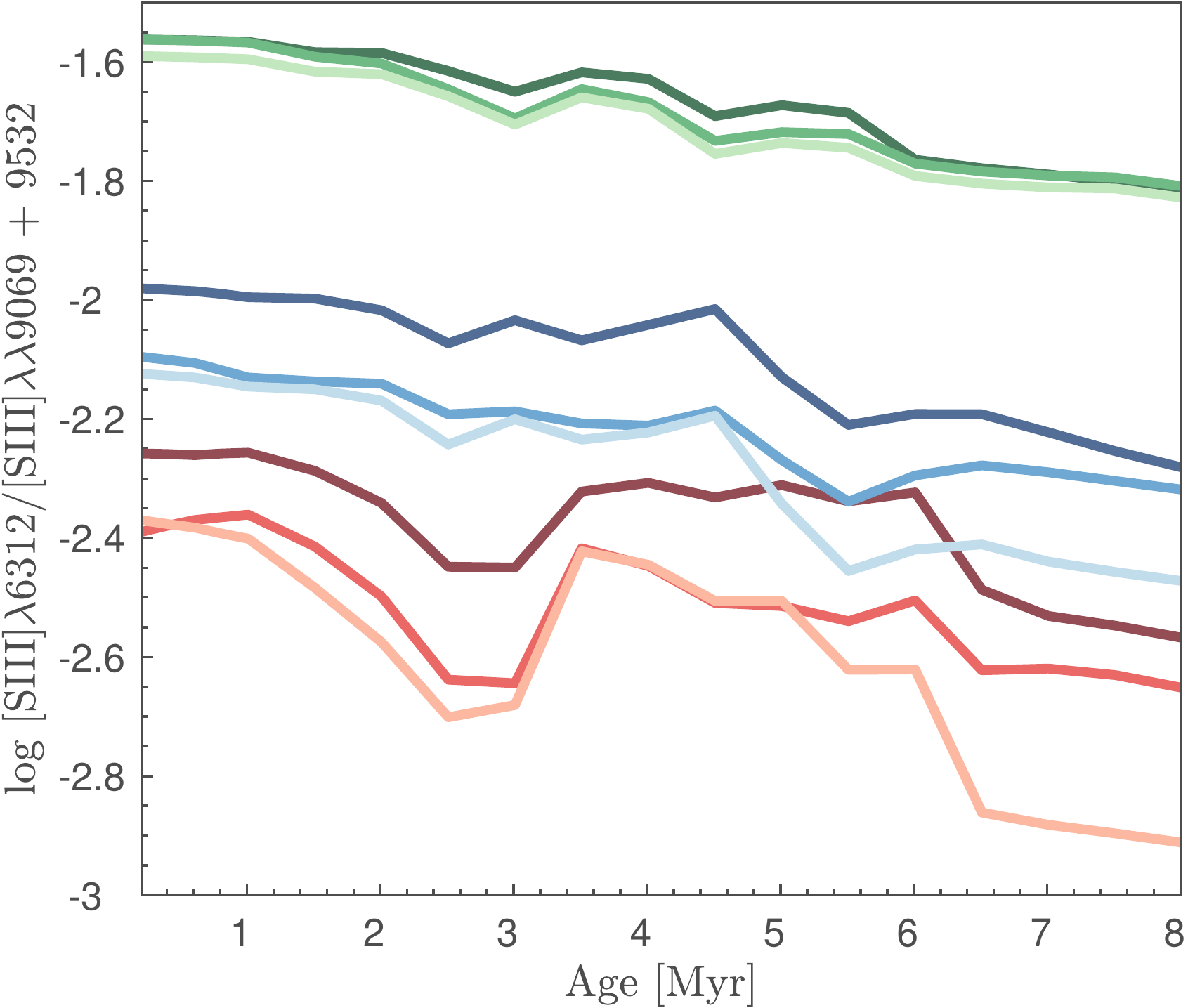}
\includegraphics[width=0.33\textwidth, trim={0.05cm 0.05cm 0.05cm 0.03cm},clip]{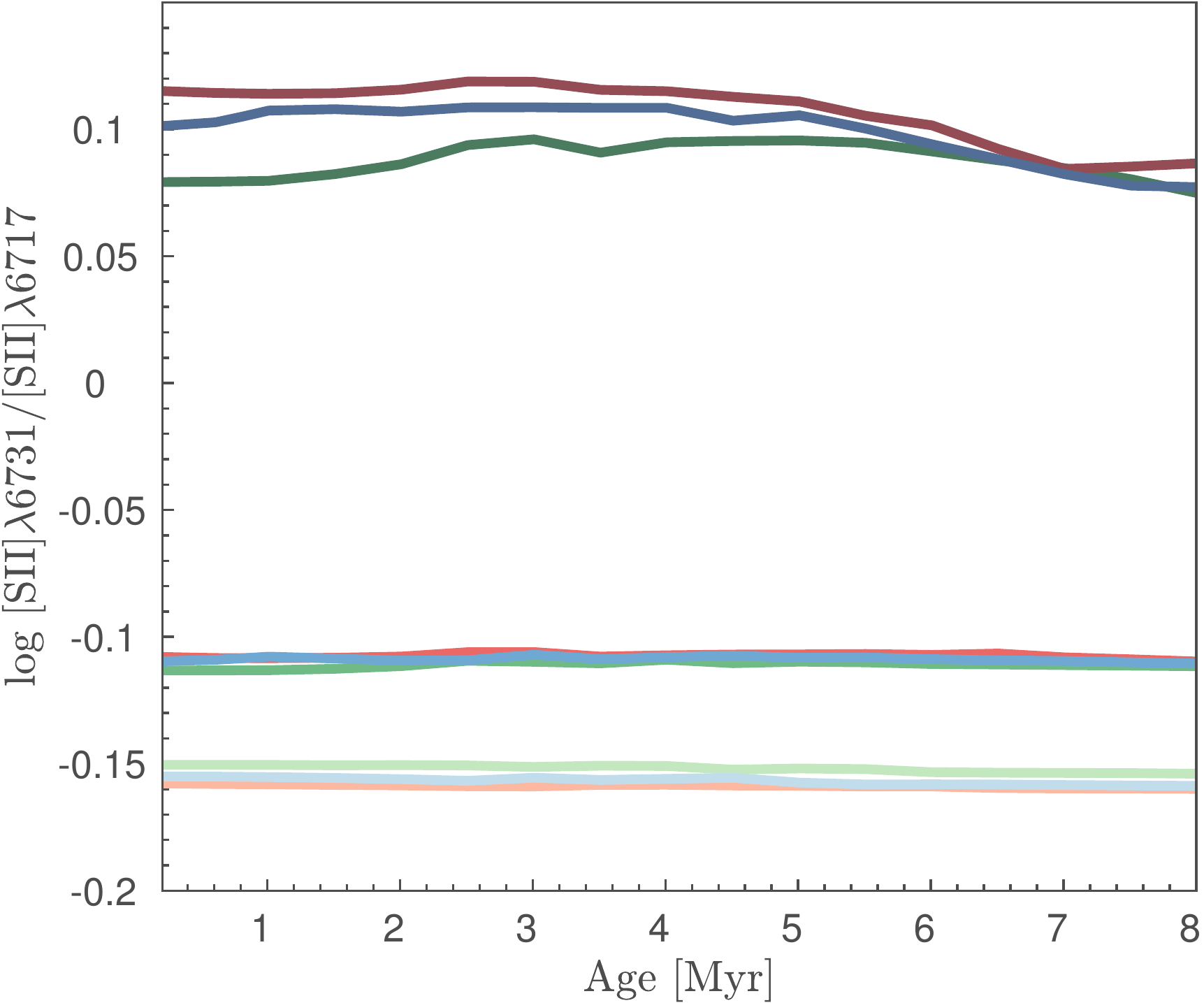}
\includegraphics[width=0.33\textwidth, trim={0.05cm 0.05cm 0.05cm 0.03cm},clip]{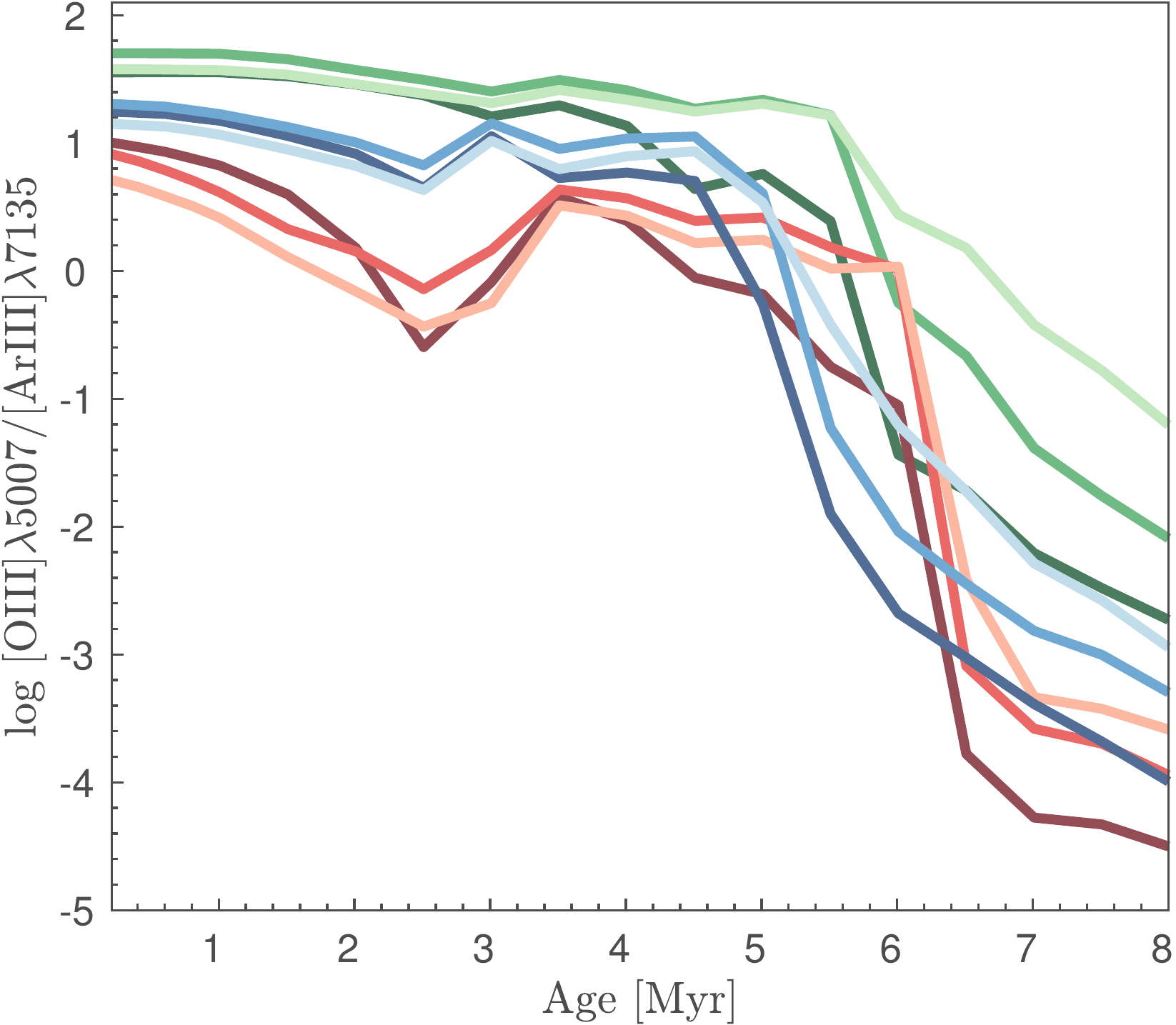}  
\caption{The evolution of different line luminosity ratios as a function of n$_{\rm{H}}$ and metallicity for a burst of star formation of strength 3000 M$_{\odot}$ for a model HII region of a fixed radius. The gas and stellar metallicities are approximately similar. The log U (a function of the burst strength, n$_{\rm{H}}$, metallicity, and time) corresponding to different times (for n$_{\rm{H}}=100$ cm$^{-3}$) is indicated along the sub- (green) and super-solar (red) tracks. The shading in panel (a) denote the approximate (overall across all three metallicities considered, the individual cases show small variations duration of the WR phase) duration of the WR phase.}
\label{fig:evolution_line_ratios}
\end{center}
\end{figure*}

\subsubsection{Control of the ionisation parameter}\label{subsubsec:controlOfU}

In \textsc{Cloudy} modelling, the intensity of the ionising continuum is set by the ionisation parameter, $U$, parameterised as the ratio of hydrogen ionising photons ($Q_{\rm{H}}$) to the total hydrogen density ($n_{\rm{H}}$), 
\begin{equation}
U = \frac{Q_H}{4\pi R^2\times n_H\times c}, 
\label{eq:ionisation_relation}
\end{equation}
where $R$ is the radius of the ionised region, and c is the speed of light. 

To produce starbursts of various strengths, for each SSP, we consider \textsc{Cloudy} models for different $U$ values in the range $-4.5\leq\log U\leq-0.5$ for different $n_{\rm{H}}$ in the $10\leq n_{\rm{H}}$[cm$^{-3}$] $\leq1000$ and gas metallicity ($Z_{\rm{g}}$) in the $-1.0\leq\log Z_{\rm{g}}/Z_{\odot}\leq+0.5$ ranges assuming a spherical ionised region of fixed $R$. 
The evolution of a number of quintessential nebular line ratios under different metallicity and density conditions for a starburst of 3000 M$_{\odot}$ is shown in Figure\,\ref{fig:evolution_line_ratios}. 

The diagnostics that have been most commonly used to classify the nature of a nebular excitation are  [\ion{O}{iii}]${\lambda}$5007/H$\beta$, [\ion{N}{ii}]${\lambda}$6584/H$\alpha$, [\ion{S}{ii}]${\lambda\lambda}$6717,31/H$\alpha$ \citep{Veilleux1987}, as well as line intensity ratios based on  [\ion{S}{iii}]$\lambda$6312/[\ion{S}{iii}]$\lambda$9069, [\ion{N}{ii}]$\lambda$5755/[\ion{N}{ii}]$\lambda$6548,84 etc. Those that use nebular lines that are close together in wavelength are generally preferred as they allow uncertainties related to reddening corrections to be minimised. For a fixed starburst, the diagnostics based on higher ionisation species (e.g.\,[\ion{O}{iii}]$\rm{\lambda}$5007/H$\beta$) show a swifter decline with time than their lower ionisation counterparts. For a fixed age, the $U$ of higher metallicity models is lower than that of lower metallicities, and likewise, the predicted ratios. The zig-zag pattern visible in the evolutionary tracks, particularly prominent in super-solar metallicity models, is caused by the brief hardening of the stellar radiation field resulting from the appearance of the Wolf-Rayet stars. The ageing tracks with low-to-high $n_{\rm{H}}$ show a low-to-high variation in line ratios, except the [\ion{O}{iii}]$\rm{\lambda}$5007/H$\beta$ track that shows the opposite trend for ages $\gtrsim5$ [Myr]. Note, however, that the predicted [\ion{O}{iii}]$\rm{\lambda}$5007/H$\beta$ for old ages is too faint to be detected observationally. 
\begin{figure}
\begin{center}
\includegraphics[width=0.45\textwidth, trim={0.05cm 0.05cm 0.01cm 0.01cm},clip]{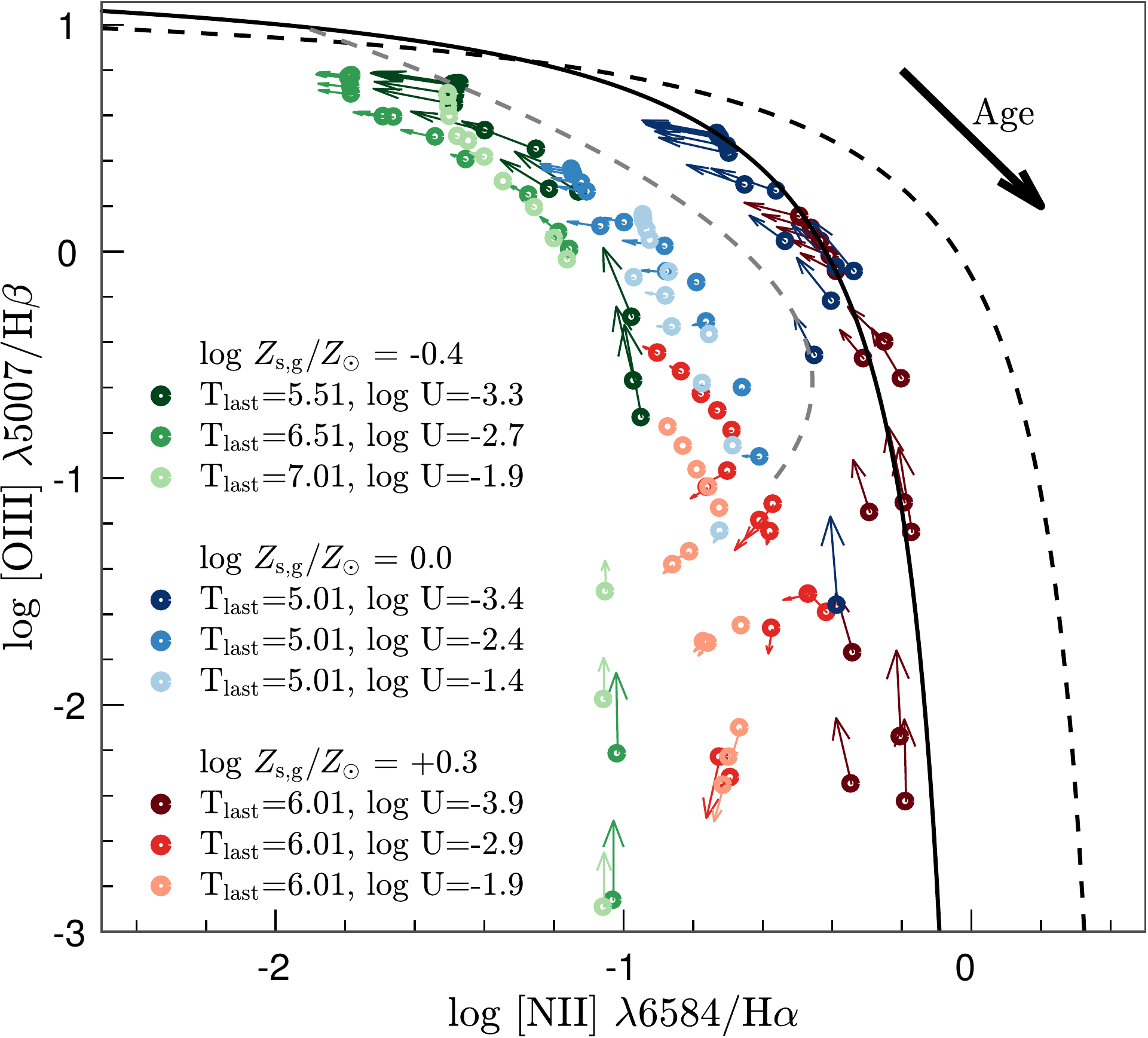}
\includegraphics[width=0.49\textwidth, trim={0.0cm 0.0cm 0.0cm 0.0cm},clip]{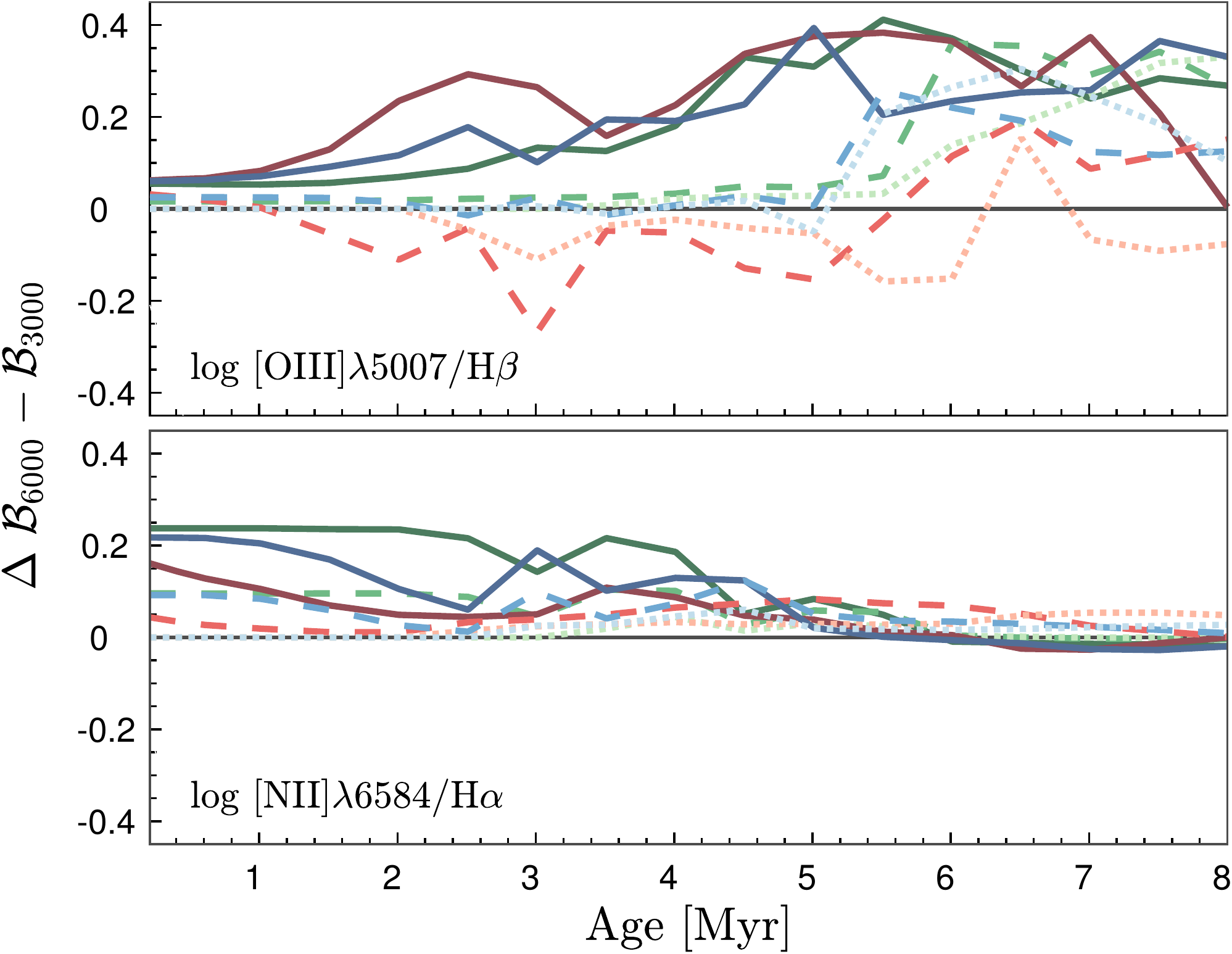}
\caption{\textit{Top panel}: BPT diagnostics from \textsc{cloudy} modelling of a starburst of 3000 M$_{\odot}$. The line ratios evolve in the direction shown by the black 'age' arrow. The different colours correspond to different metallicities, while different shadings, from dark-to-light, of the same colour, denote $n_{\rm{H}}$=1000, 100, 10 cm$^{-3}$ (associated log $U$ is shown in the legend), respectively. The legend shows the last age and log U of the data point of a given ($Z$, $n_{\rm{H}}$) still visible in the parameter space shown before moving off the diagram. The vectors show the effect of doubling the burst strength (i.e.\,from 3000 to 6000 M$_{\odot}$). The black dashed and solid lines show the \citet{Kewley2001} and \citet{Kauffmann2003} demarcations, respectively, and the dashed grey line denotes the fit to the SDSS data from \citet{Brinchmann2008}. \textit{Bottom panel}: The difference in [\ion{O}{iii}]~$\rm{\lambda}$5007/H$\beta$ and [\ion{N}{ii}]~$\rm{\lambda}$6584/H$\alpha$ between the 3000 and 6000 M$_{\odot}$ starburst as a function of time. The colours are the same as in the top panel, and in addition, the solid, dashed, and dotted lines correspond to $n_{\rm{H}}$=1000, 100, 10 cm$^{-3}$, respectively. Note that even though at earlier ages the $n_{\rm{H}}$=1000 cm$^{-3}$ tracks appear to show a progressively larger enhancement in [\ion{O}{iii}]~$\rm{\lambda}$5007/H$\beta$ with increasing metallicity, at a fixed [\ion{O}{iii}]~$\rm{\lambda}$5007/H$\beta$, the [\ion{O}{iii}]~$\rm{\lambda}$5007/H$\beta$ increases with decreasing metallicity as expected given the T$_{\rm{e}}$ dependence on metallicity. Note that the burst strengths explored here are for model HII regions of a fixed radius. }
\label{fig:evolution_in_BPT}
\end{center}
\end{figure}

The evolutionary predictions in the BPT \citep[][]{Baldwin1981} plane under different metallicity and $n_{\rm{H}}$ conditions is shown in the top panel of Figure\,\ref{fig:evolution_in_BPT}. The black solid and dashed lines denote the \citet{Kewley2001} and \citet{Kauffmann2003} demarcations separating active galactic nuclei from HII regions, with the grey dashed line indicating the excitation sequence determined by \citet{Brinchmann2008} from a sample of nearby star forming galaxies. For each track, the age increases approximately diagonally from 0.5 Myr (upper left) to T$_{\rm{last}}$, with T$_{\rm{last}}$ being the age of the last point visible in the parameter space of the figure. 

The effect of raising $n_{\rm{H}}$ is to enhance both the [\ion{O}{iii}]~$\rm{\lambda}$5007/H$\beta$ and [\ion{N}{ii}]~$\rm{\lambda}$6584/H$\alpha$ due to the increased rate of collisional excitation. At a fixed $n_{\rm{H}}$, the decline in log $U$ with increasing metallicity pushes the BPT predictions towards lower [\ion{O}{iii}]~$\rm{\lambda}$5007/H$\beta$ and higher [\ion{N}{ii}]~$\rm{\lambda}$6584/H$\alpha$ values. The low (high) metallicity and low (high) $n_{\rm{H}}$ predictions form the left (right) edge of the sequence, suggesting that it is unlikely to observe a low metallicity HII region with a low $U$ value. 

The vectors shown in Figure\,\ref{fig:evolution_in_BPT} highlight the effect of doubling the strength of the starburst with each vector indicating the magnitude and the direction of the movement of a given point. As a consequence of the increase in log $U$, the effect in large is to enhance the [\ion{O}{iii}]~$\rm{\lambda}$5007/H$\beta$ and decrease the [\ion{N}{ii}]~$\rm{\lambda}$6584/H$\alpha$. At high metallicities, however, the [\ion{O}{iii}]~$\rm{\lambda}$5007/H$\beta$ corresponding to n$_{\rm{H}}$=100 and 10 cm$^{-3}$ show a decrease with increasing log $U$. This is further illustrated in the bottom panel of Figure\,\ref{fig:evolution_in_BPT}, where we quantify the change in [\ion{O}{iii}]~$\rm{\lambda}$5007/H$\beta$ and [\ion{N}{ii}]~$\rm{\lambda}$6584/H$\alpha$ from 3000 to 6000 M$_{\odot}$ as a function of time. This behaviour of [\ion{O}{iii}]~$\rm{\lambda}$5007/H$\beta$ and  [\ion{N}{ii}]~$\rm{\lambda}$6584/H$\alpha$ can be explained by the relationship between log $U$, electron temperature (T$_{\rm{e}}$), metallicity and $n_{\rm{H}}$. 

\begin{figure*}
\begin{center}
\includegraphics[width=0.93\textwidth, trim={0.4cm 0.4cm 3.5cm 2.3cm},clip]{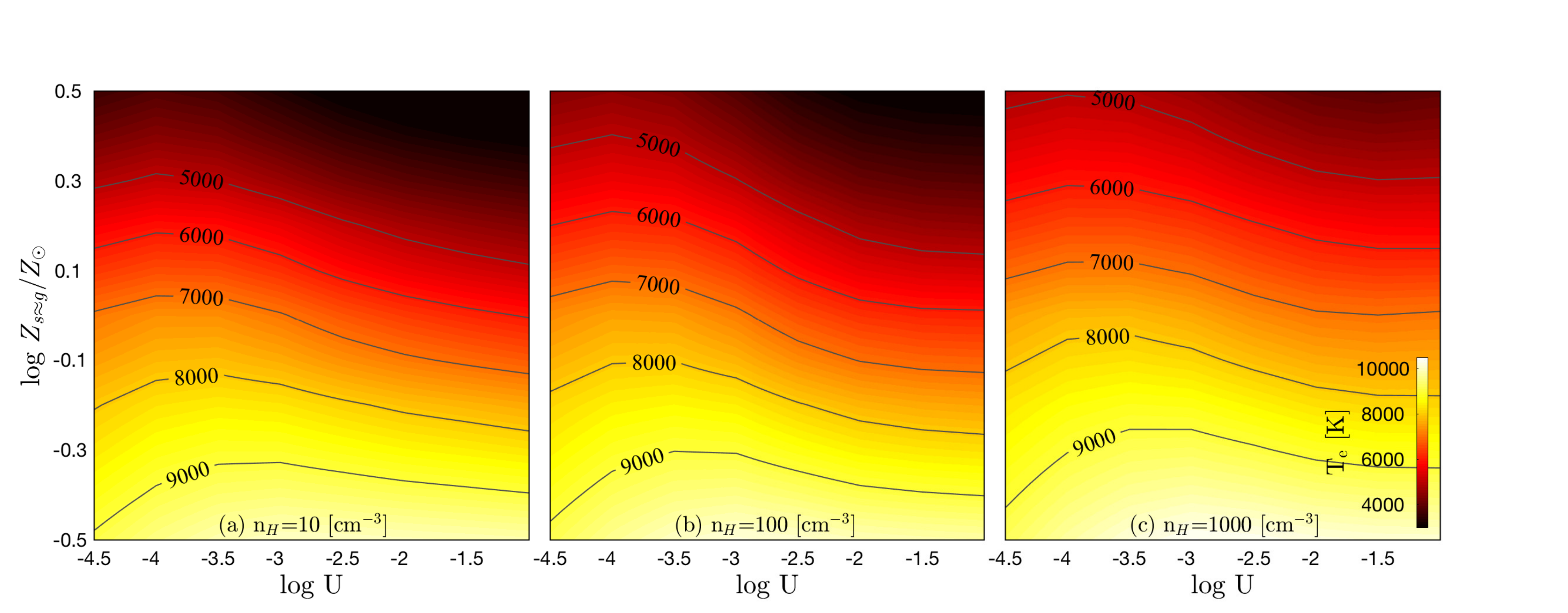}
\caption{The dependence of the volume-averaged T$_{\rm{e}}$ on metallicity and log $U$ for n$_H$=10, 100, 1000 [cm$^{-3}$] for a stellar population of aged 3 Myr. The contours denote constant T$_{\rm{e}}$ values. }
\label{fig:Te_dependence_onZ_logU_nH}
\end{center}
\end{figure*}

In Figure\,\ref{fig:Te_dependence_onZ_logU_nH}, we show the  \textsc{Cloudy} predictions of the volume-averaged T$_{\rm{e}}$ dependence on metallicity and log $U$ for n$_H=$10, 100, 1000 cm$^{-3}$, assuming that the metallicity of stars is similar to that of the star-forming gas from which they were formed.  At a fixed metallicity, T$_{\rm{e}}$ varies relatively weakly with both log $U$ and n$_H$, whereas, at a fixed log $U$ and n$_H$, T$_{\rm{e}}$ shows a significant dependence on metallicity. On average, T$_{\rm{e}}$ increases with increasing log $U$ at low metallicities and decreases with increasing log $U$ at high metallicities. T$_{\rm{e}}$ also increases with increasing $n_{\rm{H}}$ at all metallicities, relatively more steeply at high metallicities than at low metallicities. In comparison to the T$_{\rm{e}}$ dependence on metallicity, however, its dependence on n$_H$ is relatively weak. 

The T$_{\rm{e}}$-metallicity-log $U$ relationship is primarily driven by the cooling efficiency in an HII region, which is largely a function of gas metallicity. At very low log $Z_g/Z_{\odot}$ (e.g.\,$\lesssim-1$), the contribution of collisionaly excited metal lines to cooling is negligible, thus T$_{\rm{e}}$ largely increases as a function of log $U$.  At high log $Z_g/Z_{\odot}$, however, the cooling efficiency increases as a function of log $U$, resulting in a decrease in T$_{\rm{e}}$ as a function of increasing log $U$. As the cooling efficiency is, on average, an inverse function of density \footnote{This is partly a result of many of the strong coolants been suppressed above their critical density} ($n_{\rm{H}}$), T$_{\rm{e}}$ increases with increasing gas density \citep{Ferland1999, Byler2017}.
 
 As such, T$_{\rm{e}}$ does not always increase as a function of log $U$. This means that even though doubling the strength of a starburst increases log $U$, it leads to a decrease in T$_{\rm{e}}$ at high metallicities. At $n_{\rm{H}}$ of 100 cm$^{-3}$, in particular, an increase in log $U$ corresponds to a larger decline in T$_{\rm{e}}$ (i.e.\,the constant T$_{\rm{e}}$ contours are closer together at high metallicities than at low metallicities, Figure\,\ref{fig:Te_dependence_onZ_logU_nH} middle panel). Given the sensitivity of [\ion{O}{iii}] to T$_{\rm{e}}$, the decline in T$_{\rm{e}}$ with increasing log $U$ (at high metallicities) and $n_{\rm{H}}$=100 cm$^{-3}$ leads to a decrement in [\ion{O}{iii}]~$\rm{\lambda}$5007/H$\beta$. The decline in T$_{\rm{e}}$ with increasing log $U$ is still high enough at $n_{\rm{H}}$=10 cm$^{-3}$ to cause a decrease in [\ion{O}{iii}]~$\rm{\lambda}$5007/H$\beta$, albeit smaller in comparison to $n_{\rm{H}}$=100 cm$^{-3}$ (Figure\,\ref{fig:evolution_in_BPT} bottom panel). At $n_{\rm{H}}$=1000 cm$^{-3}$, the difference in T$_{\rm{e}}$ caused by doubling the burst strength is sufficiently small as the T$_{\rm{e}}$ contours are spanned out (Figure\,\ref{fig:Te_dependence_onZ_logU_nH} right panel), resulting in an enhancement in [\ion{O}{iii}]~$\rm{\lambda}$5007/H$\beta$ with increasing log $U$ as expected. 

\subsubsection{The importance of the nebular continuum}
 \begin{figure*}
\begin{center}
\includegraphics[scale=0.48, trim={2.0cm 11.9cm 2.1cm 1.6cm},clip]{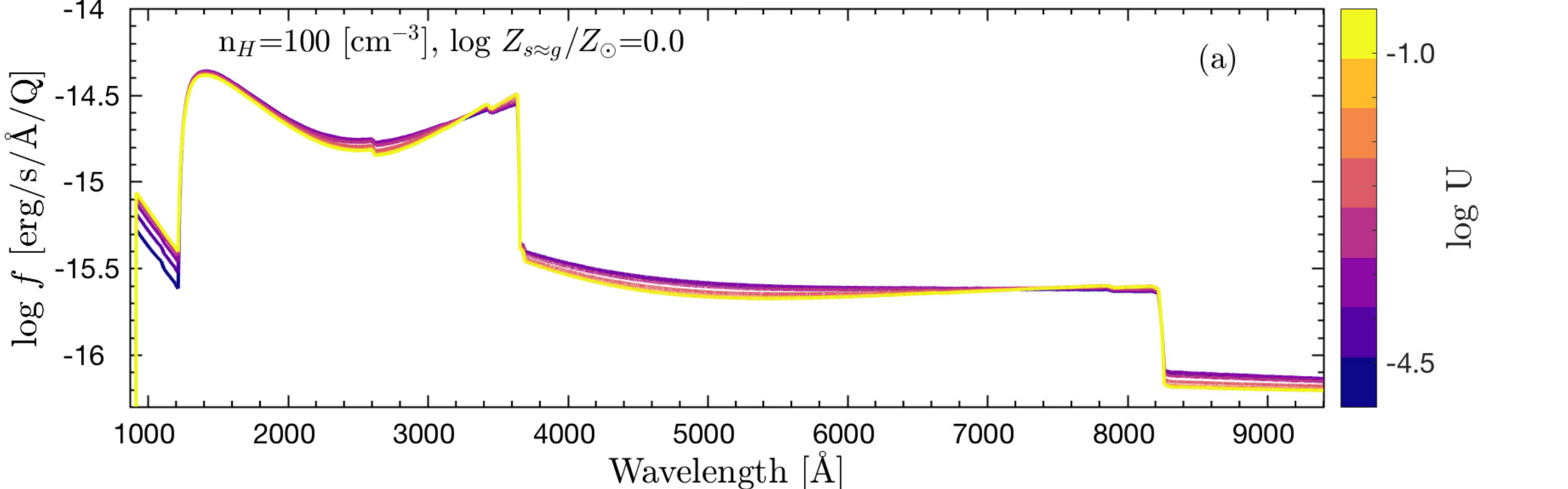}
\includegraphics[scale=0.48, trim={2.0cm 11.9cm 2.1cm 1.6cm},clip]{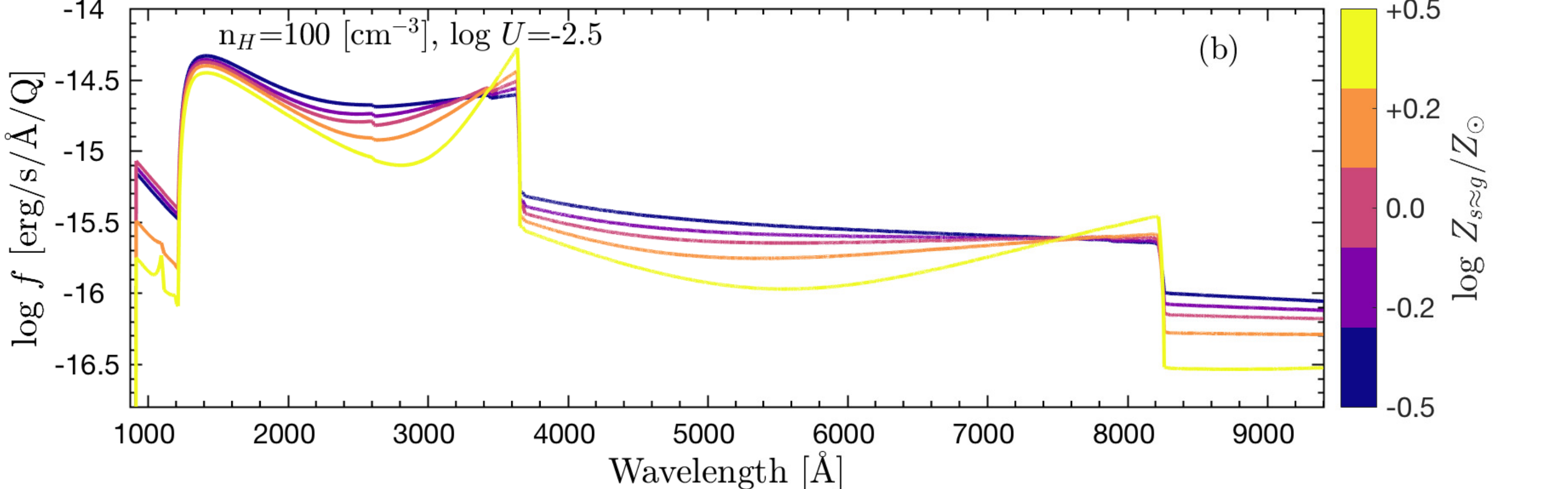}
\includegraphics[scale=0.48, trim={2.0cm 11.9cm 2.1cm 1.6cm},clip]{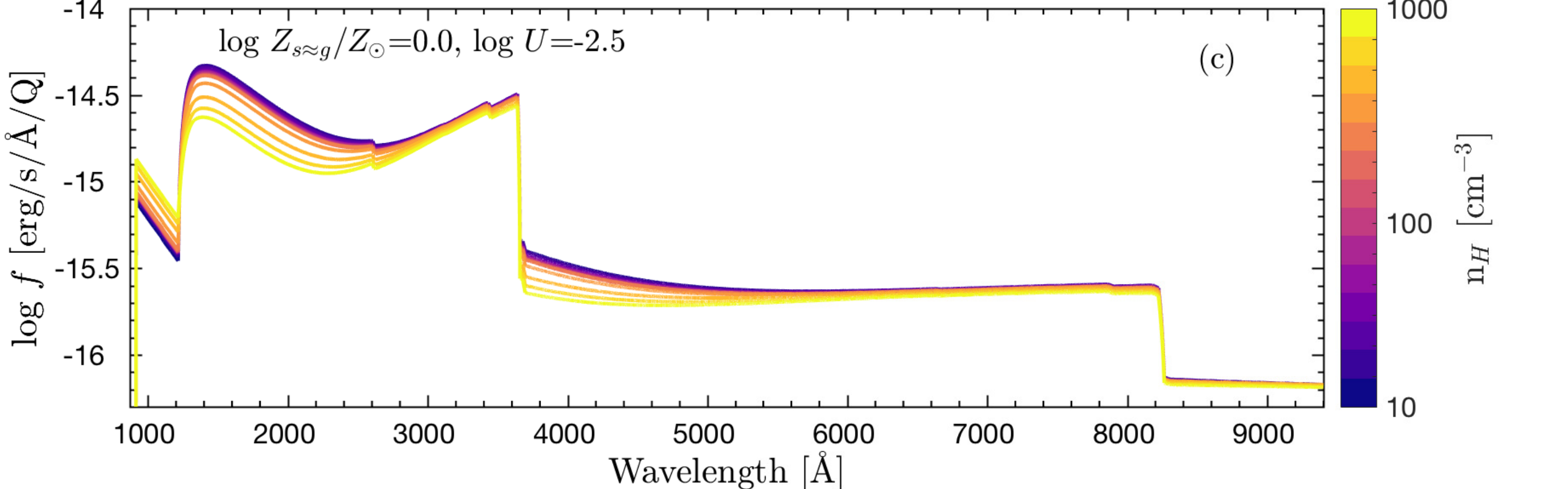}
\caption{The behaviour of the nebular continuum of a stellar population aged 3 Myr with respect to three different properties. (a) The nebular continuum as a function of the ionisation parameter ($U$) at n$_H$=100 cm$^{-3}$ and log $Z_{s\approx g}/Z_{\odot}$ of 0.0 ($s\approx g$ means that the stellar metallicity is assumed to be the same as the gas metallicity). While the shape of the nebular continuum shows a little dependence on log $U$, the absolute normalisation depends strongly on the number ionising photons ($Q_{\rm{H}}$). Since we show the nebular continuum in the unit of erg s$^{-1}$ \AA$^{-1}$ Q$^{-1}$, the absolute normalisation dependence on the number of ionising photons has been removed. (b) The nebular continuum as a function of log $Z_{s\approx g}/Z_{\odot}$ at n$_H$=100 cm$^{-3}$ and log $U$=-2.5. The shape of the nebular continuum is primarily informed by the free-bound and free-free continua dependence on T$_{\rm{e}}$. (c) The nebular continuum as a function of n$_H$ at log $Z_{s\approx g}/Z_{\odot}$ of solar and log $U$=-2.5. The effects of n$_H$ is largely confined to the bluer wavelengths, highlighting the dependence of the 2$\gamma$ continuum on n$_H$. }
\label{fig:cloudy_nebularC_dependence_onU_ZgZs_nH}
\end{center}
\end{figure*}
\begin{figure*}
\begin{center}
\includegraphics[scale=0.5, trim={2.5cm 3.8cm 2.0cm 3.8cm},clip]{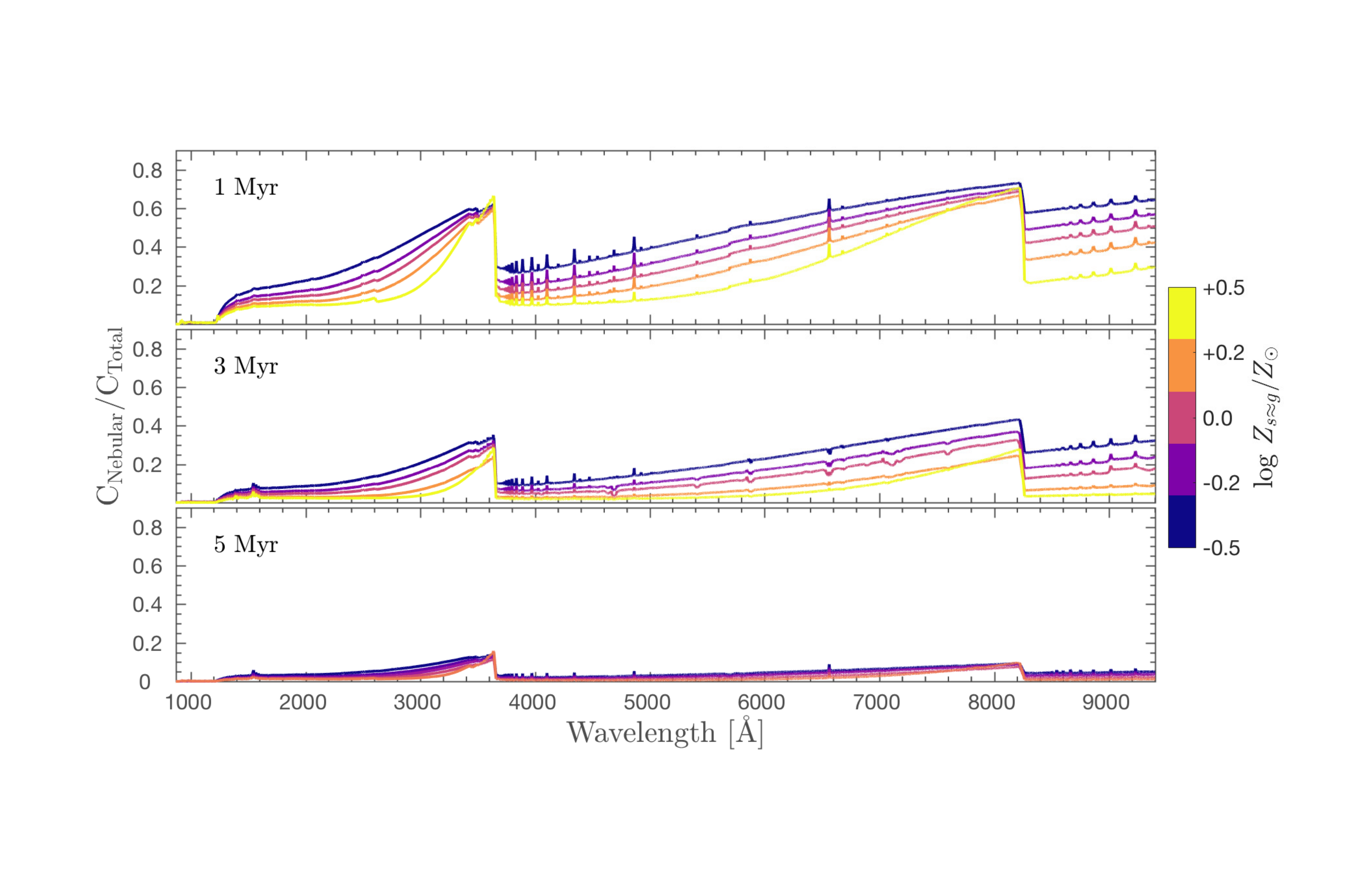}
\caption{The ratio of nebular continuum to total (stellar + nebular) continuum as a function of metallicity for fixed log $U$ of -2.5 and $n_{\rm{H}}$ of 100 cm$^{-3}$, and  for SSPs of aged 3 and 5 Myr. Five different metallicities are considered, where the metallicity of stars is assumed to be similar to that of gas from which they are formed. }
\label{fig:nebular_to_total}
\end{center}
\end{figure*}

The continuous emission spectrum produced by the ionised gas can contribute significantly to the observed spectrum. The total nebular continuum is dominated by the emission from the free-bound recombination processes of hydrogen and helium ions and electrons, free-free (Bremsstrahlung) transitions in the Coulomb fields of H$^+$, He$^+$ and He$^{2+}$ and two-photon (bound-bound) decay of the 2$^2$S$_{1/2}$ level of HI and HeII, and 2$^1$S$_0$ level of HeI, which is relatively less important \citep{Ercolano2006}.

The free-bound continuum shows discrete jumps due to the discrete nature of the lower energy levels at the ionisation energy, followed by continuous emission to higher energies. It is also a dominant contributor to the total nebular continuum over optical wavelengths, and, as a product of radiative recombination processes, its absolute intensity depends strongly on the ionisation parameter, with the shape only showing a small dependence (Figure\,\ref{fig:cloudy_nebularC_dependence_onU_ZgZs_nH}a).
The free-free continuum is power-law like ($\propto$$\nu^{-2}$) in shape and progressively becomes a dominant contributor to the total nebular continuum towards the infrared wavelengths. The total energy radiated increases as the electron temperature (T$_{\rm{e}}$) is raised. 

The dependence of the nebular continuum on metallicity, again highlighting the underlying dependence on T$_{\rm{e}}$, is shown in Figure\,\ref{fig:cloudy_nebularC_dependence_onU_ZgZs_nH}b. The lower metallicity models (higher T$_{\rm{e}}$) show higher continuum emission, and lower magnitudes for discontinuities at 3646\AA\,and 8207\AA\,than the higher metallicity models. The free-bound continuum contribution to the total nebular continuum dominates at higher metallicities (lower T$_{\rm{e}}$), producing a total continuum with more pronounced saw-tooth like jumps at 3646\AA\,and 8207\AA. With increasing T$_{\rm{e}}$, the free-free continuum becomes progressively more significant, and the power-law like free-free continuum starts to dilute the saw-tooth structure of the free-bound continuum. As the relative contribution of the free-free contribution at infrared wavelengths is higher than at bluer wavelengths, the amplitudes of the recombination edges at longer wavelengths will be more affected.      

The two-photon (or 2$\gamma$) continuum is a result of the radiative decay from the 2$^2$S to1$^2$S level of hydrogenic species (HI and HeII). A radiative decay to the 1$^2$S level is strongly forbidden, however, a transition can take place if two photons with sum of energies equal to that of Ly$\alpha$ is emitted. The decay of 2$^1$S level of HeI also produces, albeit a weaker, two-photon continuum \citep{Nussbaumer1984, Drake1986, Draine2011}. Hence the 2$\gamma$ continuum has a peak at half the Ly$\alpha$ frequency and a natural cut-off at the Ly$\alpha$ frequency, highlighting the importance of the 2$\gamma$ continuum for near-, and far-ultraviolet observations. 

The density of the plasma can have a significant impact on the 2$\gamma$ continuum. As the radiative lifetime of the 2$^2$S level is long, the 2$^2$S level can be de-populated through angular momentum changing collisions (2$^2$S$\rightarrow$2$^2$P) with other ions and electrons before a two-photon decay can occur. The radiative decay to 1$^2$S level from 2$^2$P level may then occur by emitting a Ly$\alpha$ photon. Therefore, while at low densities, the 2$\gamma$ continuum dominates the free-bound continuum, it can be suppressed in HII regions with high densities ($\sim10^5$ cm$^{-3}$) \citep{Aller1984}. The dependence of the 2$\gamma$ continuum on n$_H$ is shown in Figure\,\ref{fig:cloudy_nebularC_dependence_onU_ZgZs_nH}c. The `bump' at $\sim$1500\AA\,is primarily caused by the 2$\gamma$ continuum, with lower n$_H$ models showing higher continuum emission over the UV wavelength regime than their higher density counterparts.  

In young star forming regions, the nebular continuum can dominate the total continuum, especially around the wavelengths of the Balmer and Paschen discontinuities. The ratio of the nebular to total (nebular + stellar) continuum with respect to different metallicities for stellar populations of 1, 3, and 5 Myr age is shown in Figure\,\ref{fig:nebular_to_total}. For stellar populations of ages between 1 to 4 Myrs, the nebular continuum contributes between 20 to 80\% of the emission.        

\section{Self-consistent modelling of stellar and nebular features}\label{sec:method}
In this section, we present the properties of the model libraries constructed to model the HII regions in the Antennae galaxy (\S\,\ref{subsec:stellar_nebular_library}), the development of the fitting procedures to take both stellar and nebular information into account in constraining the best-fitting parameters for a given spectrum (\S\,\ref{subsec:continuum_fitting} and \S\,\ref{subsec:line_fitting}) and address the potential biases and degeneracies that can impact the derived best-fitting solutions for an HII region (\S\,\ref{subsec:double_solutions}). A detailed discussion of the MUSE observations of the Antennae galaxy is presented in \S\,\ref{sec:data}. 

\subsection{The stellar \& nebular model library}\label{subsec:stellar_nebular_library}
\begin{table*}
\caption{The model libraries used in the fitting of the spectra of HII regions in the Antennae galaxy.}
\begin{threeparttable}[t]
\centering
\begin{tabular}{l|c|c}
\hline
 Parameter						& Geneva library					& Parsec library \\
\hline
Ages [Myr] 						& 0.21 Myr $-$ 9.3 Gyr\tnote{a}\, (67 different values)	        & Same \\
Stellar metallicity [$Z_{\rm{s}}$] 		&  0.008 $-$  0.04\tnote{b}\,  (8 different values)       &  Same\tnote{b}\,  \\
IMF upper mass cut [M$_{\odot}$]             & 100   							& 120 \\
Gas metallicity [$Z_{\rm{g}}$] 		        &  0.0006 $-$  0.063 (7 different values)    & 0.002 $-$  0.063 (6 different values) \\
$n_{\rm{H}}$ [cm$^{-3}$]				&  10 $-$  1000\tnote{c}  				& 10 $-$  600\tnote{c}    \\			
Burst strengths [log M$_{\odot}$]  		& 2 $-$  4\tnote{d}\,  (in 0.1 dex steps) 	& Same\tnote{d}  \\
\hline
\end{tabular}
\begin{tablenotes}
     \item[a] {The ages are variably sampled.  In the models, the young ages (i.e.\,$<10$ Myr) are sampled at 0.5 Myr steps in order to fully capture the evolutionary effects of massive stars, particularly the effects of the crucial short-lived WR phase. }
     \item[b] {As also noted earlier, the range in stellar metallicity probed by our model libraries is currently limited as our intention is to build a comprehensive model library for the HII regions in the Antennae galaxy, which are known to host stellar populations of around solar-like metallicities. We intend to extend the libraries to lower stellar metallicities in the future.}
     \item[c] {The $n_{\rm{H}}$ is variably sampled. In Geneva models, we use 14 different values between 10 and 1000 cm$^{-3}$, and $n_{\rm{H}}$ around 100 cm$^{-3}$ is finely sampled. In Parsec models there are only 10 values between 10 and 600 cm$^{-3}$.}
     \item[d] {These burst strengths are for model HII regions of a fixed radius. For an HII region of a different radius, these strengths need to be scaled following Eq.\,\ref{eq:ionisation_relation}. }  
\end{tablenotes}
    \end{threeparttable}%
\label{tab:model_libraries}
\end{table*}%

We construct two separate model libraries using the Geneva and Parsec isochrones. Each library is parameterised by stellar metallicity ($Z_{\rm{s}}$), gas metallicity ($Z_{\rm{g}}$), age, $n_{\rm{H}}$ and strength of the starburst, with the Kroupa IMF high-mass cut-off assumed to be fixed. We describe the ranges and the sampling of each parameter in detail in Table\,\ref{tab:model_libraries}. 

The model continua of young stellar populations show a substantial evolution. Over the $<5$ Myr timescale, the nebular continuum contribution to the total continuum can be significant (see Figure\,\ref{fig:nebular_to_total}). As the nebular continuum is largely featureless except for the jumps, the overall continuum shape of the $<5$ Myr stellar populations appear to be devoid of weak absorption features. The strong absorption features like the Balmer features, on the other hand, are clearly apparent in these young stellar populations. The nebular continuum contribution to the total continuum diminishes with increasing age, so the model continua of $5-10$ Myr stellar populations show the appearance more prominent continuum structures, such as the TiO bands around $7000$\AA, as well as different absorption features. The model continua of $>10$ Myr stellar populations, as expected, are rich in various absorption features and with increasing age, these features progressively become more prominent.  

\begin{figure*}
\begin{center}
\includegraphics[width=0.33\textwidth, trim={0.0cm 0.cm 0.0cm 0.0cm},clip]{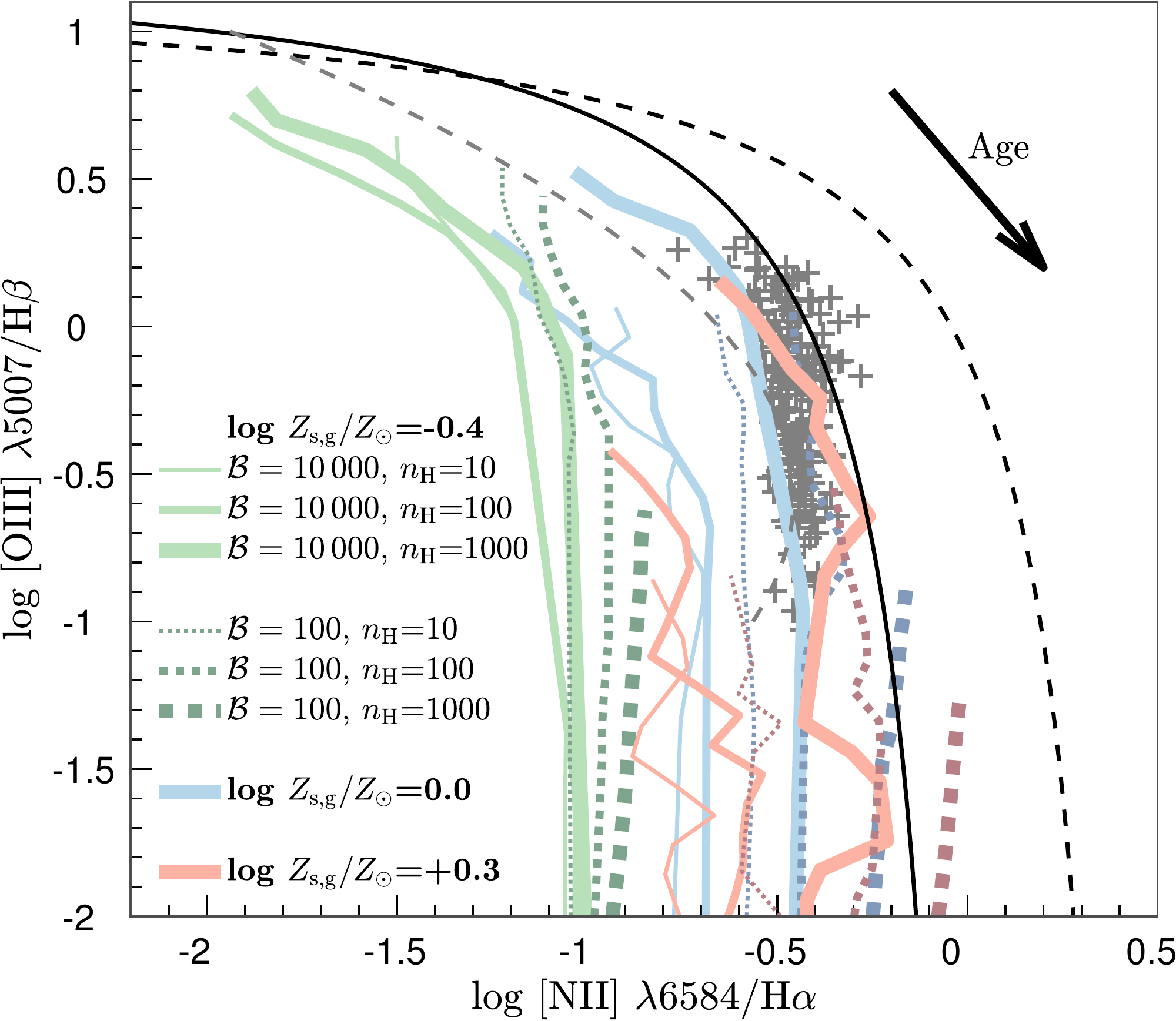}
\includegraphics[width=0.33\textwidth, trim={0.0cm 0.cm 0.0cm 0.0cm},clip]{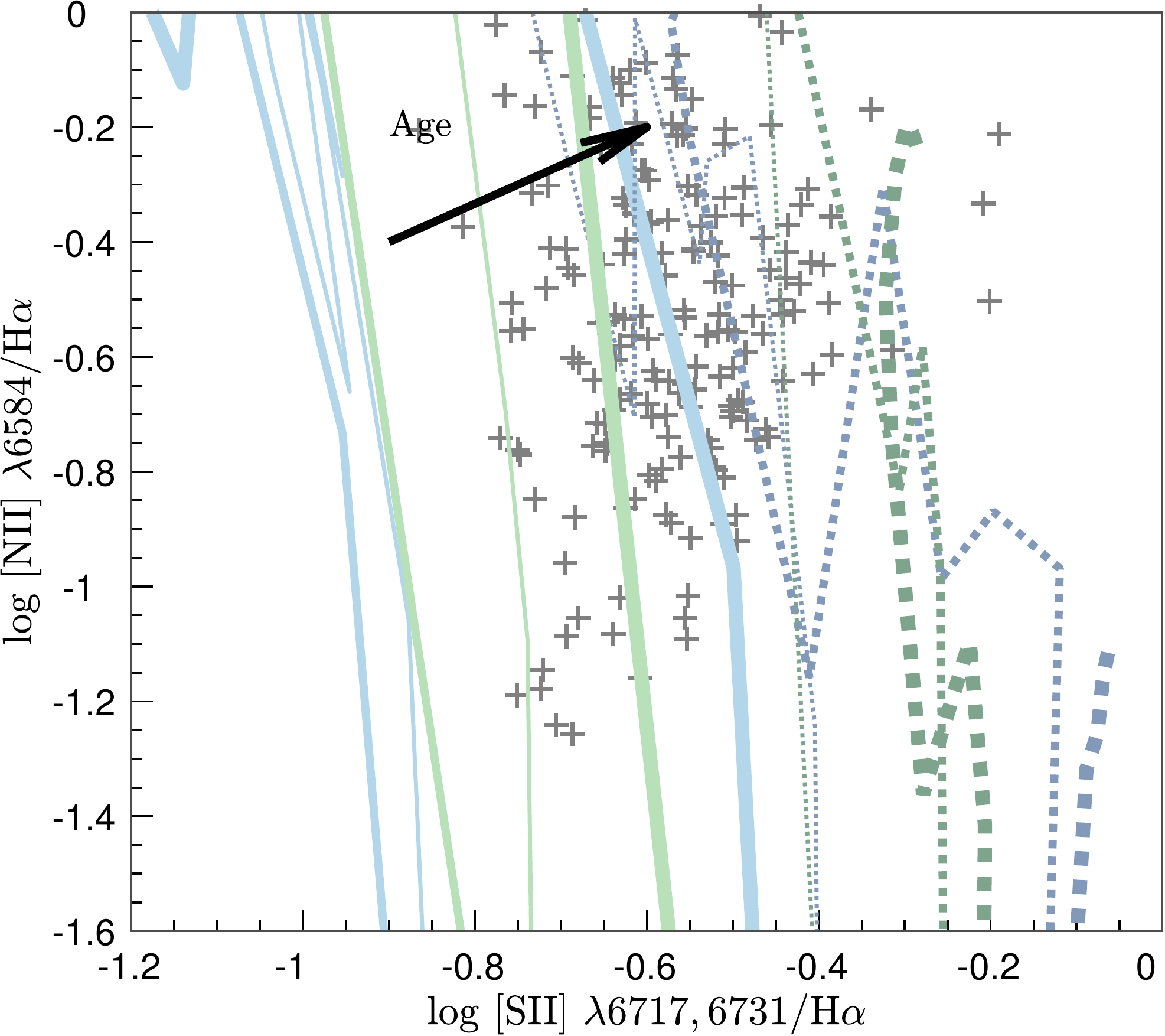}
\includegraphics[width=0.33\textwidth, trim={0.0cm 0.cm 0.0cm 0.0cm},clip]{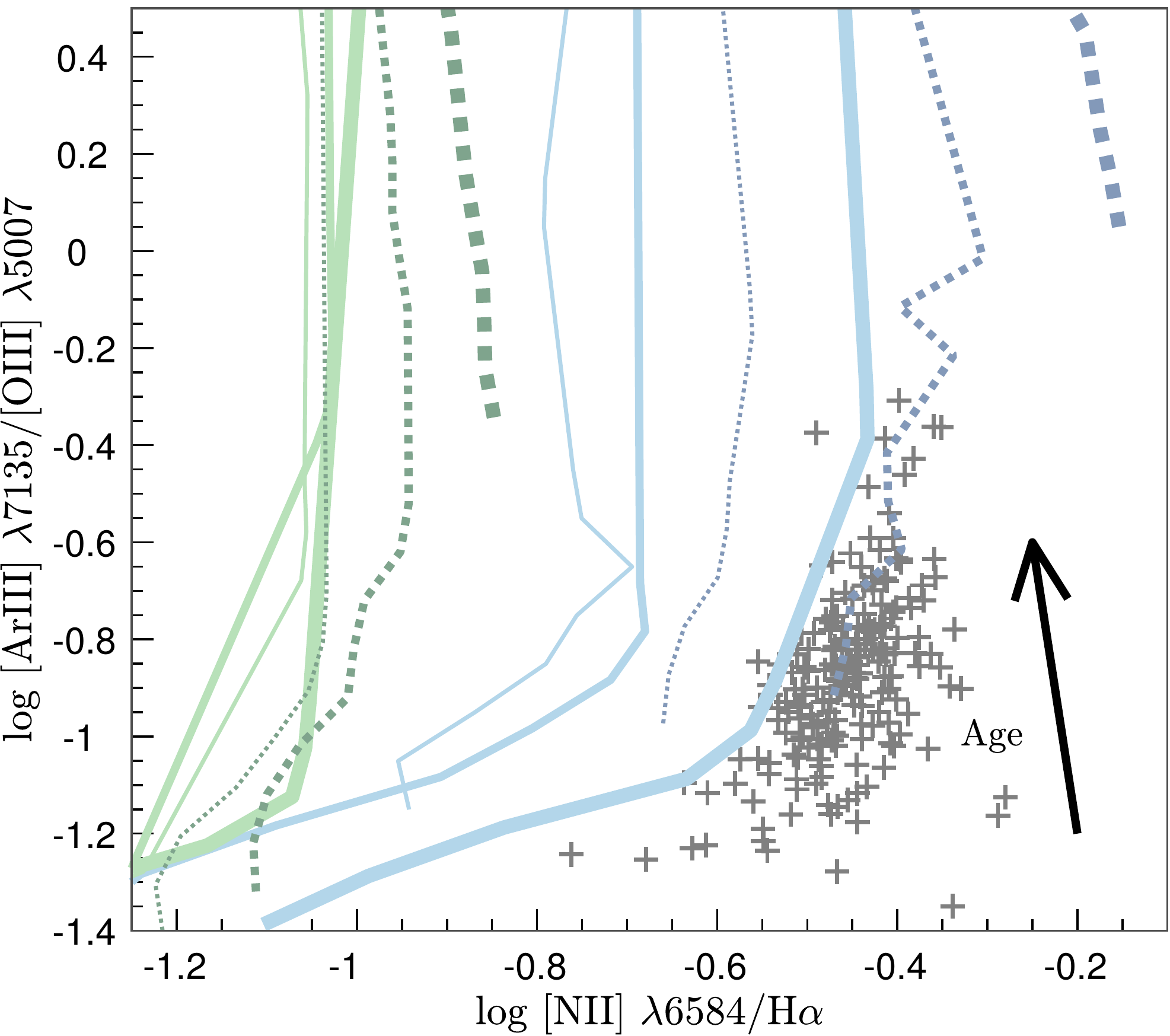}
\includegraphics[width=0.33\textwidth, trim={0.0cm 0.cm 0.0cm 0.0cm},clip]{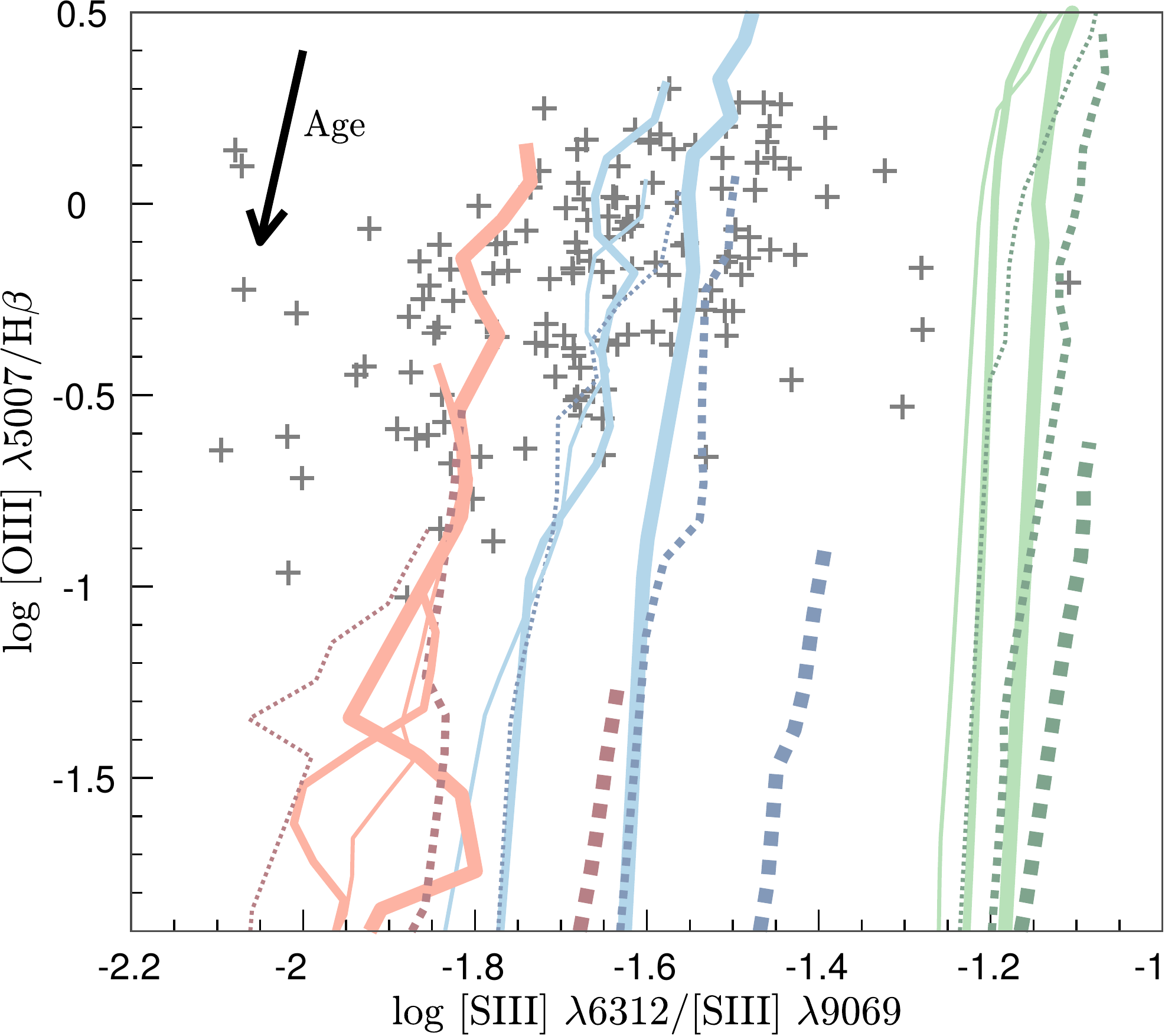}
\includegraphics[width=0.33\textwidth, trim={0.0cm 0.cm 0.0cm 0.0cm},clip]{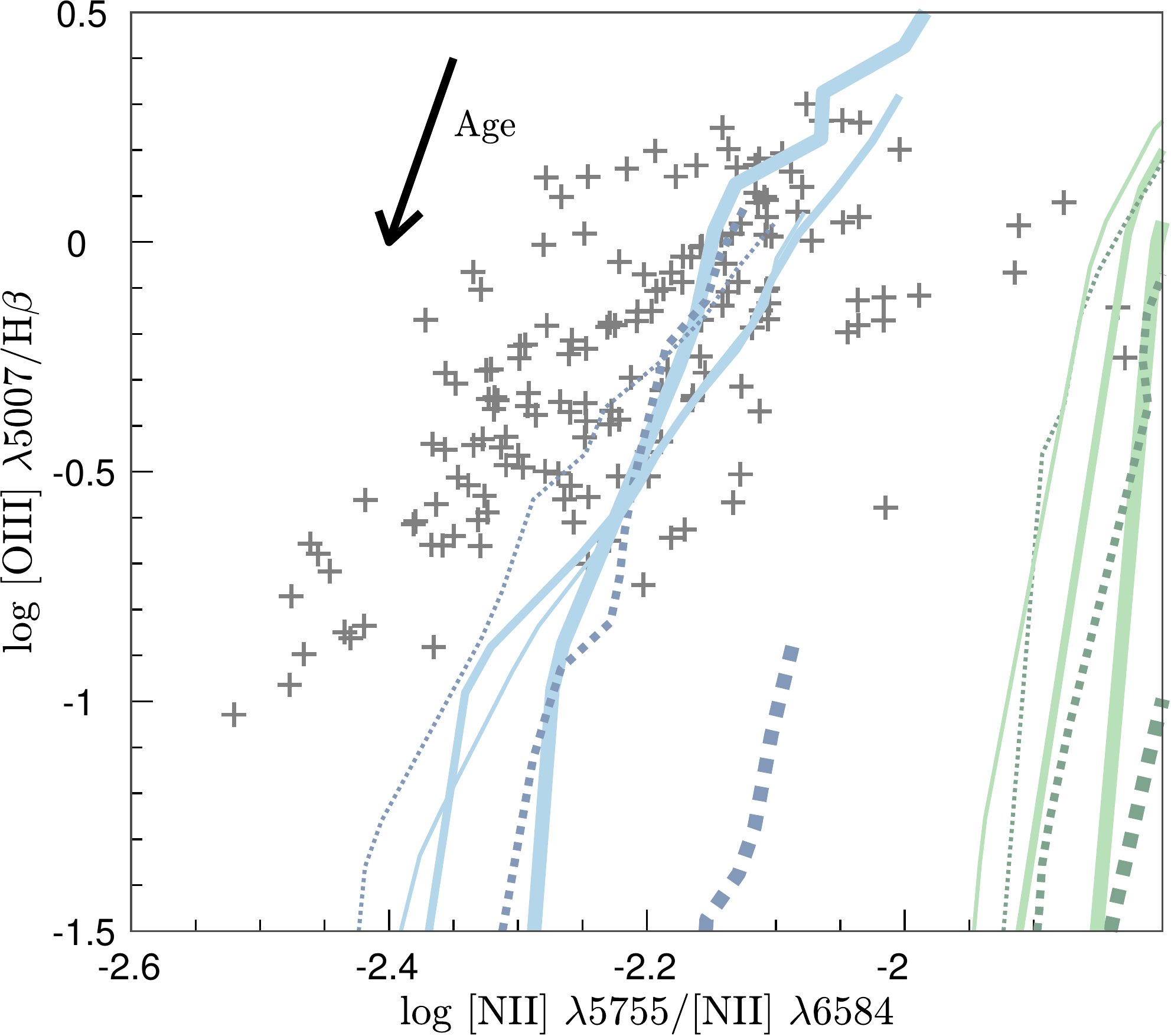}
\includegraphics[width=0.33\textwidth, trim={0.0cm 0.cm 0.0cm 0.0cm},clip]{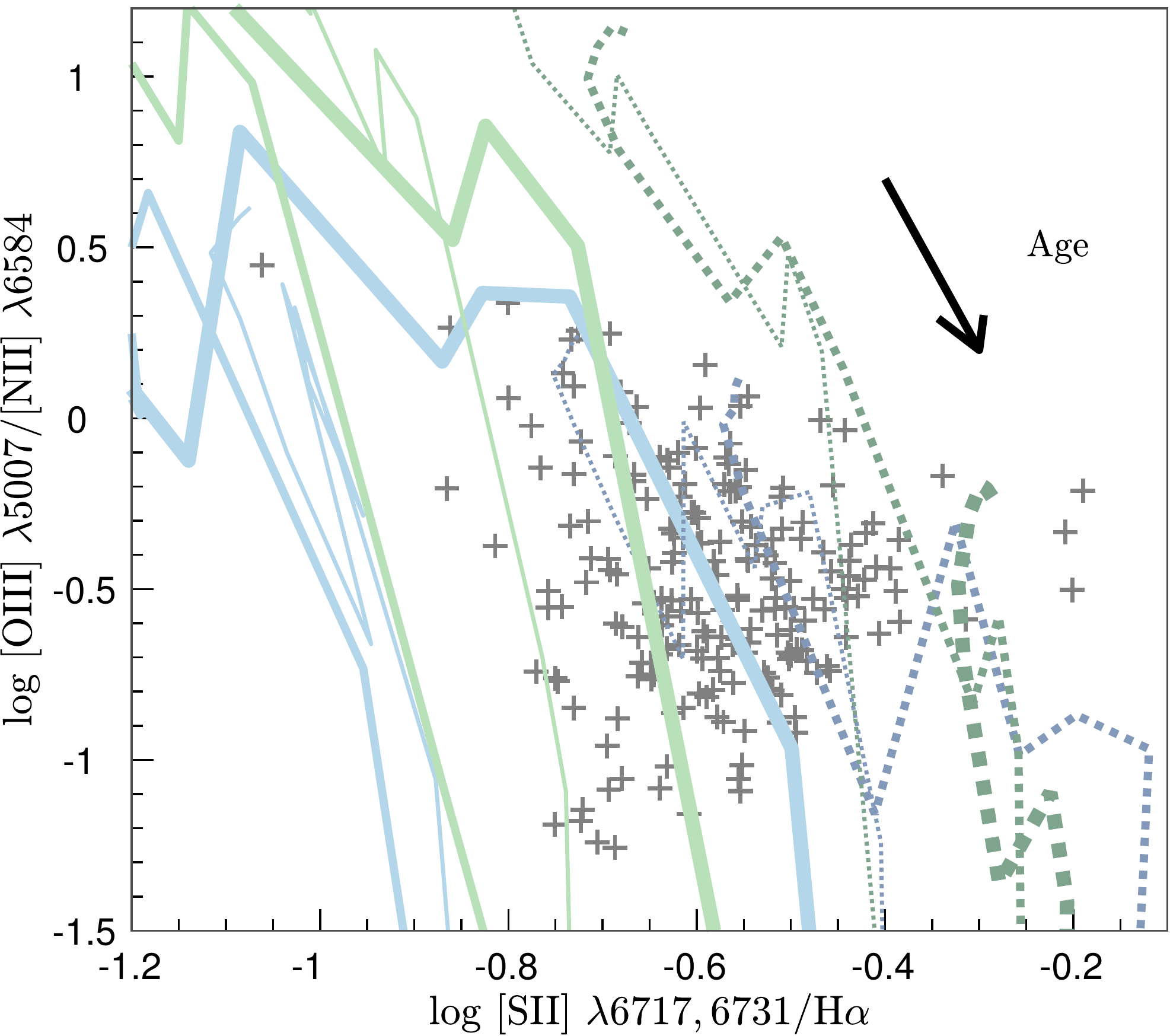}
\caption{The behaviour of predicted nebular lines in different diagnostic diagrams. The different colours correspond to different metallicities, while the line thickness and styles denote the $n_{\rm{H}}$ values and burst strengths, respectively. The ages increase along each track with arrows pointing to the general direction of increase. The zig-zag patterns evident are a result of the appearance of WR stars. For clarity, we have only shown the super-solar metallicity predictions for the two left-most diagnostics diagrams. The crosses indicate the distribution of the HII regions in the Antennae galaxy. A signal-to-noise cut is applied to the observations based on the signal-to-noise of the weakest line in a given set of emission line ratio diagnostics. Note that the burst strengths ($\mathcal{B}$) explored here are for model HII regions of a fixed radius.}
\label{fig:diagnostics_vs_obs}
\end{center}
\end{figure*}

The behaviour of the model emission line ratio diagnostics is shown in Figure\,\ref{fig:diagnostics_vs_obs}. The colour-coding is the same as in Figure\,\ref{fig:evolution_line_ratios}, the thick-to-thin lines denote high-to-low $n_{\rm{H}}$, and the dotted versus solid lines of the same colour and thickness correspond to starbursts of 100 M$_{\odot}$ and 10000 M$_{\odot}$, respectively. The arrow points to the general direction of the increase in age of the stellar population along each model track. The underlying data points denote the distributions of the observed emission line ratios of the HII regions in the Antennae galaxy from \citet{Weilbacher2018}, with the cross to the left indicating the average errors associated with the observed fluxes.  

The most substantial changes evident in the model evolutionary tracks appear to be driven by metallicity. In the BPT plane (top-left),  the sub-solar $Z$ tracks are closer together than other metallicity models such that an increase in burst strength affect [\ion{O}{iii}]/H$\beta$, a diagnostic sensitive to both $U$ and the age of the stellar population, more significantly than [\ion{N}{ii}]/H$\alpha$, a diagnostic sensitive to abundance. Note, however, that this assertion is largely true for later ages, as the changes in both [\ion{O}{iii}]/H$\beta$ and \ion{N}{ii}]/H$\alpha$ appear to be significant and somewhat similar at earlier ages. In contrast, [\ion{N}{ii}]/H$\alpha$ show a greater change than [\ion{O}{iii}]/H$\beta$ with increasing metallicity.   

The behaviour of [\ion{O}{iii}]/H$\beta$ in low metallicity models can largely be explained by T$_{\rm{e}}$ (see also \S\,\ref{subsubsec:controlOfU}). As shown in Figure\,\ref{fig:evolution_logQ_logU}, at a given burst and age, the central stars of low-metallicity HII regions have higher effective temperatures, thus higher $U$ than high-metallicity regions (top-left panel of Figure\,\ref{fig:evolution_line_ratios}). Therefore, in these regions, which are already characterised by higher excitation species than high-metallicity regions, an enhanced burst will further act to amplify the degree of excitation of [\ion{O}{ii}]. Furthermore, Nitrogen has a substantial secondary nucleosynthesis contribution over a large range in metallicity, except at very low metallicities, \citep{Matteucci1986, Dopita2000} such that the abundance ratio of a secondary to a primary element, e.g.\,N/O, will systematically increase with nebular abundance. Therefore the lack of a change in  [\ion{N}{ii}]/H$\alpha$ in low metallicity models is likely due to low N$^{+}$ abundances.  

In high-metallicity HII regions, which are characterised by low T$_{\rm{e}}$, the higher excitation energy transitions (e.g.\,O$^{+}$ to O$^{++}$) are suppressed, while the relatively lower energy species (e.g.\,N$^{++}$) are continuously excited \citep{Dopita2006}. This process along with the relatively high Nitrogen abundances likely drive the increase in enhancement (decrement) evident in [\ion{N}{ii}]/H$\alpha$ ([\ion{O}{iii}]/H$\beta$). In fact, it is this sensitivity of [\ion{N}{ii}] to abundance that makes the [\ion{N}{ii}]~$\lambda6584$/H$\alpha$ a useful abundance diagnostic.     

Similarly, the observed trend of [\ion{Ar}{iii}]~$\lambda7135$/[\ion{O}{iii}]~$\lambda5007$ with respect to [\ion{N}{ii}]~$\lambda6584$/H$\alpha$ (top-right) for a large part is due to the inverse relationship between metallicity and T$_{\rm{e}}$, which translates into an increase of the [\ion{Ar}{iii}]~$\lambda7135$ and [\ion{O}{iii}]~$\lambda5007$ emissivity ratios. According to \cite{Stasinska2006}, the decrease in $U$ with increasing metallicity also adds to the observed trend in [\ion{Ar}{iii}]~$\lambda7135$/[\ion{O}{iii}]~$\lambda5007$, however, it does not play a dominant role.  

The [\ion{O}{iii}], [\ion{N}{ii}] and [\ion{S}{iii}] have high-energy structures with considerably different excitation energies, such that the ratio between the lines originating from different excitation energy levels can be utilised as temperature diagnostics \citep{Osterbrock, Peimbert2017}. So in Figure\,\ref{fig:diagnostics_vs_obs} bottom-left panel, we show the model predictions\footnote{The intensity ratios are defined as [\ion{O}{iii}] $(I(4959) + I(5007))/I(4363)$, [\ion{N}{ii}] $I(5755)/(I(6548) + I(6583))$ and [\ion{S}{iii}] $I(6312)/(I(9532)+I(9069))$ \citep{Peimbert2017}, for the reason that, for example in [\ion{O}{iii}], an excitation to the $^1D$ results in the emission of a photon in either $\lambda5007$\AA\,or$\lambda4959$\AA\,with a relative probability of 3 to 1 \citep{Osterbrock}. However, the [\ion{O}{iii}]~$\lambda4363$\AA\,and [\ion{S}{iii}]~$\lambda9532$\AA\,are outside the MUSE wavelength coverage, and in the fitting, we tie $I(6548)$ to $I(6584)$, so only the $I(5755)/I(6583)$ and $I(6312)/I(9069)$ are shown.} for the ratio of [\ion{S}{iii}]~$\lambda6312$\AA, which originates from the excitation energy of level $^1S_0$ with 3.37 eV, to [\ion{S}{iii}]~$\lambda9069$, the line resulting from the lower excitation level of $^1D$ with 1.4\,eV. Similarly, in the bottom-middle panel of Figure\,\ref{fig:diagnostics_vs_obs}, we show the ratio of [\ion{N}{ii}]~$\lambda$5755\AA, corresponding to the excitation energy of the level $^1S_0$ with 4.05 eV, to [\ion{N}{ii}]~$\lambda$6584\AA, corresponding to the excitation energy of the level $^1D$ of 1.9 eV \citep[][]{Luridiana2015}. Overall, the model evolutionary tracks show a large spread both as a function of metallicity, emphasising their underlying sensitivity to T$_{\rm{e}}$, and $n_{\rm{H}}$. 

Finally, as is evident in Figure\,\ref{fig:diagnostics_vs_obs}, a single diagnostic alone cannot accurately trace an entire HII region, which is a complex ionisation/thermal structure.  The high ionisation lines (e.g.\,[\ion{O}{iii}], [\ion{Ar}{iii}], [\ion{S}{iii}]) are strongly sensitive to the ionisation parameter, which itself depends on the age. Therefore these species will be mostly produced in very young HII regions and, with increasing age, will be limited to tracing the innermost of the HII regions \citep{Byler2017}. Consequently, the T$_{\rm{e}}$ estimated from high ionisation line based diagnostics like [\ion{O}{iii}] and [\ion{S}{iii}], and $n_{\rm{H}}$ measured from [\ion{Cl}{iii}] and [\ion{Ar}{iv}] will only be representative of the innermost regions of a HII region. In contrast, the [\ion{N}{ii}],  [\ion{O}{ii}] and [\ion{S}{ii}] lines are stronger in the outer parts of an HII region, where the ionisation is low. In modelling integrated spectra of star-forming galaxies, the lower ionisation species will also get a much greater weighting from the aged HII regions than the higher excitation species like O$^{++}$. The temperature and density diagnostics based on these species, likewise, mostly trace the outermost of HII regions. In concluding this section, it is worthwhile to note that the model line ratio diagnostics sample the parameter space of the observed emission line ratios, suggesting that our model libraries are capable of (approximately) reproducing the observations. 

\subsection{Fitting for the stellar \& nebular continua}\label{subsec:continuum_fitting}
Our model fitting routines are built on \textsc{Platefit}, a code originally written to perform a non-negative least-squares fit with dust attenuation modelled as a free parameter to find the best-fitting stellar continuum model for a given spectrum \citep{Tremonti2004, Brinchmann2004}. In this study, we update and equip \textsc{platefit} with additional routines to allow self-consistent modelling of stellar and nebular continua, and nebular emission line ratios using the libraries described in Table\,\ref{tab:model_libraries}.      

For the continuum fitting, the dust attenuation is modelled as a free parameter using a simple attenuation curve, $\tau(\lambda)\propto\lambda^{-1.3}$. The exponent of $-1.3$ is recommended by \citet{Charlot2000}, which was observed to correspond to the middle range of the optical properties of dust grains between the Milky Way, the LMC and the SMC, and was found to be appropriate for modelling the attenuation of birth clouds by \citet{daCunha2008}. We have tested other common attenuation laws in the literature \citep[e.g.][]{Prevot1984, Calzetti2000}, and their impact on the outcomes of the fitting appear to be minimal. In this analysis, we assume that the young and old stellar populations are attenuated to the same extent. 

The SFHs of starbursting HII regions are likely rather episodic than constant or exponential-like. \citet{Wilson1995}, for example, find evidence for burst-like SFHs in local luminous HII regions. Therefore, following the underlying assumption that the SFH can be approximated as a sum of discrete bursts, we use \textsc{platefit} to fit the underlying stellar and nebular continua of a given spectrum to extract the individual models most likely contributing to its SFH.  

Strictly speaking, to preserve self-consistency, the stellar and nebular continua must be modelled as a function of all the parameters of the model library (Table\,\ref{tab:model_libraries}). The model libraries described in \S\,\ref{subsec:stellar_nebular_library} are comprehensive in both the range in parameters probed and the sampling of the parameter space. As such, the use of the full model library in the fitting can be computationally expensive. There are, however, specific model parameters that do not contribute to describing the shape of the stellar and nebular continua over the MUSE rest wavelength coverage of the Antennae spectra. These parameters, therefore, can be set to fiducial values without sacrificing self-consistency.  For example, recall that the variations in $n_{\rm{H}}$ affect mostly the $2\gamma$ continuum, whereas the shape of the nebular continuum over the $4600-9300$\AA\,range only show a small dependence (see Figure\,\ref{fig:cloudy_nebularC_dependence_onU_ZgZs_nH}c). Similarly, the shape of the nebular continuum shows no dependence on $U$ (see Figure\,\ref{fig:cloudy_nebularC_dependence_onU_ZgZs_nH}a), which we use to trace the strength of the starburst. Consequently, we can perform the \textsc{platefit} stellar and nebular continua fitting as a function of three model parameters (i.e\,$t$, $Z_{\rm{g}}$, $Z_{\rm{s}}$), instead of five, to extract the individual light-weighted underlying stellar populations likely contributing to the SFH as a function of $Z_{\rm{g}}$ and $Z_{\rm{s}}$. The best-fit velocity dispersion is determined from the continuum fitting, assuming trial velocity dispersions to converge to the solution that minimises the chi-squared of the fit. 

\begin{figure*}
\begin{center}
\includegraphics[width=0.7\textwidth, trim={0.cm 0.cm 0.cm 0.0cm},clip]{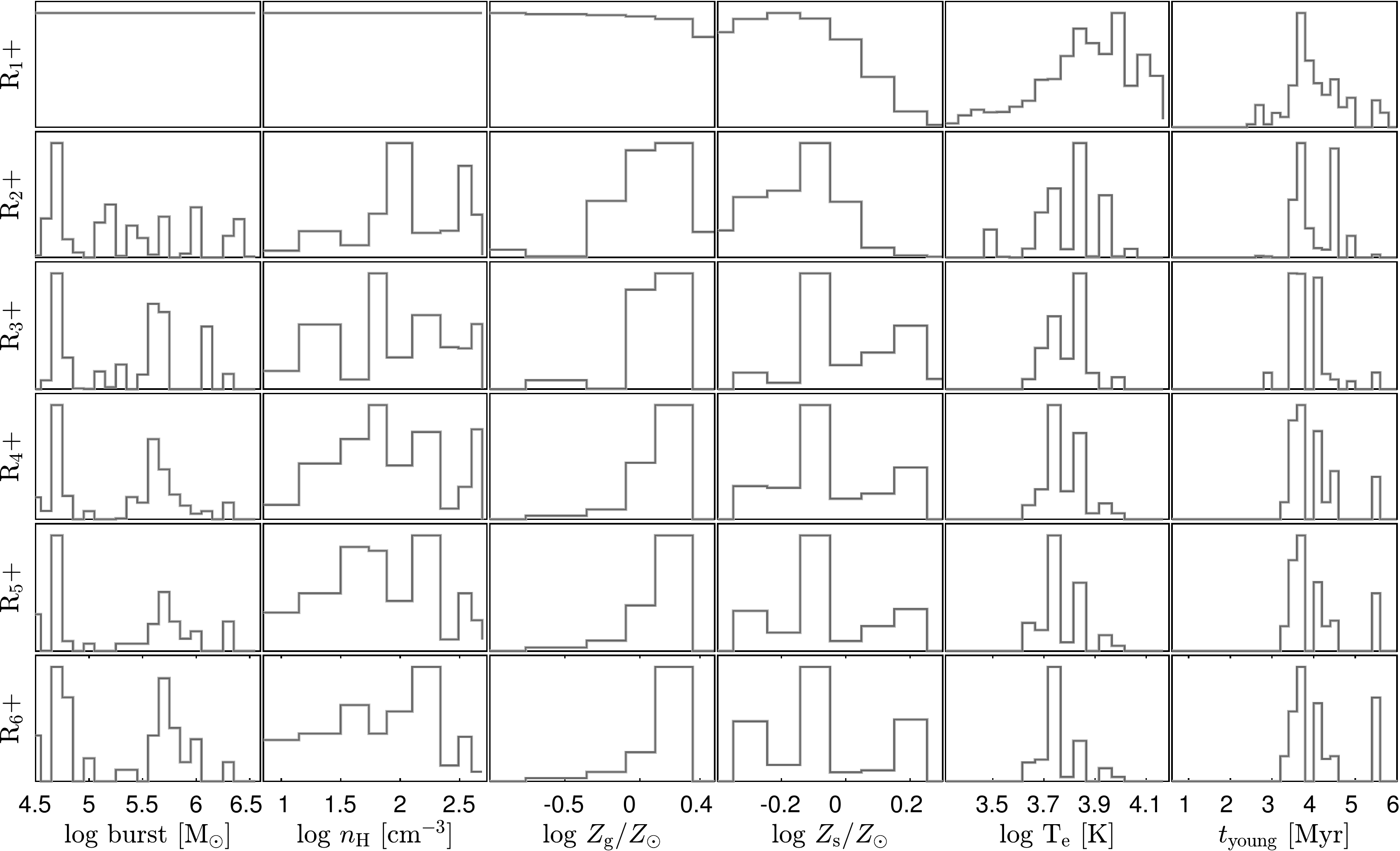}
\caption{The result of the fit to the Antennae HII region A901 using Parsec-based model library. Each column corresponds to the PDFs for one derived property as progressively more emission line ratios are added. The columns from left-to-right shows the strength of the burst, $n_{\rm{H}}$, gas and stellar metallicities with respect to solar, T$_{\rm{e}}$, and light weighted age of the $<10$ Myr stellar population. The top row shows the PDFs after fitting the total continuum, which adds constraints on gas and stellar metallicities, T$_{\rm{e}}$ and light weighted age of the young stellar populations.   
The strength of the burst and $n_{\rm{H}}$ are primarily constrained through nebular line luminosity ratios, therefore their PDFs essentially show the flat priors adopted.}
\label{fig:parsec_fitting_example}
\end{center}
\end{figure*}
\begin{figure*}
\begin{center}
\includegraphics[width=0.7\textwidth, trim={0.cm 0.cm 0.cm 0.0cm},clip]{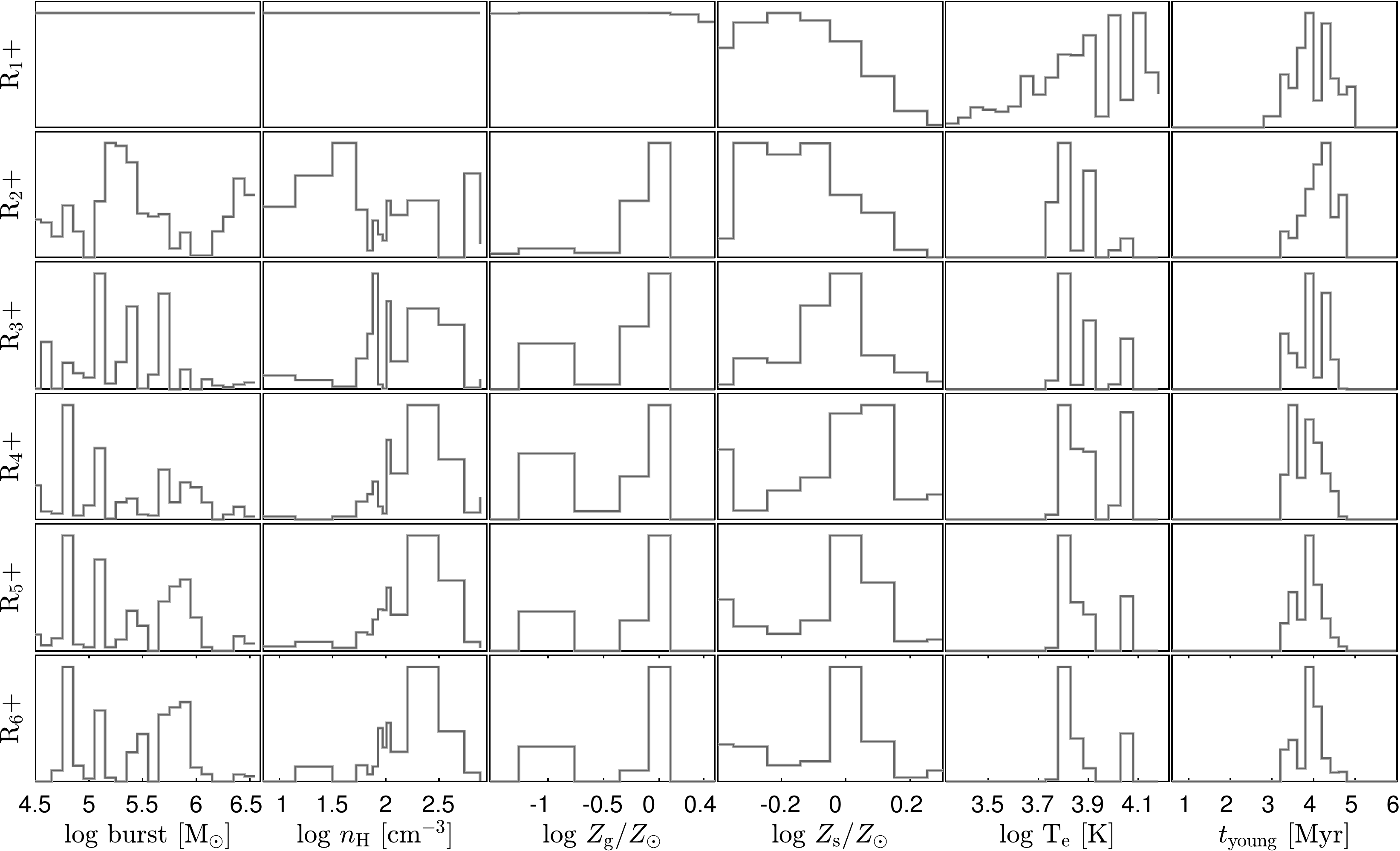}
\caption{Similar to Figure\,\ref{fig:parsec_fitting_example}, but using the Geneva models. }
\label{fig:geneva_fitting_example}
\end{center}
\end{figure*}
After subtracting the best-fitting stellar and nebular continuum (derived from adding the individual light-weighted stellar population models) in each $Z_{\rm{g}}$ and $Z_{\rm{s}}$ from the observed spectrum, we model all the emission lines with Gaussians simultaneously. We were able to model the majority of the Antennae spectra this way. A subset of spectra, however, required removing remaining residuals by fitting a Legendre polynomial of degree 15 before modelling the emission lines. The residuals, in most part, can be used to inform the improvements needed in population models, in attenuation models and fitting routines, as well as the issues related to data reduction. Ideally, the uncertainties from the continuum fitting should be folded-in to the emission line flux errors without adjusting the best-fitting model. It is, however, difficult to ascertain the true uncertainties without duplicate observations of the Antennae galaxy. Encouragingly, the best-fitting models in each $Z_{\rm{g}}$ and $Z_{\rm{s}}$ show a good agreement with the data over the wavelength ranges of the strong emission lines (e.g.\,H$\alpha$, H$\beta$, [\ion{O}{iii}]~$\lambda$5007\AA, [\ion{N}{ii}]~$\lambda 6548,84$\AA, [\ion{S}{ii}]~$\lambda 6717,31$\AA\,and red-wards of the Paschen jump) without requiring a fit to the residuals. On the other hand, the weak nebular lines (e.g.\,[\ion{Cl}{iii}]~5517,37\AA, [\ion{O}{ii}]~7323,32\AA\,and weak He lines) are affected by any mismatch between the best-fitting models and the data, thus benefit from a fit to the residuals.   

\subsection{Fitting for the nebular emission line intensity ratios}\label{subsec:line_fitting}
\begin{figure*}
\begin{center}
\includegraphics[width=0.9\textwidth, trim={0.cm 1.5cm 0.cm 0.0cm},clip]{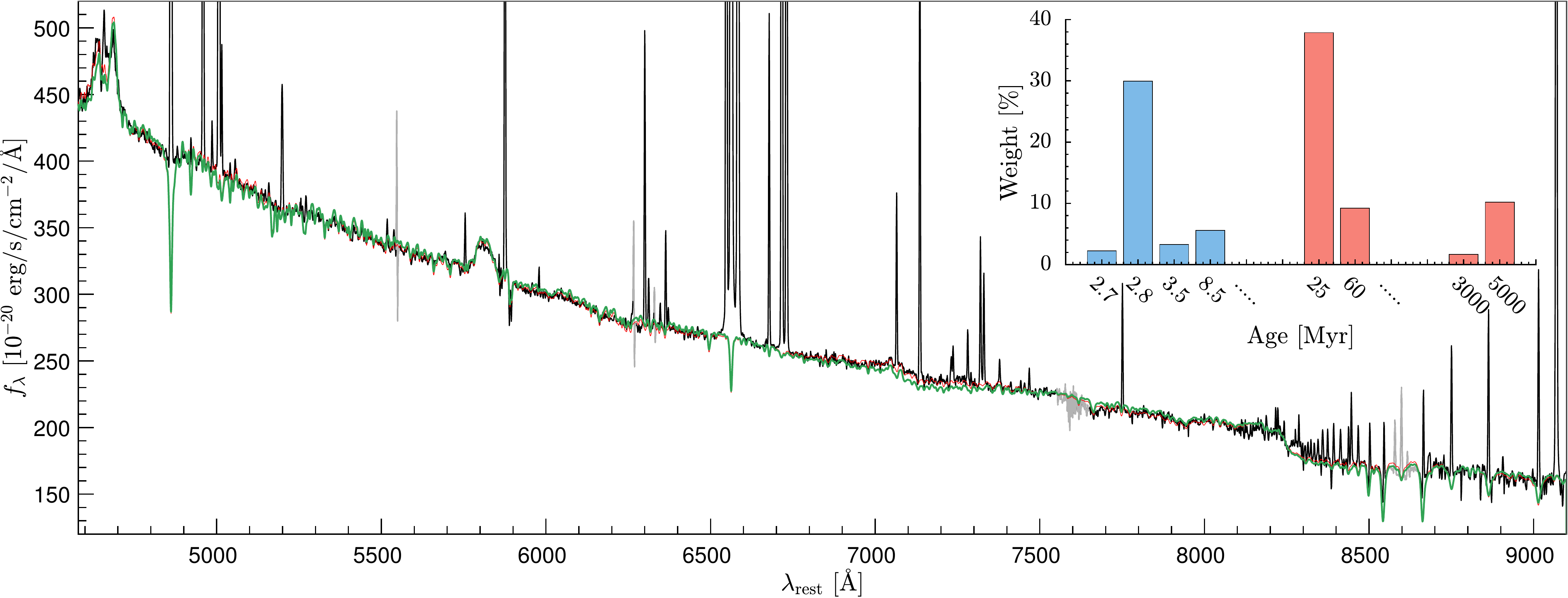}
\includegraphics[width=0.9\textwidth, trim={0.cm 0.cm 0.cm 0.0cm},clip]{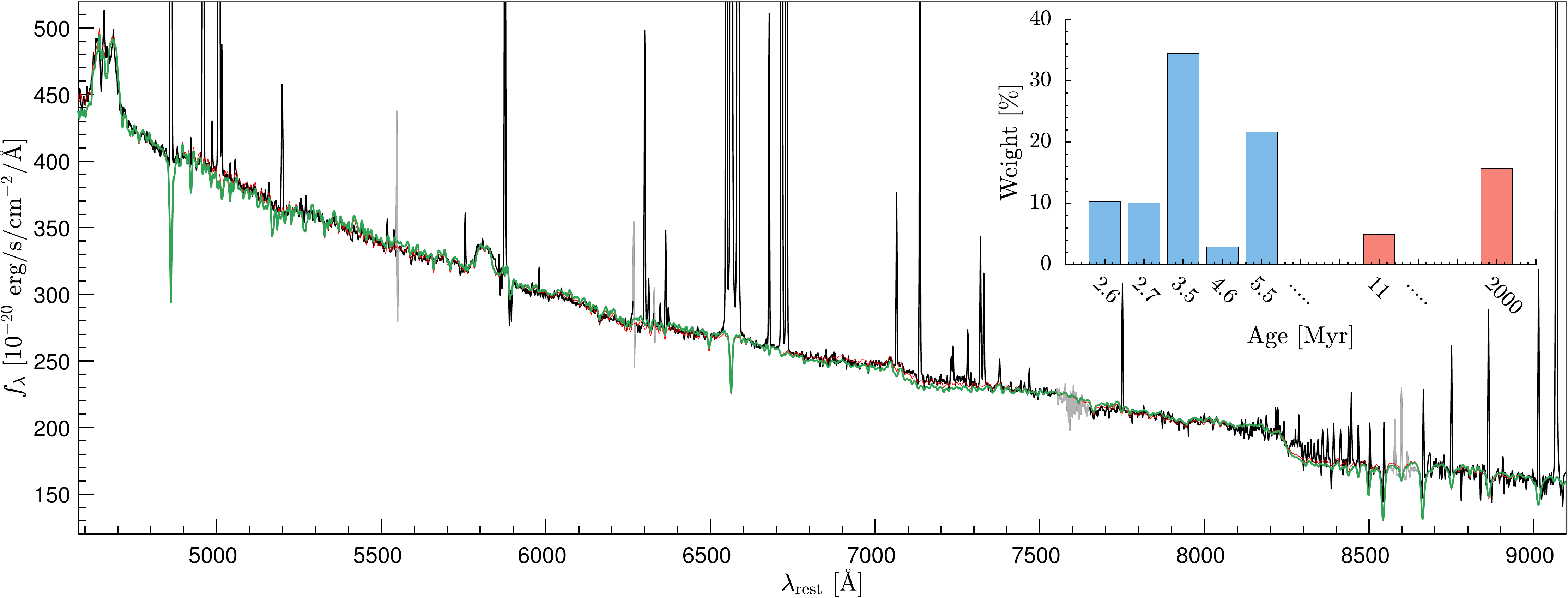}
\caption{The best-fitting Parsec (top) and Geneva (bottom) models corresponding to the constrained PDFs of the Antennae HII region A901 shown in Figures\,\ref{fig:parsec_fitting_example} \& \ref{fig:geneva_fitting_example}, respectively. In each panel, the best-fitting model of the continuum is plotted in green, the best-fitting model adjusted with a fit to the residuals in red, and the spectrum of A901 in black. The wavelength regions affected by sky lines/telluric residuals (shown in gray) are masked out in the fitting. The best-fitting model provides a good fit to the observed spectrum without requiring much adjustment. The offset between the green and red models highlight the likely wavelength regions where the underlying models may require further improvement in their treatment of old stellar populations. The wavelength windows where the difference between the green and red models the largest are used for the discrimination between various solutions as described in \S\,\ref{subsec:double_solutions}. The insets show predicted star formation histories with the $<$ and $>$ 10 Myr populations denoted by blue and red, respectively. Note that the region centred around $8600$\AA\,shows a slight dip in $F_{\lambda}$. This may be caused by the second order spectral effects associated with the MUSE extended mode observations \citep{Weilbacher2015} and/or by the poorly subtracted sky features around 8600\AA, and is masked out in the fitting. The slight excess in continuum flux at $\sim$8250\AA, which we are unable to model with the current libraries, is maybe a real feature that could be associated with WR stars. This excess is also apparent in some of the SDSS spectra of HII regions in the low metallicity emission line galaxies \citep{Guseva2006}, although its exact origin is uncertain.}
\label{fig:A901_fits}
\end{center}
\end{figure*}
\begin{table*}
\caption{The physical properties derived using the Parsec versus Geneva spectral libraries for the Antennae HII regions A901 (shown in Figure\,\ref{fig:A901_fits}) and A893. }
\begin{threeparttable}[t]
\centering
\begin{tabular}{l|c|c}
\hline
 Property								& Geneva 					  & Parsec  \\
\hline
\multicolumn{3}{c}{A901} \\
\hline
Light weighted age of the $<10$ Myr population [Myr]	& $3.86\pm0.32$      			&  $4.05\pm{0.87}$  \\
Dominant old stellar population 	[Myr]				&  2000 						&  25  \\
$\log\,Z_{\rm{s}}/Z_{\odot}$              				& $0.01\pm{0.06}$ 			     &  $-0.1\pm0.046$ \\
$\log\,Z_{\rm{g}}/Z_{\odot}$					& $0.0\pm0.09$ 				&  $0.25\pm{0.15}$ \\
$\log\,n_{\rm{H}}$ [cm$^{-3}$]				         & $2.49\pm0.27$  				&  $2.00\pm{0.46}$   \\			
log T$_{\rm{e}}$ 	[K]						& $3.807\pm{0.043}$ 			&  $3.710\pm{0.0227}$ \\
Burst strength   							& $>10^4$				      	&  $>10^4$\\
\hline
\multicolumn{3}{c}{A893} \\
\hline 
Light weighted age of the $<10$ [Myr] population [Myr]	& $3.05\pm{0.57}$     &  $2.62\pm{0.21}$\\
Dominant old stellar population [Myr]					&  5000\tnote{a} 	 &  80, 3000\tnote{b} \\
$\log\,Z_{\rm{s}}/Z_{\odot}$              					& $-0.25\pm{0.089}$   &  $0.13\pm{0.054}$\\
$\log\,Z_{\rm{g}}/Z_{\odot}$						& $-0.09\pm{0.13}$ 	 & $0.01\pm0.084$ \\
$\log\,n_{\rm{H}}$ [cm$^{-3}$]				                 & 2.43$\pm{0.26}$ 	  &  $2.34\pm{0.29}$ \\		
log T$_{\rm{e}}$ [K]					        			& $3.827\pm{0.0456}$&  $3.816\pm{0.022}$\\  
Burst strength [M$_{\odot}$]  						& $>10^4$		&  $>10^4$\\
\hline
\end{tabular}
\begin{tablenotes}
     \item[a] {The last template of the library is also given a significant weight.  } 
     \item[b] {Both 80 Myr and 3000 Myr templates appear to have similar weights.  } 
\end{tablenotes}
\end{threeparttable}%
\label{table:geneva_vs_parsec}
\end{table*}%
\begin{figure*}
\begin{center}
\includegraphics[width=\textwidth, trim={1.5cm 0.cm 0.cm 0.0cm},clip]{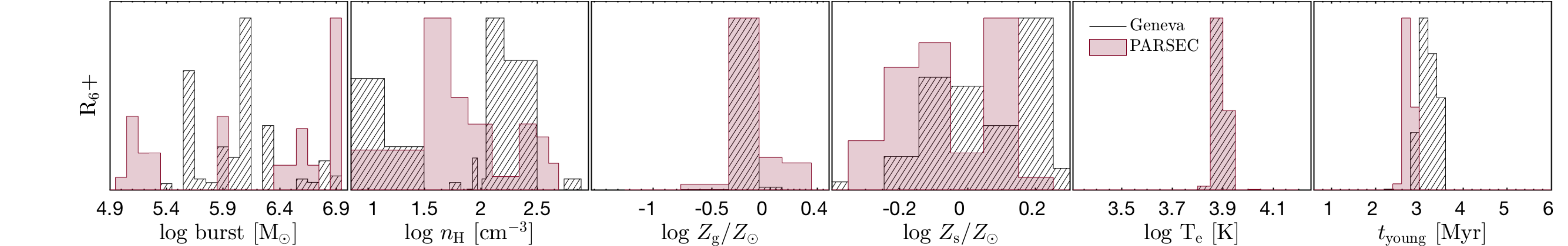}
\includegraphics[width=\textwidth, trim={0.cm 1.5cm 0.cm 0.0cm},clip]{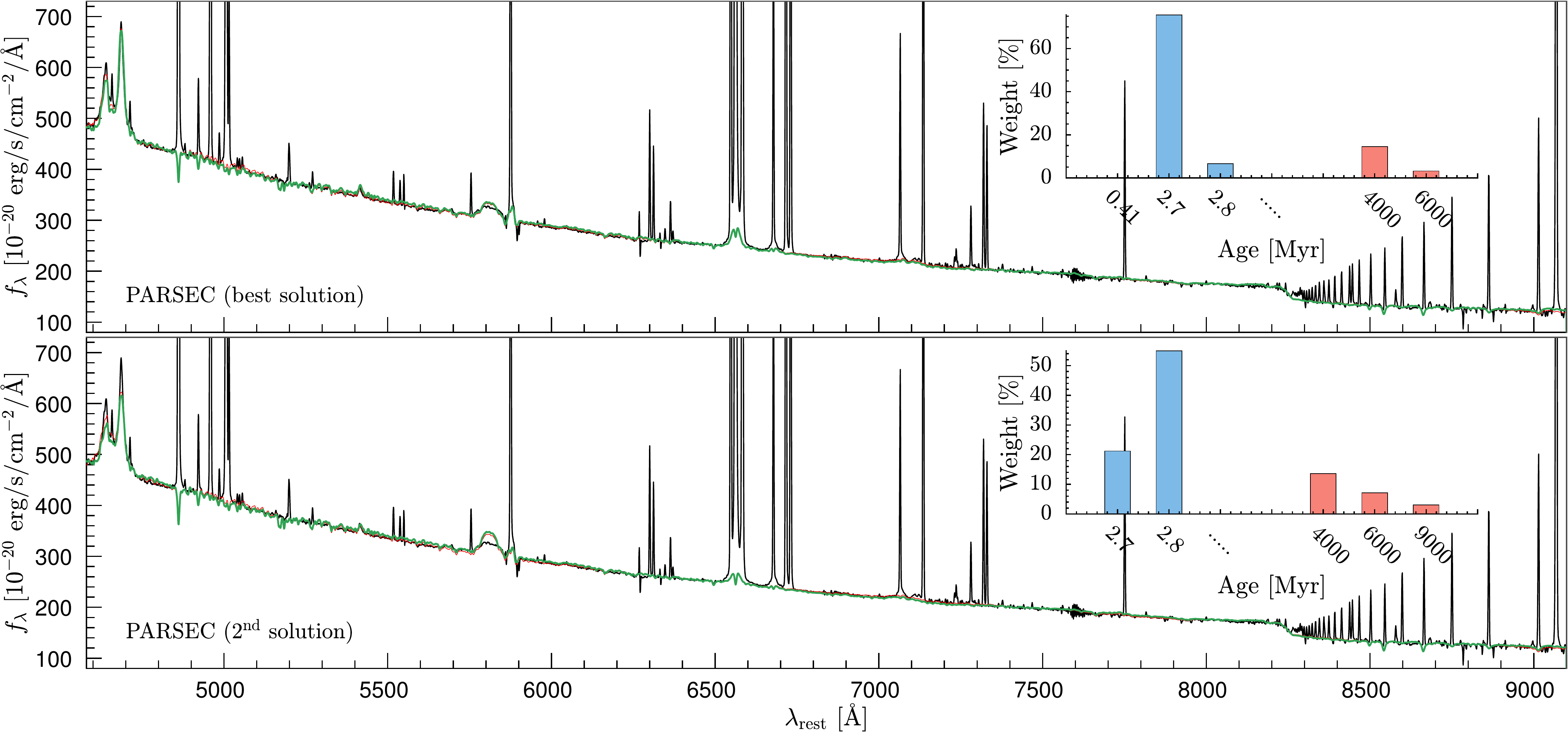}
\includegraphics[width=\textwidth, trim={0.cm 0.cm 0.cm 0.0cm},clip]{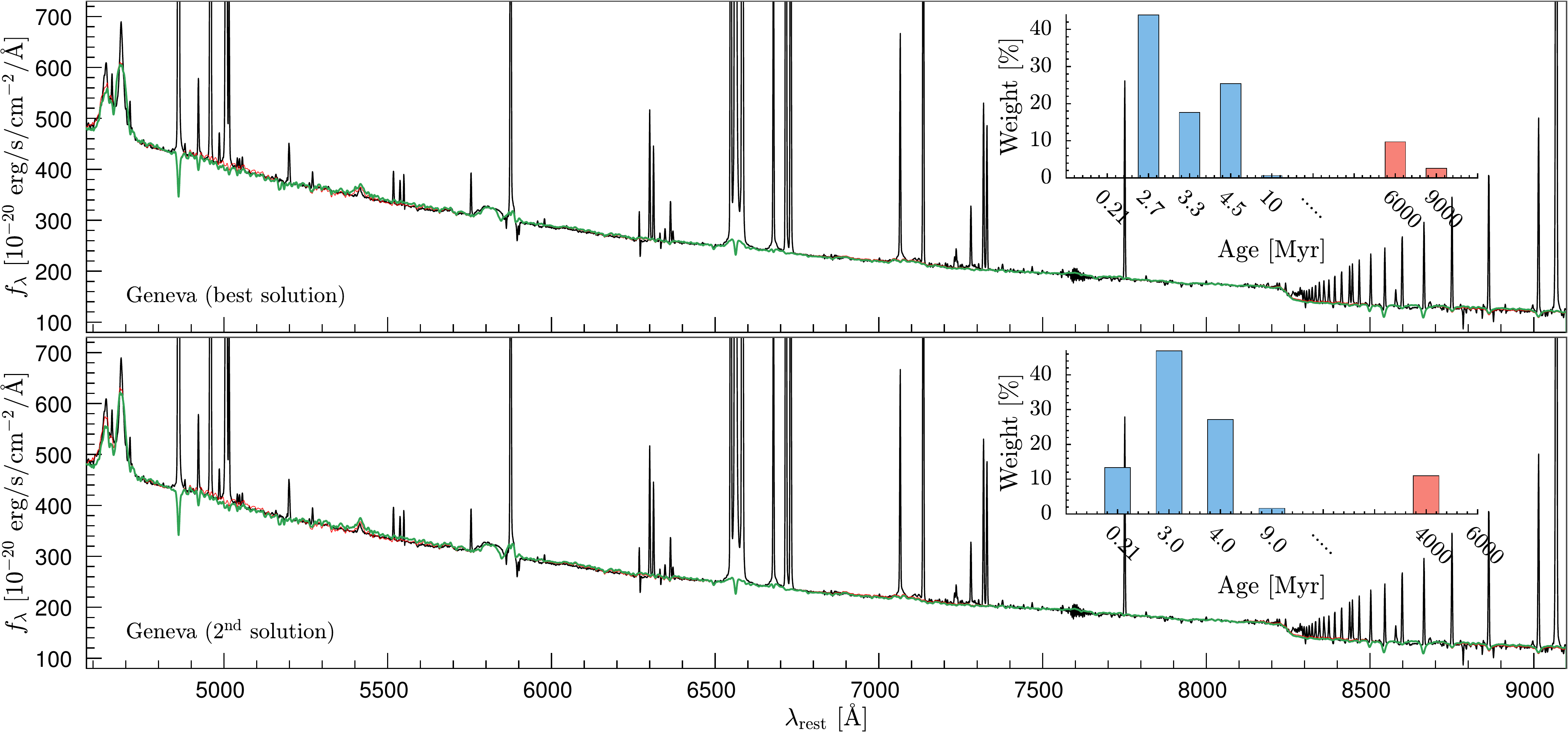}
\caption{The double $Z_{\rm{s}}$ solutions obtained from fitting Parsec and Geneva models to the Antennae HII region A927. From top-to-bottom, the final PDFs of Parsec (red) and Geneva (black) solutions, the Parsec best-fitting spectral model corresponding to the slightly super-solar $Z_{\rm{s}}$ as shown in the respective PDF (in the top row), the Parsec second solution corresponding to a $Z_{\rm{s}}$ of slightly sub-solar metallicity as shown in the respective PDF, the Geneva best-fitting spectral model corresponding to the slightly super-solar $Z_{\rm{s}}$ as shown in the respective PDF, and the Geneva second solution corresponding to a Z$_{\rm{s}}$ of slightly sub-solar metallicity. The insets show the age distributions corresponding to each solution.}
\label{fig:double_solutions}
\end{center}
\end{figure*}

To match the nebular model predictions with the individual light-weighted underlying stellar populations extracted during the continuum fitting process, we weight the respective nebular model by the lightweight of the respective stellar population. The best-fitting nebular model is, then, given by the sum of individual light-weighted nebular models. The nebular models exist only for the $<10$ Myr stellar populations, and we assume that these young populations are responsible for all observed nebular emission. 

For the fitting, we adopt the Bayesian approach outlined in \citet{Brinchmann2004, Brinchmann2013} and derive the probability density functions (PDF) for the physical properties of interest. The overall fitting process is illustrated in Figures\,\ref{fig:parsec_fitting_example} \& \ref{fig:geneva_fitting_example}, where we show how the likelihood distributions of six different properties are constrained progressively with the addition of emission line ratios. The columns, from left-to-right, correspond to the strength of the burst in units of M$_{\odot}$, $n_{\rm{H}}$ in units of cm$^{-3}$, gas ($Z_{\rm{g}}$) and stellar ($Z_{\rm{s}}$) metallicities in units of solar metallicity, $e^-$ temperature (T$_{\rm{e}}$) in unit of K, and the light-weighted age of the $<10$ Myr stellar population ($t_{\rm{young}}$) in unit of Myr. The top row of panels shows the likelihood distributions after the continuum fitting process and with one emission line ratio added. Already at this stage, the likelihoods of $Z_{\rm{g}}$, $Z_{\rm{s}}$, T$_{\rm{e}}$, and $t_{\rm{young}}$ reflect the constraints imposed by the continuum fitting process, and as we do not fit for $U$ and $n_{\rm{H}}$ during the continuum fitting (\S\,\ref{subsec:continuum_fitting}), the burst strength and $n_{\rm{H}}$ PDFs are not constrained at all. Between the first (i.e.\,top) and second row, we add another line ratio, and so on, until most of the properties of interest are strongly constrained. We consider a range of line ratios (e.g.\,see Figure\,\ref{fig:diagnostics_vs_obs} for some examples), particularly those based on strong lines. 

The PDFs shown in Figures\,\ref{fig:parsec_fitting_example} and \ref{fig:geneva_fitting_example} are derived from fitting the Parsec and Geneva libraries, respectively, to the Antennae HII region, A901. Except for the [\ion{S}{ii}]~$\lambda6717,31$\AA\,and the BPT diagnostics that correspond to R$_1$ and R$_2$, the rest include various line ratios involving [\ion{O}{iii}], [\ion{S}{iii}], [\ion{Cl}{iii}], [\ion{Ar}{iii}], [\ion{S}{ii}], [\ion{O}{ii}], [\ion{N}{ii}], and Balmer lines (see Figure\,\ref{fig:diagnostics_vs_obs} for the behaviours of some of the line ratios considered with respect to models), and are included in the fits in no particular order. Note also that we have considered combinations of low and high ionisation line ratios to extract all potential metallicity, $n_{\rm{H}}$ and age solutions. 

The first column shows the PDF of the burst strength, which is somewhat poorly constrained. As discussed earlier, the presence of different ionisation structures within an HII region means that different ionisation species dominate different regions. The lower ionisation species (e.g.\,[\ion{O}{ii}], [\ion{S}{ii}] and [\ion{N}{ii}]) tend to mostly occur in outer parts, while higher excitation lines, such as [\ion{O}{iii}], [\ion{S}{iii}], [\ion{Cl}{iii}] and [\ion{Ar}{iii}], dominate the innermost parts of an HII region. As we determine the burst strength that can produce observed emission line ratios using $U$, with lower versus higher ionisation lines pointing to low and high $U$ values, the burst strength can indirectly be sensitive to the ionisation structure of an HII region. Consequently, the PDF of the burst strength can yield multiple solutions as shown. 

The $n_{\rm{H}}$ is shown in the second column. Immediately following the inclusion of the [\ion{S}{ii}]~$\lambda6717,31$\AA\,ratio, a well-known gas density diagnostic, the PDF appear to converge. The [\ion{S}{ii}]~$\lambda6717,31$\AA\,is, however, a probe of the low-density, low ionisation zones of an HII region, therefore, with the inclusion of additional line diagnostics, especially those with higher excitation energies, the PDF either converges to provide an average $n_{\rm{H}}$ or two potential $n_{\rm{H}}$ solutions. Note also that there is a degeneracy between $n_{\rm{H}}$ and burst strength, where high $n_{\rm{H}}$ imply low ionisation parameters, and hence, low burst strengths, and vice versa.  

In the third column, we show the PDFs of the gas metallicity in units of solar metallicity. The continuum fitting, principally the Paschen jump, provides some constraints on $Z_{\rm{g}}$ as evident in the first PDF. The strongest constraints, however, appear to come from the emission line ratio diagnostics, which can be expected given the dependence of some emission lines (e.g.\,[\ion{N}{ii}], [\ion{S}{ii}]) on abundance. 

The PDFs of the stellar metallicity is presented in the fourth column. As apparent in the first PDF, the continuum model fitting, informed in particular by the WR bumps and the Paschen jump, add strong constraints on $Z_{\rm{s}}$. The inclusion of the emission line ratio diagnostics appears to strengthen the constraints on $Z_{\rm{s}}$ further. Overall, we find that constraining $Z_{\rm{s}}$ is generally more challenging than constraining $Z_{\rm{g}}$ -- see \S\,\ref{subsec:double_solutions} for a discussion on the possible ways of improving the robustness of $Z_{\rm{s}}$ derivations. 

We show the likelihood distributions of the T$_{\rm{e}}$ in the fifth column. Since T$_{\rm{e}}$ is a strongly decreasing function of metallicity (Figure\,\ref{fig:Te_dependence_onZ_logU_nH}), the continuum fitting process notably constrains the PDF. Recall that at solar-like metallicities, the T$_{\rm{e}}$ dependence on the burst strength (i.e.\,$U$) is weak (see Figure\,\ref{fig:Te_dependence_onZ_logU_nH}), therefore, the lack of strong constraints on the burst strength parameter does not significantly influence the PDFs of the T$_{\rm{e}}$ and $n_{\rm{H}}$.  

The final column shows the likelihood distribution for the light weighted age of the young (i.e.\,$<10$ Myr) stellar population. Owing mainly to the presence of WR features, the PDF of the young ages is tightly constrained by the continuum fitting process, with relatively little constraint imposed by the emission line ratio diagnostics. 

The best-fitting spectral models corresponding to the final PDFs shown in Figures\,\ref{fig:parsec_fitting_example} \& \ref{fig:geneva_fitting_example} are presented in in the top and bottom panels of Figure\,\ref{fig:A901_fits}, respectively. The best-fitting spectral model is shown green, the best-fitting model adjusted with a Legendre polynomial fit to the residuals in red and the observed spectrum in black.  The inset within each panel shows the distribution of ages of the underlying stellar populations contributing to the best-fitting spectral model (i.e.\,the best-fitting SFH).

The offset between the best-fitting spectral model (green) and the best-fit with residuals removed (red) is relatively small, suggesting that the best-fit model has adequately captured the properties of the dominant underlying stellar populations residing in these HII regions. The variations in the shape of the continuum over certain wavelength regimes carry information about the types of old stellar populations present in a given HII region. Therefore, provided that any issues related to data reduction have not impacted the shape of the observed spectrum, the offsets between the models and data are indicative of the improvements required in the modelling of older stellar populations to fully capture their evolution. 

The characteristic properties obtained for A901 (Figure\,\ref{fig:A901_fits}) and for A893 by fitting the Parsec and Geneva model libraries is presented in Table\,\ref{table:geneva_vs_parsec}. In general, the best-fitting parameters are largely consistent between the Geneva and Parsec model libraries. As discussed in \S\,\ref{subsec:props_wr}, the WR phase appear to occur at earlier times in stellar models generated using the Parsec isochrones. This could be a result of the stellar metallicity -- age degeneracy given the difference in Parsec versus Geneva stellar metallicity estimates for A901. Also, the WR features, the blue bump, in particular, appear to be sharper in Parsec models than in Geneva ones. As a result, the light-weighted ages derived from fitting the Parsec library is somewhat biased towards younger ages than those derived from fitting the Geneva library for spectra displaying prominent WR features\footnote{We emphasis that this is only a slight bias and is largely seen in the fitting of spectra that show prominent WR features. Otherwise, the young ages derived from fitting Parsec models appear to be biased toward older ages. See Appendix\,\ref{appB} for a discussion.}.  

Finally, we present a more thorough comparison and a detailed discussion on the differences between Parsec and Geneva solutions in \S\,\ref{sec:props}.

\subsection{Discriminating between double solutions}\label{subsec:double_solutions}
\begin{figure*}
\begin{center}
\includegraphics[scale=0.55, trim={3.5cm 0.7cm 0.8cm 2.1cm},clip]{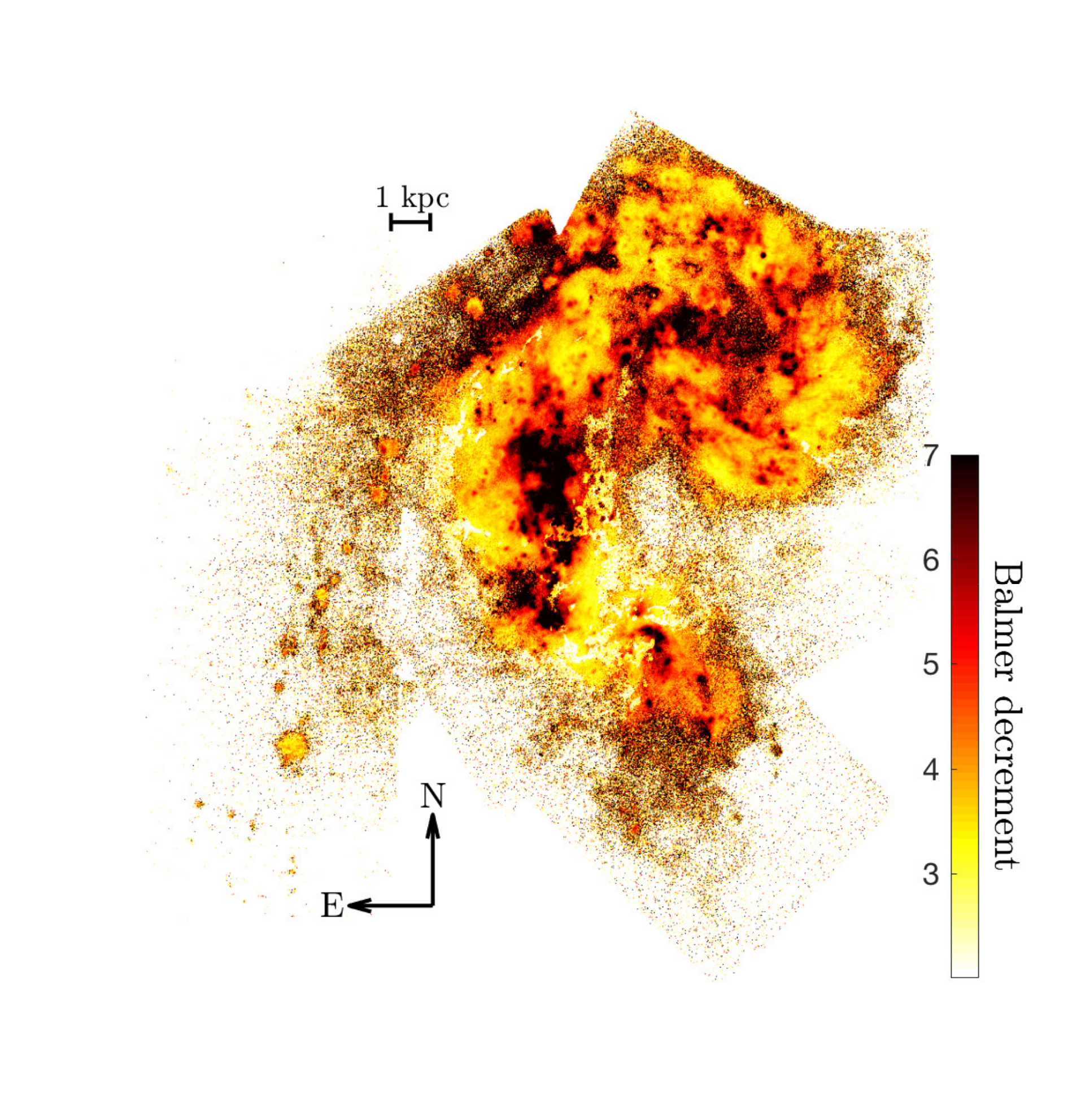}
\includegraphics[scale=0.55, trim={3.5cm 0.7cm 0.8cm 2.1cm},clip]{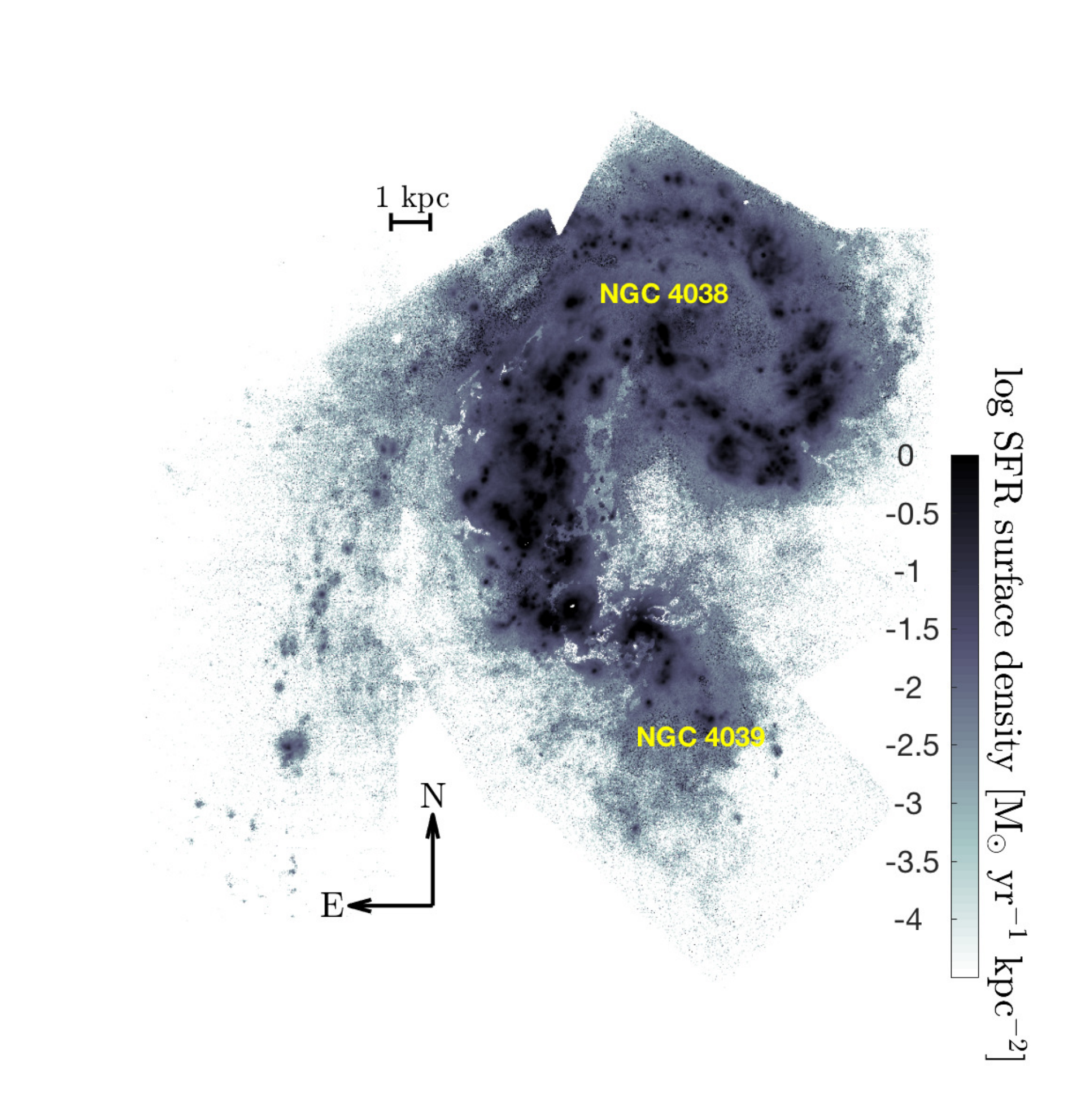}
\caption{The Balmer decrement (left) and log intrinsic H$\alpha$ SFR surface density (right) maps for the central regions of the Antennae galaxy, illustrating the wide-range range in dust obscurations and SFR surface densities sampled by the young, massive stellar clusters in the galaxy. The loop-like region west of the northern galaxy NGC 4038 is called `the western loop', and the region between NGC 4038/39 that show significantly enhanced SFR surface densities is called `the overlap region'. }
\label{fig:Antennae_data_demons}
\end{center}
\end{figure*}

Several degeneracies, like the age--stellar metallicity and n$_{\rm{H}}$--burst strength, can affect the fitting outcomes, which generally lead to PDFs displaying multiple peaks, i.e.\,potential solutions. Figure\,\ref{fig:double_solutions} demonstrates the double solutions obtained for stellar metallicity in the fitting of the spectrum A927. The top panel shows the PDFs obtained from fitting the Geneva (black dashed) and Parsec (brown filled) model libraries, with the PDFs of stellar metallicity displaying a broad distribution with two distinct peaks. The panels that follow show the best-fitting model spectra corresponding to each stellar metallicity solution. 

The age--stellar metallicity degeneracy arises from the fact that with increasing stellar metallicity, the stellar continuum features start to appear at earlier ages. Therefore, there can be multiple age and stellar metallicity solutions to a given spectrum, with a higher metallicity solution predicting younger stellar ages than a lower metallicity solution. In the case of A927, both Geneva and Parsec models predict a best-fitting stellar metallicity that is slightly super-solar, as well as a second stellar metallicity solution that is slightly sub-solar. While the modelling of the WR features can be used to distinguish between the Parsec solutions clearly, the same cannot be used to discriminate between the Geneva solutions. In these cases, we assess `the best-fit model' by considering the level of agreement between the best-fitting model and the observations over the wavelengths regimes containing the blue (4584--4811\AA) and red (5280--5861\AA) WR features, and the Paschen Jump (7827--8240\AA) as well as the telluric residuals prone 6875--7551\AA\,region. These are the wavelength ranges that generally show notable residuals. Another method of constraining multiple stellar metallicity solutions is through the best-fitting gas metallicity as the metallicity of the stars is unlikely to be significantly different from the metallicity of the gas from which they were formed. Therefore, the best-fitting gas metallicity can be used as a prior in the determination of the stellar metallicity.  In the case of A927, however, all the best-fitting solutions appear to be largely similar in their distributions of age, metallicity, T$_{\rm{e}}$, and $n_{\rm{H}}$.
  
The degeneracy between n$_{\rm{H}}$ and burst strength is mostly related to the lower versus higher ionisation nature of the emission line ratios used in the fitting. The line luminosity constraints based on higher ionisation emission lines (e.g.\,[\ion{O}{iii}]~$\lambda5007$\AA, [\ion{Ar}{iii}]~$\lambda7135$\AA, [\ion{S}{iii}]~$\lambda6312$\AA, and [\ion{S}{iii}]~$\lambda9069$\AA) tend to bias n$_{\rm{H}}$ (burst strength) towards higher (lower) values, and vice versa with lower ionisation emission line constraints (e.g.\,[\ion{N}{ii}]~$\lambda6584$, $\lambda5755$\AA, [\ion{S}{ii}]~$\lambda$6717,31\AA). Therefore, this apparent degeneracy is perhaps a tracer of the ionisation structure of an HII region. 

\section{The MUSE Antennae Project}\label{sec:data}
\begin{figure*}
\begin{center}
\includegraphics[scale=0.5, trim={3.5cm 2.1cm 0.5cm 2.1cm},clip]{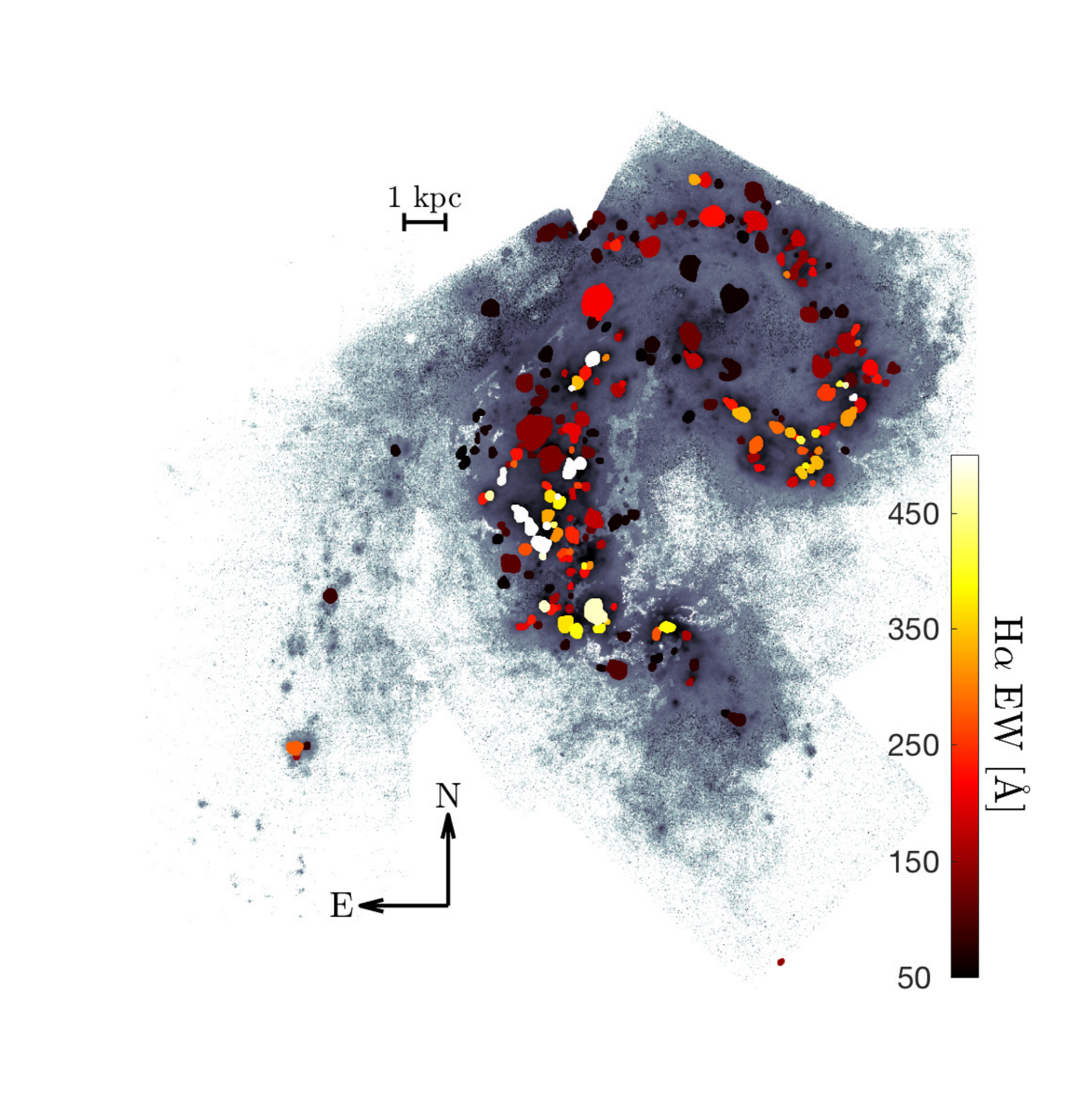}
\includegraphics[scale=0.5, trim={1.6cm 2.1cm 1.0cm 2.1cm},clip]{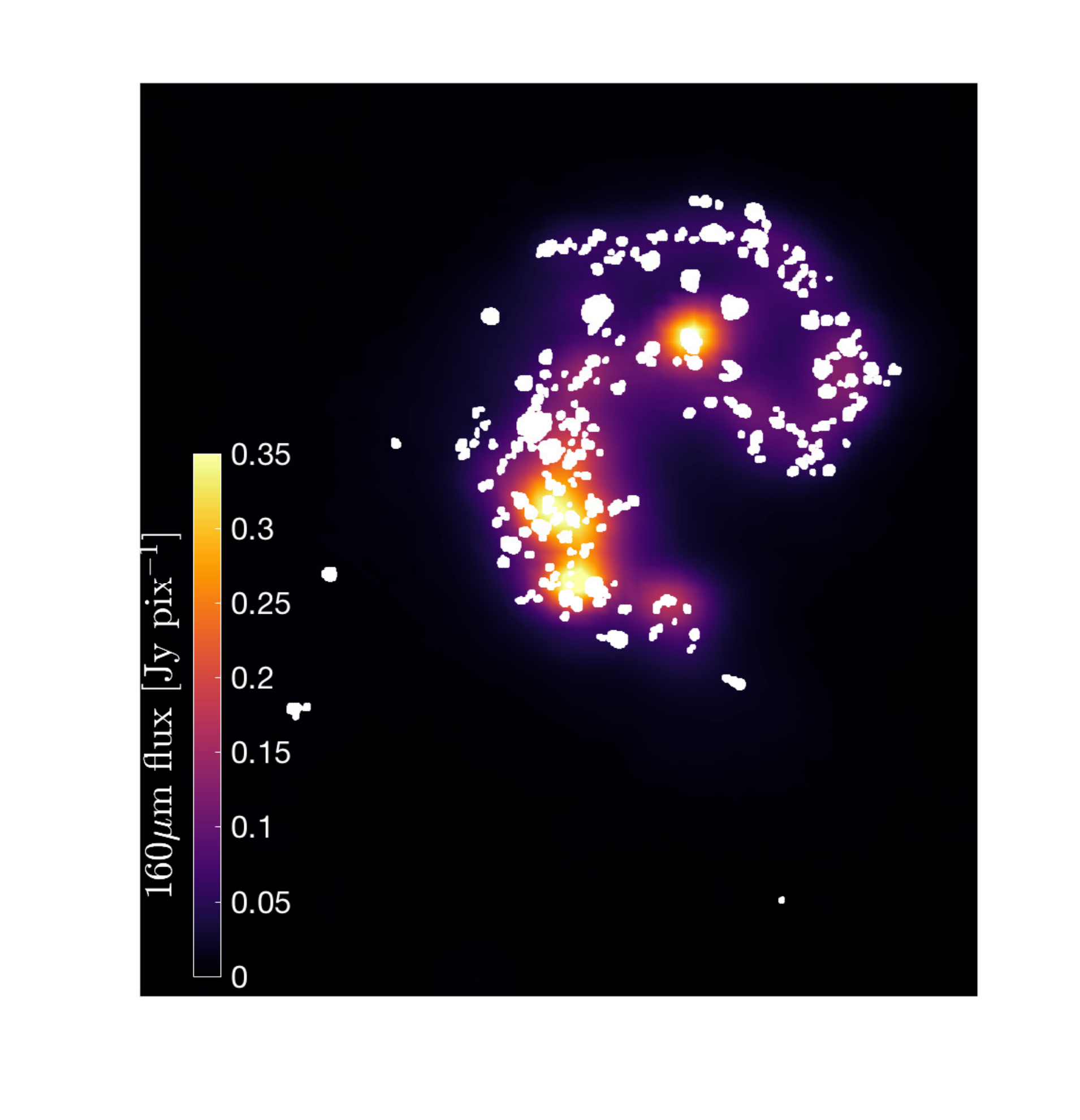}
\caption{The HII regions in a `starburst' phase (i.e.\,H$\alpha$ EW>50\AA) over-plotted on the SFR surface density map of the central regions of the Antennae galaxy and colour-coded by their average H$\alpha$ EW (left panel), and the same regions over-plotted on the \textsc{Herschel}160$\mu$m map of the Antennae galaxy (right panel). The maker size denotes the relative size of the HII regions. }
\label{fig:Antennae_HIIregions}
\end{center}
\end{figure*}
\begin{figure*}
\begin{center}
\includegraphics[scale=0.342, trim={0.cm 0.cm 1.75cm 1.2cm},clip]{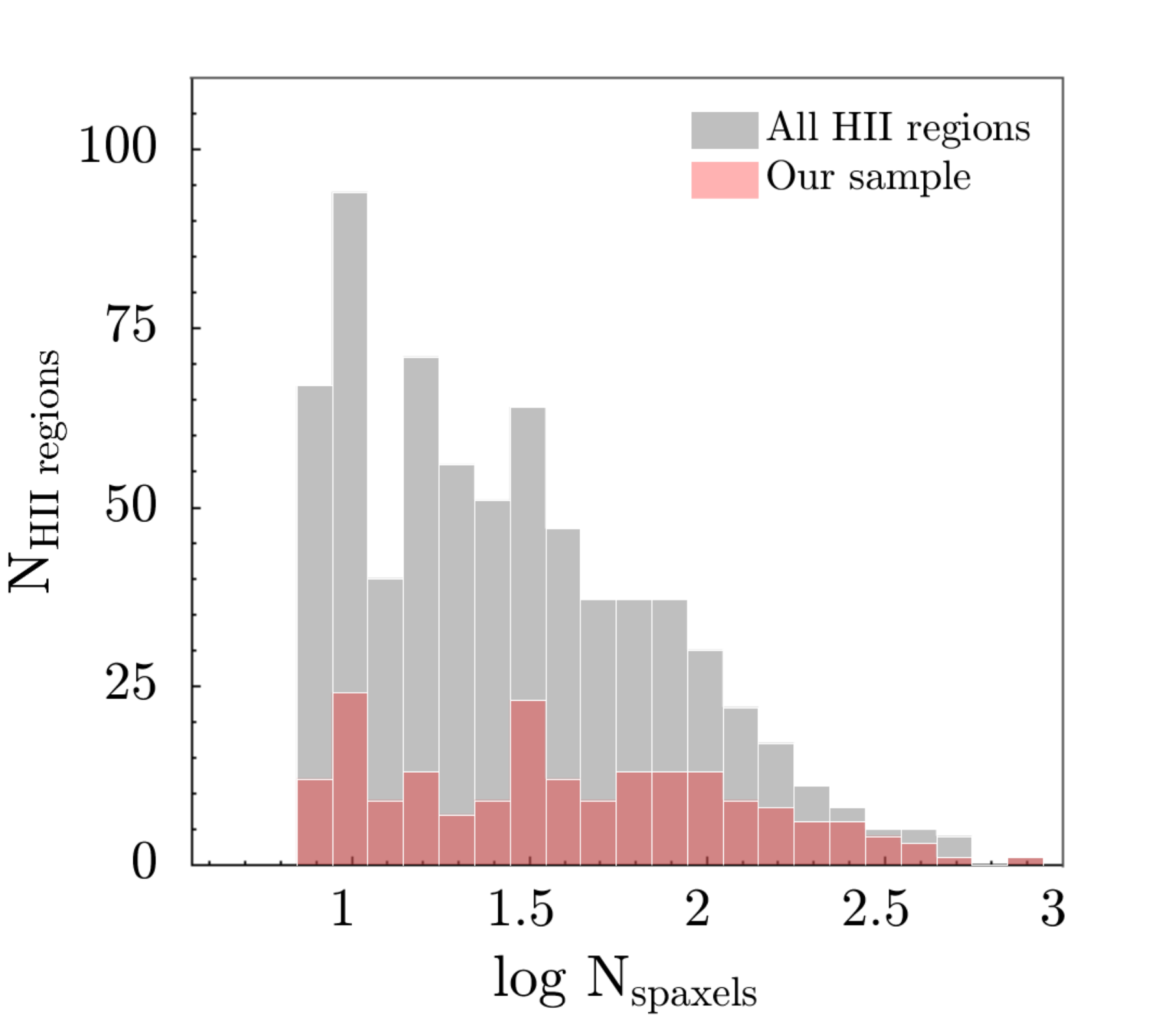}
\includegraphics[scale=0.34, trim={0.cm 0.cm 1.75cm 1.2cm},clip]{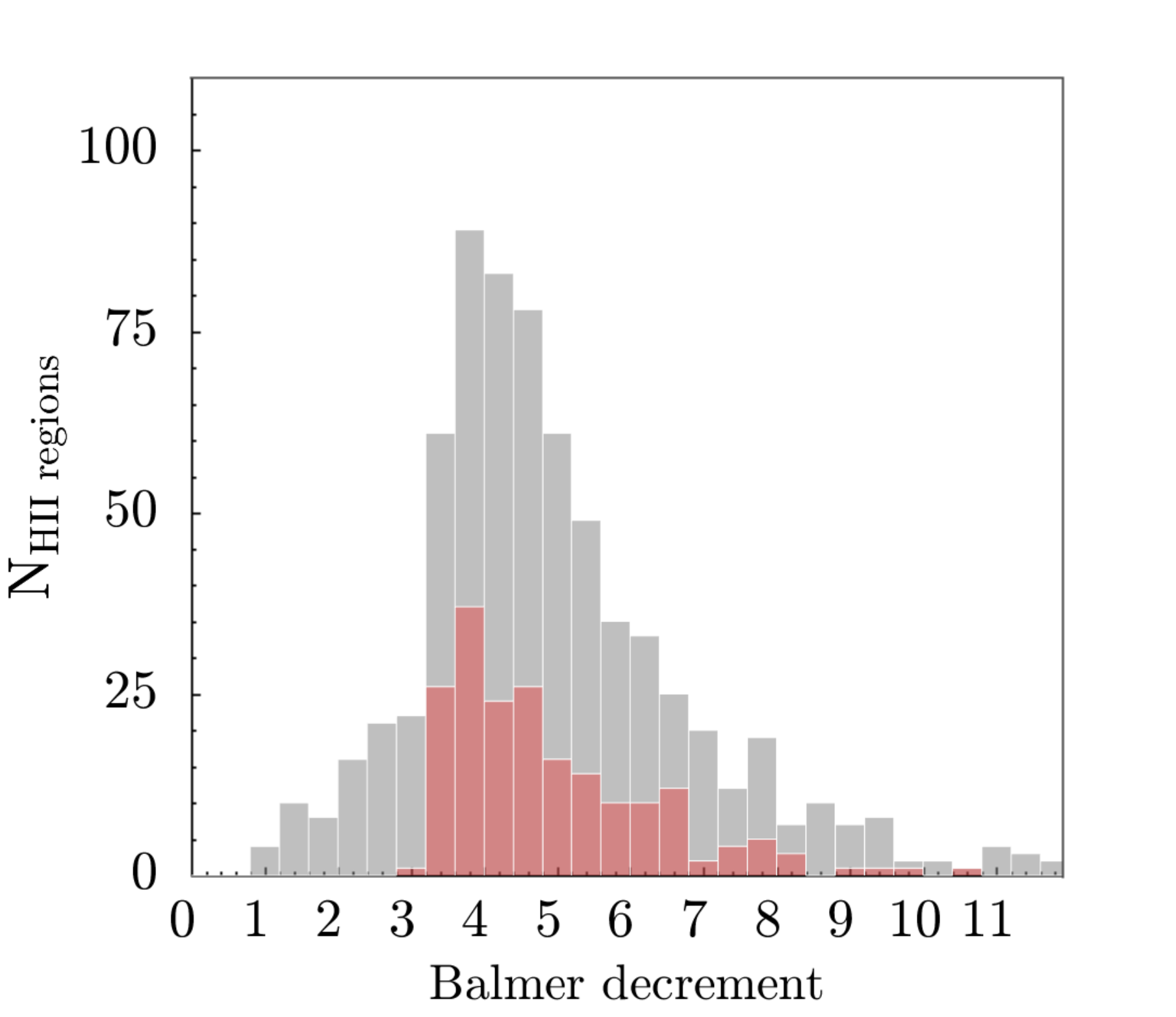}
\includegraphics[scale=0.34, trim={0.cm 0.cm 1.75cm 1.2cm},clip]{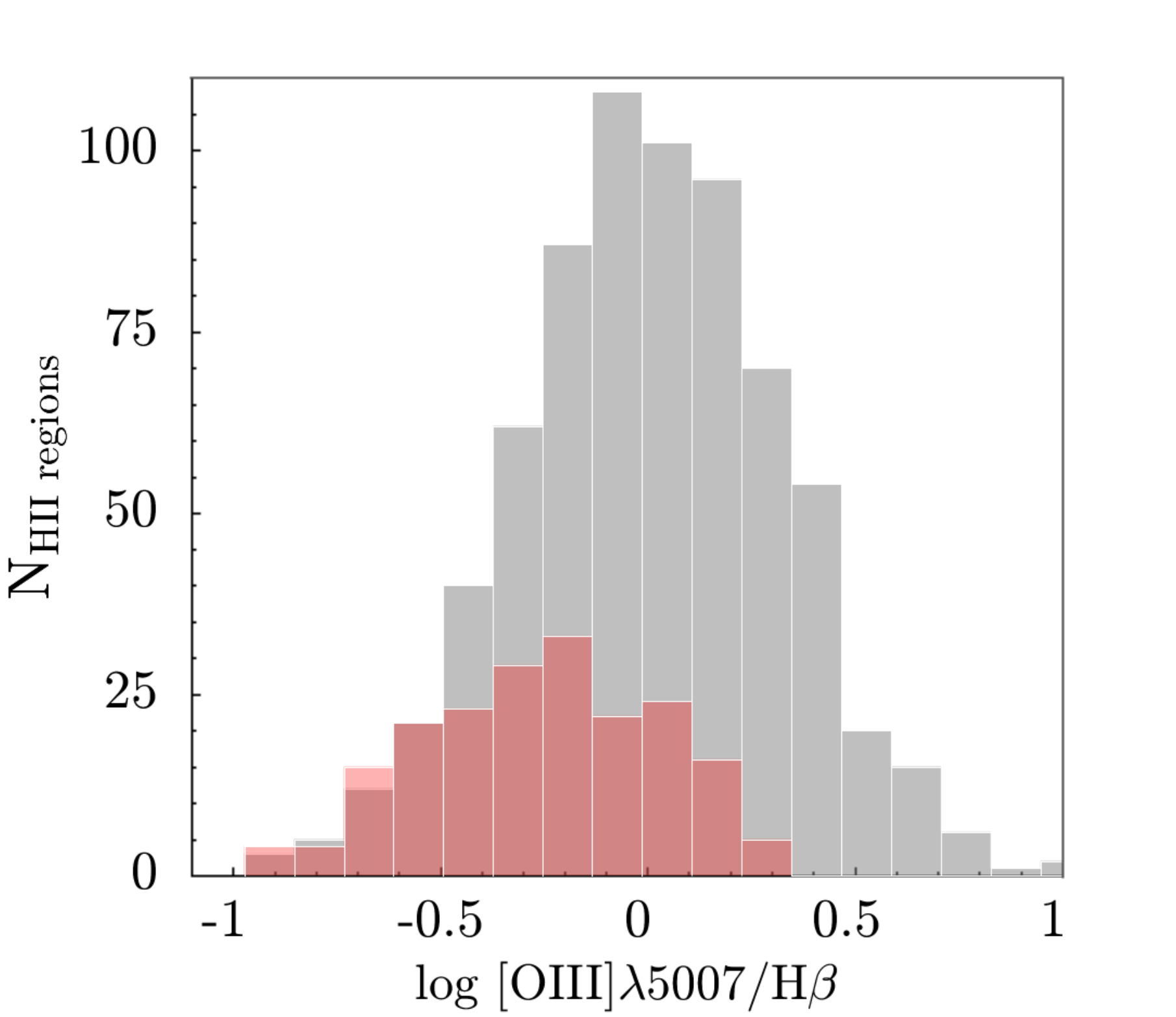}
\caption{From left-to-right, the distribution of all HII regions in the Antennae galaxy compared that of the H$\alpha$ EW>50\AA\,regions as a function of size (indicated in terms of the total number of spaxels defining a given region), Balmer decrement, and [\ion{O}{iii}]$\lambda5007$/H$\beta$ flux ratio. }
\label{fig:Antennae_HIIregions_compare_withFull}
\end{center}
\end{figure*}

The observations of the Antennae galaxy were taken with the Multi-Unit Spectroscopic Explorer \citep[MUSE;][]{Bacon2012} at the 8m Very Large Telescope (VLT), with a $1'\times1'$ field-of-view between April-May 2015 and February-May 2016. The wide-field mode set up with extended wavelength configuration ($\sim4600-9350$ \AA\,spectral coverage at relatively high resolution of R$\sim$3000) and $0.2"$ sky sampling was employed to observe the central regions and the tip of the southern tidal tail of the Antennae galaxy.

The MUSE observation strategy of the Antennae galaxy and the data processing are described in detail in \cite{Weilbacher2018}. Briefly, most pointings composed on 1350s exposures taken with a spatial dithered pattern at fixed position angle, with several shallow extra pointings obtained at different position angles. Additionally, the observations of the central Antennae regions were supplemented with 200s sky exposures. All exposures, except for one shallow central observation, were taken in clear or photometric conditions. The overall variation in seeing is approximately $0.5"$ to $1.2"$, with an average in the sub-arcsecond regime. The depth of the observations vary across the central regions \citep[Figure\,2 of][]{Weilbacher2018}, and is taken into account through the relative weights assigned during the data processing, done consistently using the MUSE pipeline as outlined in \cite{Weilbacher2018}, to produce data cubes of log and linear samplings. For the analysis presented in this study, we use the data cube with (linear) sampling of $0.2"\times0.2"\times1.25$ \AA\,voxel$^{-1}$.  

In Figure\,\ref{fig:Antennae_data_demons}, we show the Balmer decrement (i.e.\,H$\alpha$ to H$\beta$ flux ratio) and intrinsic star formation rate (SFR) surface density maps for the central regions of the Antennae galaxy. The Balmer decrement is an obscuration sensitive parameter, and its departure from the Case B theoretical value of 2.86 at an electron temperature of 10$^4$ K and an electron density of 100 cm$^{-3}$ \citep{Osterbrock} is an indication of the dust extinction along the line-of-sight. To construct the maps, we impose a signal-to-noise cut of 3 on both H$\alpha$ and H$\beta$ fluxes, and for the subset of cases where the spaxels show Balmer decrements $<$2.86, which can result from intrinsically low-reddening combined uncertainties in line flux calibration and measurement, we assume no dust obscuration. The HII regions in the Antennae galaxy sample a wide range in dust obscurations and SFR surface densities, making the Antennae data ideal for studying the environments of young, massive stellar clusters formed in a violent gas-rich merger. 

\subsection{The HII regions in the Antennae galaxy}
For this study, we use the HII regions discussed in \cite{Weilbacher2018}, which are extracted using the \textsc{dendrograms} tool\footnote{Part of the \textsc{astrodendro} package from \url{http://dendrograms.org}.}. Each HII region represents a peak in the H$\alpha$ flux map, which is then extracted down to the surface brightness level where the corresponding contours join, thus defining the size of the extracted region. 

For the present analysis, we exclude the southern tidal tail and focus on the HII regions in the central merger undergoing a `starburst' following the H$\alpha$ equivalent width (EW)$>50$ \AA\,definition of \cite{Rodighiero2011}. The 236 HII regions meeting the EW criteria of \cite{Rodighiero2011} are shown in Figure\,\ref{fig:Antennae_HIIregions}, over-plotted on the SFR surface density map of the central region and colour-coded by their log H$\alpha$ EW (left panel), and over-plotted on the 160 $\mu$m map of the Antennae galaxy (right panel).

The right panel of Figure\,\ref{fig:Antennae_data_demons} and the left panel of Figure\,\ref{fig:Antennae_HIIregions}, together, show that the EW$>$50\AA\,regions are generally characterised by log H$\alpha$ SFR surface densities$>-1$ M$_{\odot}$ yr$^{-1}$ kpc$^{-2}$, highlighting the extreme environments inhabited by young, massive stellar clusters.  Similarly, the EW$>50$ \AA\,regions show a good correlation with the 160 $\mu$m distribution, particularly in the overlap region, demonstrating that the HII regions selected for this analysis samples all the significant star-formation in the central regions of the Antennae galaxy. 
We also show the distribution of the HII regions selected for this study compared to that of all HII regions with signal-to-noise in H$\alpha>10$ as a function three different properties in Figure\,\ref{fig:Antennae_HIIregions_compare_withFull}. Overall, the size and Balmer decrement distributions of the starbursting HII regions largely overlaps with that of the full sample, suggesting the selected starbursting regions probe a wide range in environments. The distribution of [\ion{O}{iii}]~$\rm{\lambda}$5007/H$\beta$ of starbursting regions is, however, skewed towards lower values with respect to the full distribution, which suggests that the selected starbursts are somewhat biased against lower metallicities (see also Figure\,\ref{fig:diagnostics_vs_obs}).

\subsection{The characterisation of spatially correlated noise in the construction of integrated spectra}\label{subsec:noise_characterisation}

For each HII region shown in Figure\,\ref{fig:Antennae_HIIregions}, we create a signal-to-noise weighted integrated spectrum from the individual spectra (i.e.\,spatial pixels) defining that region. Likewise, we also create an integrated error spectrum by analytically propagating the error arrays of all the contributing spatial pixels. The analytically propagated error spectrum is, however, likely to under-represent the true error due to noise in the adjacent pixels being spatially correlated, a consequence of the interpolation procedure used to obtain a regular grid of 0.2"$\times$0.2" in the final data cube. 
Therefore, to characterise the effect of correlated noise, we adopt the error analysis based on the full spectral fitting method described in \cite{Garcia-Benito2015} and \cite{Husemann2013}. 
\begin{figure}
\begin{center}
\includegraphics[scale=0.4, trim={0.1cm 0.cm 0.cm 0.cm},clip]{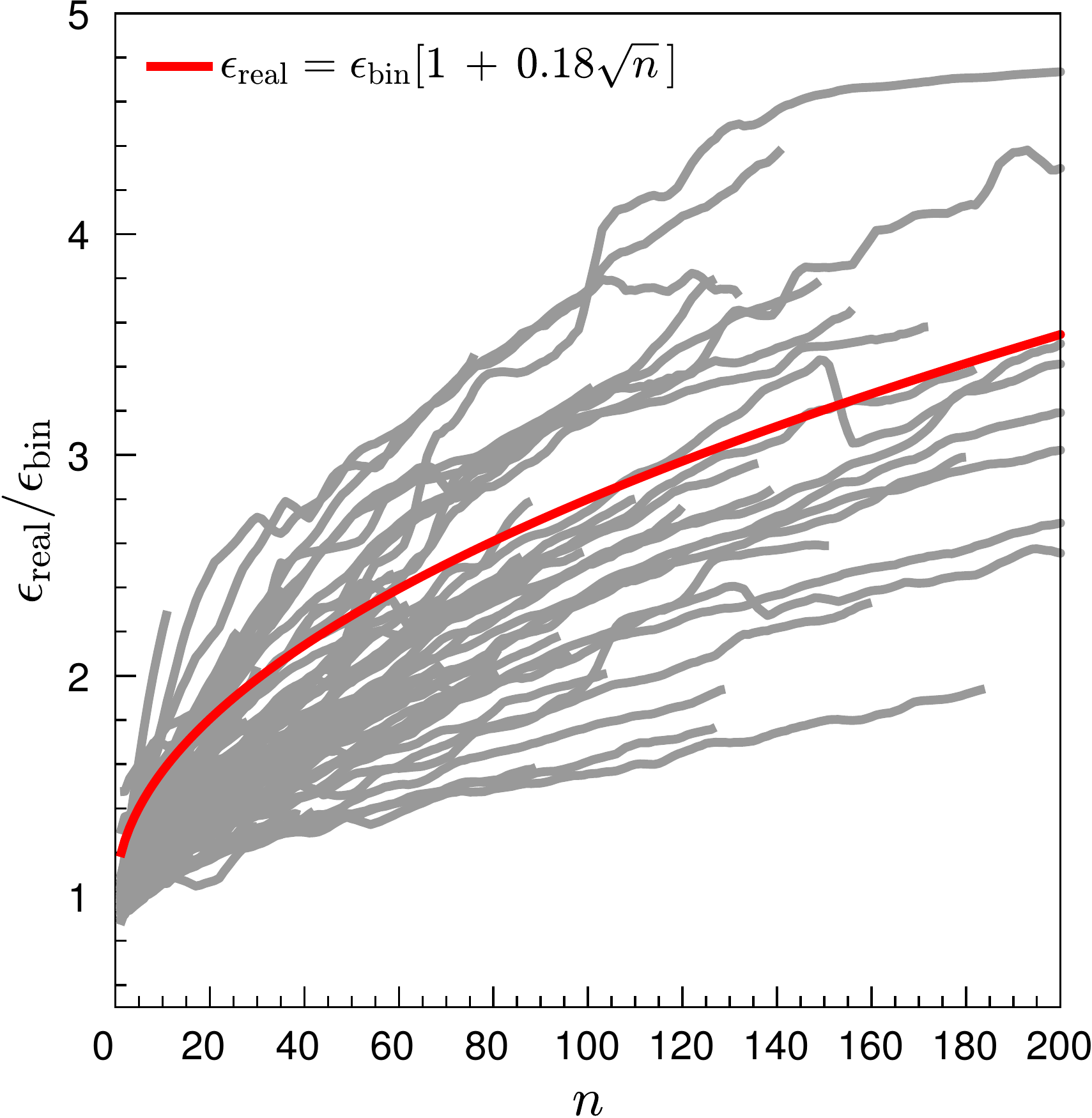}
\caption{The characterisation of the spatially correlated noise. The grey lines denote the individual relationships between the ratios of the real-to-analytically propagated noise (i.e.\,$\epsilon_{\rm{real}}$ to $\epsilon_{\rm{binned}}$) versus the number of spaxels (i.e.\,$n$) obtained for the HII regions considered in this study. The relation that best-describe the data is shown in red. Note that this is not a best-fitting relation. It well-describes the data in the high $n$ region, but slightly overestimates the error in the low $n$ regime.}
\label{fig:noise_character}
\end{center}
\end{figure}

Neglecting any uncertainties or systematic deviations in models, the full spectral fitting analysis can provide a fair assessment of the accuracy of errors of a spectrum. In Figure\,\ref{fig:noise_character}, we show the ratio the real noise ($\epsilon_{\rm{real}}$) to the analytically propagated noise ($\epsilon_{\rm{binned}}$) as a function of the spatial pixels contributing to each HII region (grey lines) considered in this study. The ratio of $\epsilon_{\rm{real}}$ to $\epsilon_{\rm{binned}}$ is determined over the $4750-5600$ and $6600-7600$ \AA\,windows. On average, each relation shows an initial increase in $\epsilon_{\rm{real}}$ to $\epsilon_{\rm{binned}}$, followed by a flattening. We find that this behaviour is best described by the $\epsilon_{\rm{real}} = \epsilon_{\rm{binned}} (1 + 0.18\sqrt{n})$ (indicated by the red line), which we use to scale the analytically propagated error spectrum of each region. 

\section{The characteristic properties of the Antennae HII regions} \label{sec:props}
\begin{figure*}
\begin{center}
\includegraphics[width=0.43\textwidth, trim={3.5cm 3.5cm 1.3cm 1.7cm},clip]{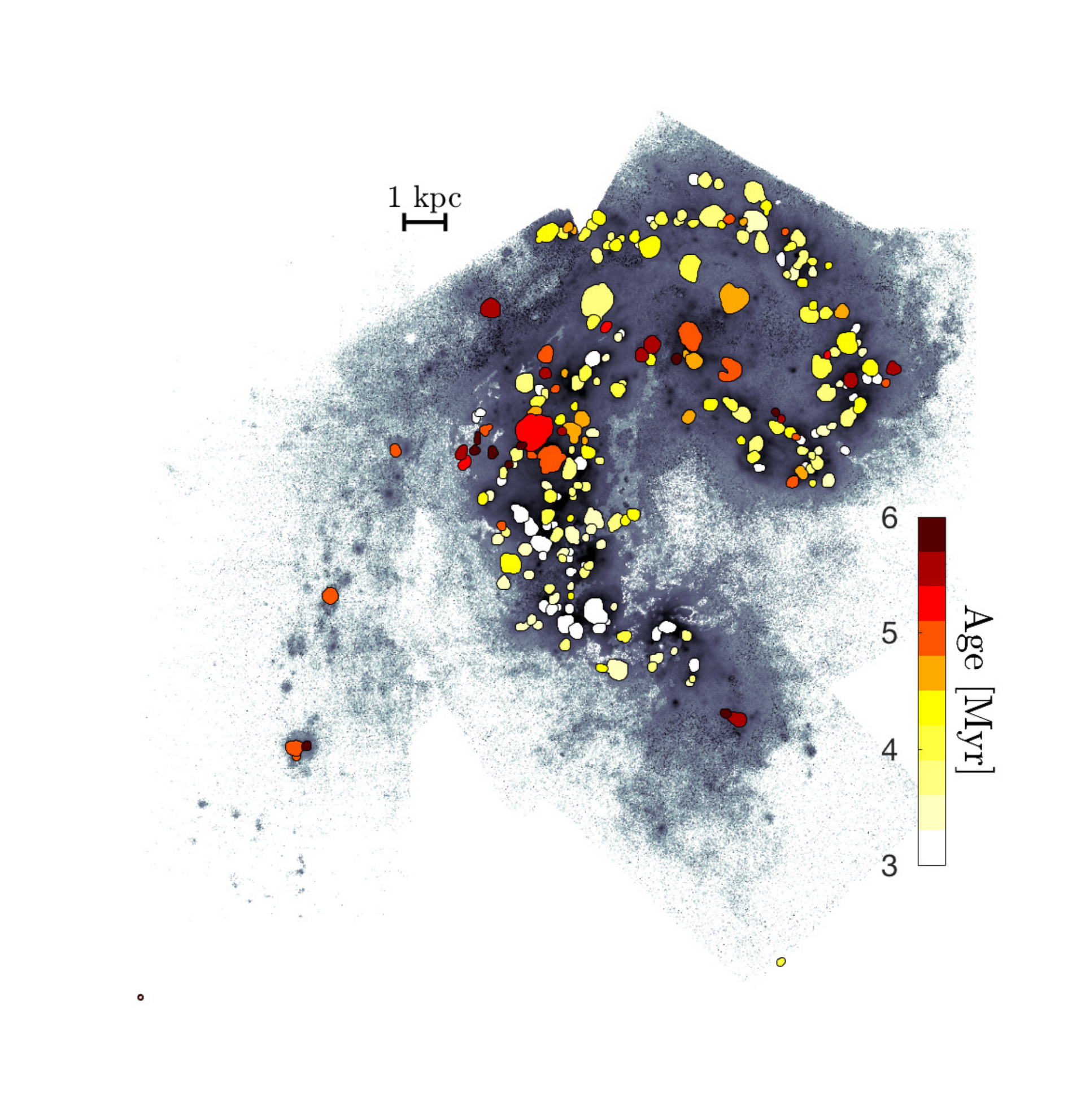} \vspace{-0.0cm}
\includegraphics[width=0.43\textwidth, trim={3.5cm 3.5cm 1.3cm 1.7cm},clip]{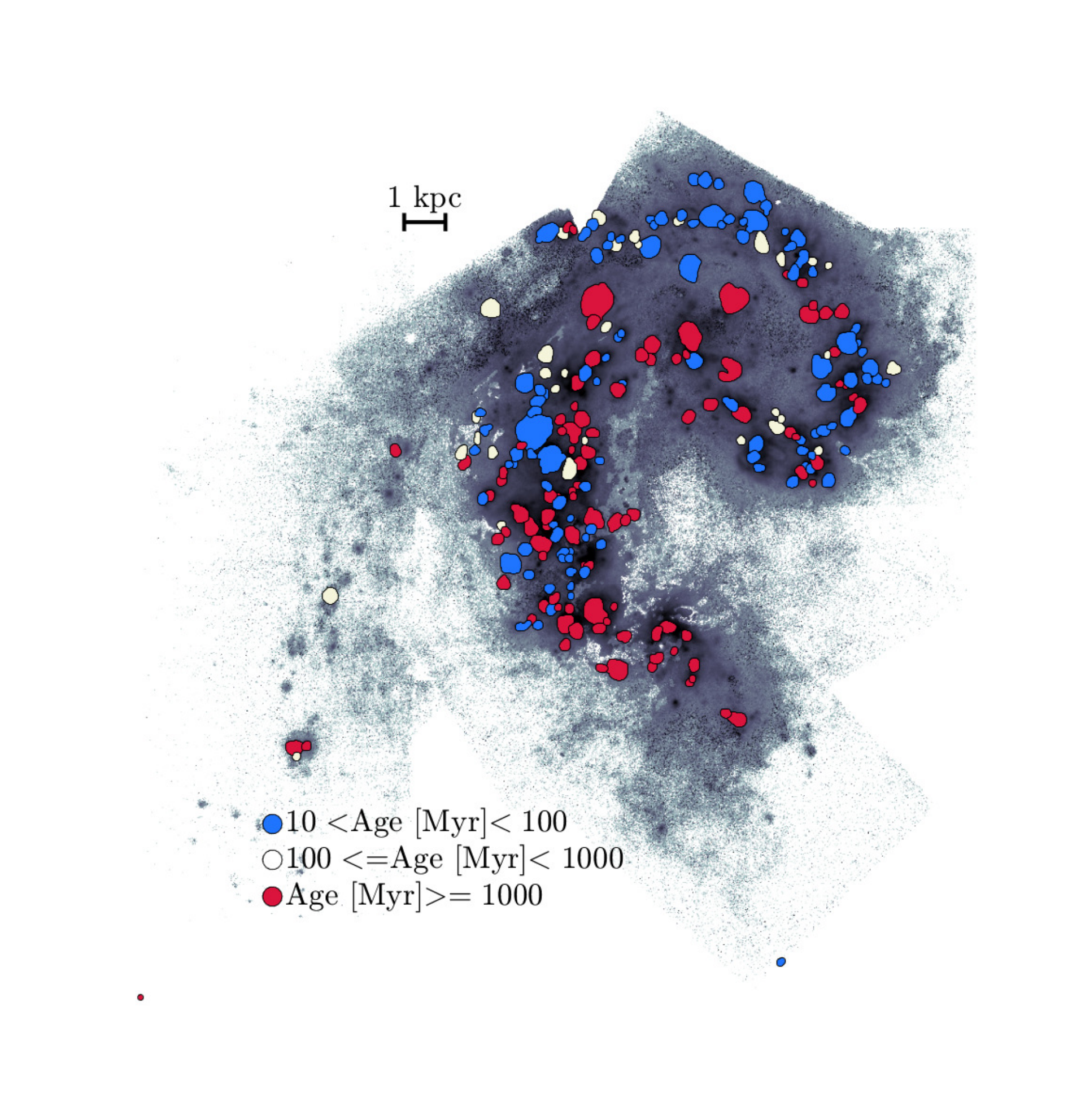} \vspace{-0.0cm}
\includegraphics[width=0.43\textwidth, trim={3.5cm 3.5cm 1.3cm 1.7cm},clip]{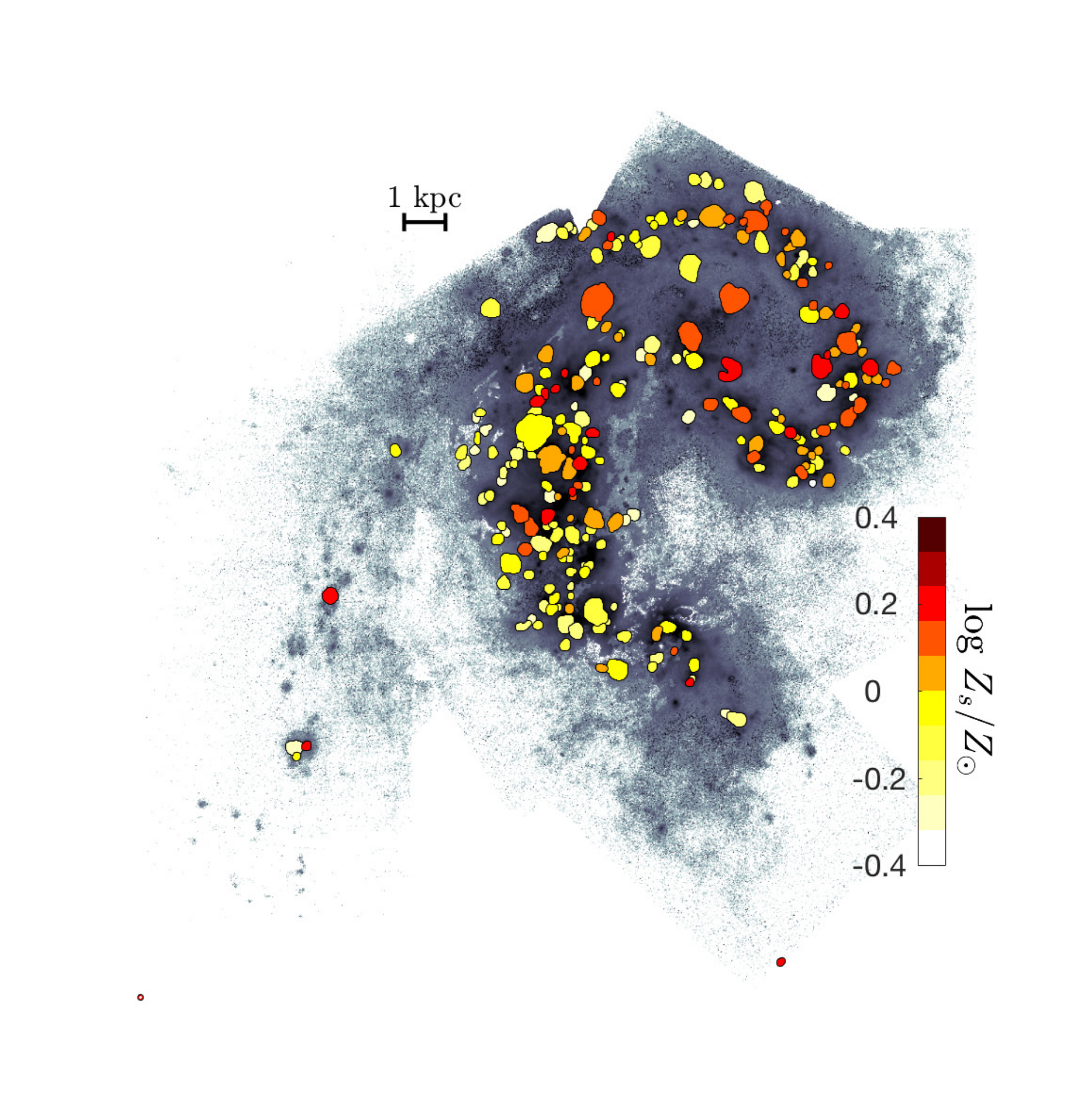} \vspace{-0.0cm}
\includegraphics[width=0.43\textwidth, trim={3.5cm 3.5cm 1.3cm 1.7cm},clip]{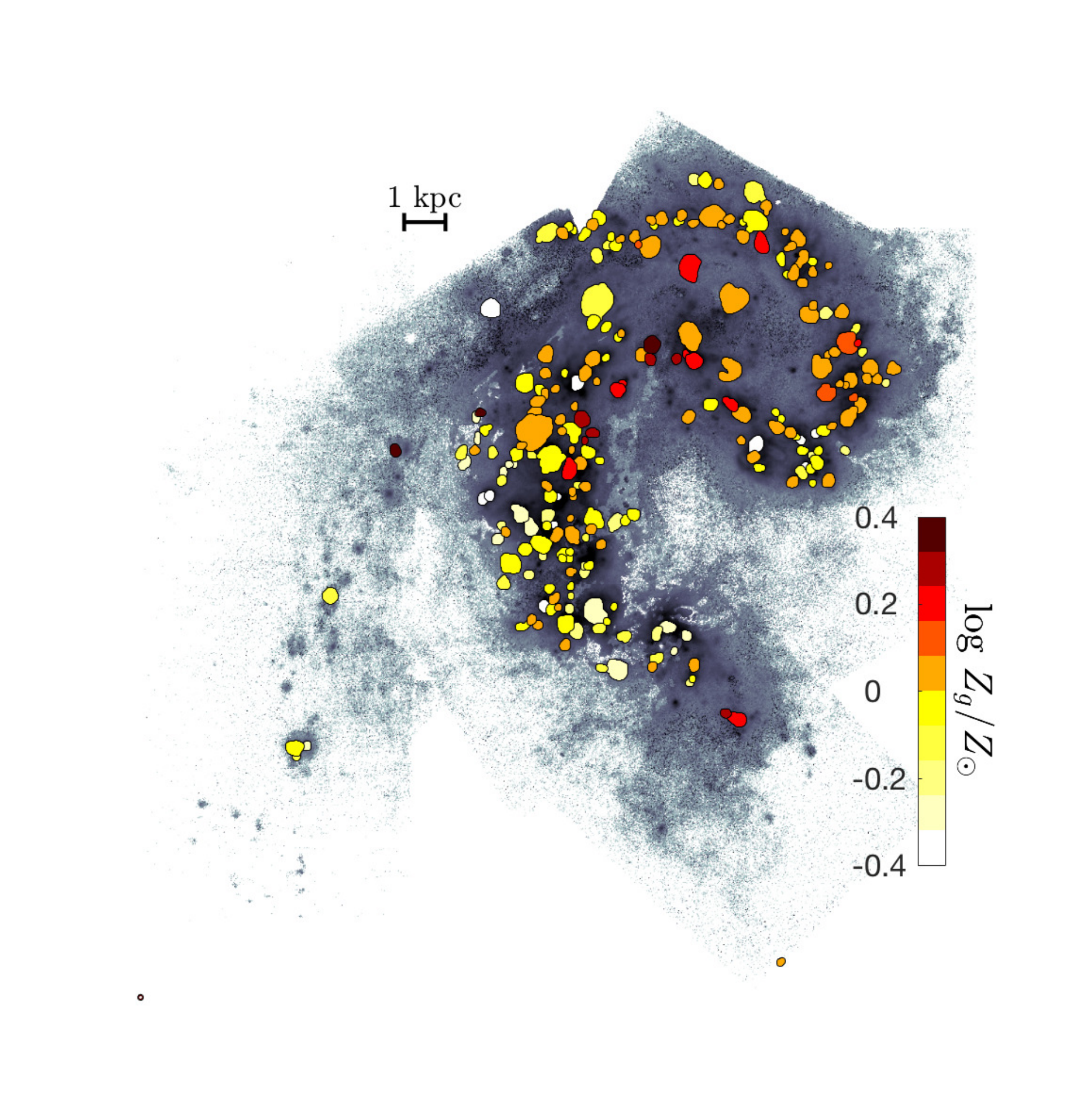} \vspace{-0.0cm}
\includegraphics[width=0.43\textwidth, trim={3.5cm 3.5cm 1.3cm 1.7cm},clip]{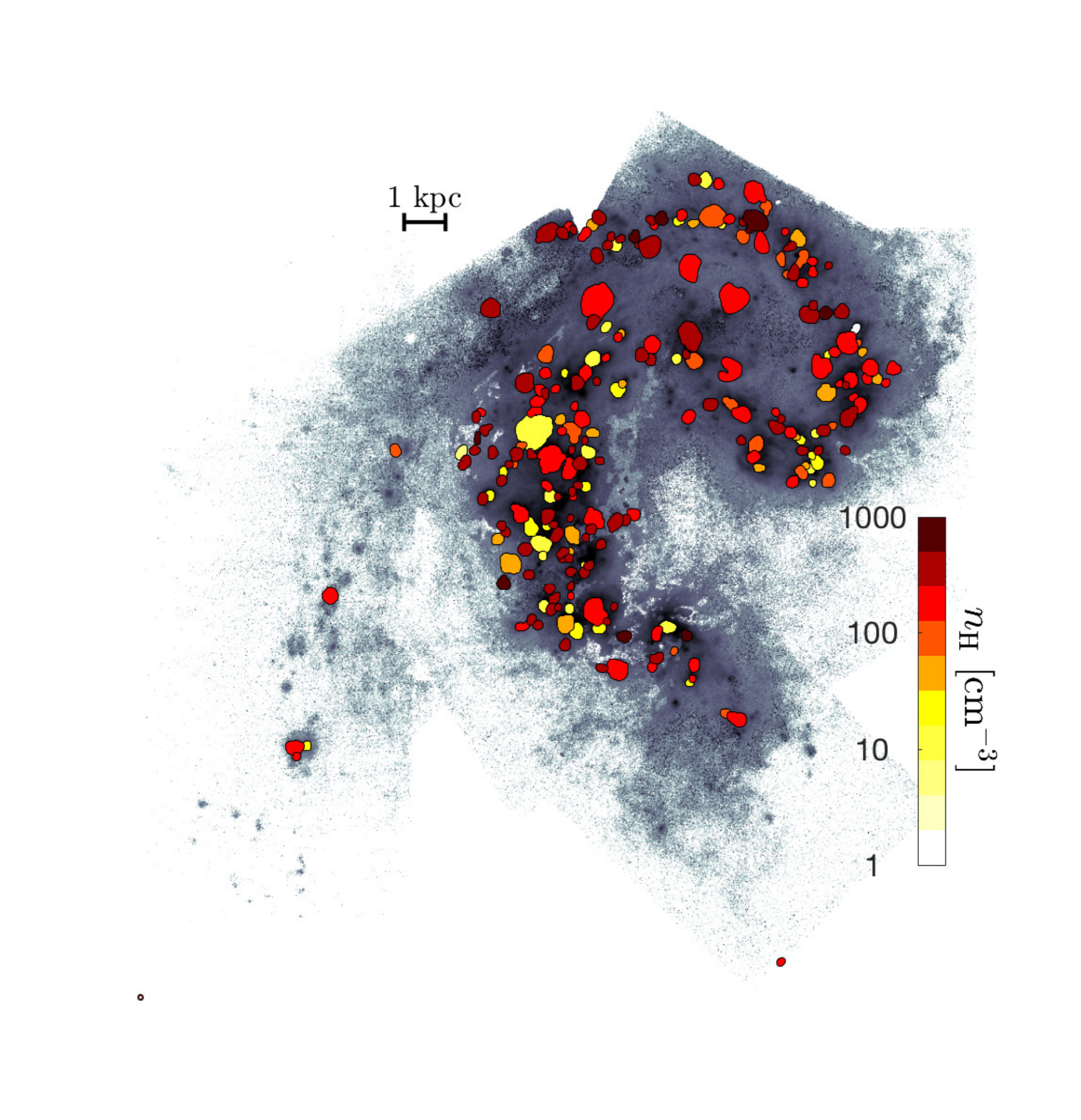} 
\includegraphics[width=0.43\textwidth, trim={3.5cm 3.5cm 1.3cm 1.7cm},clip]{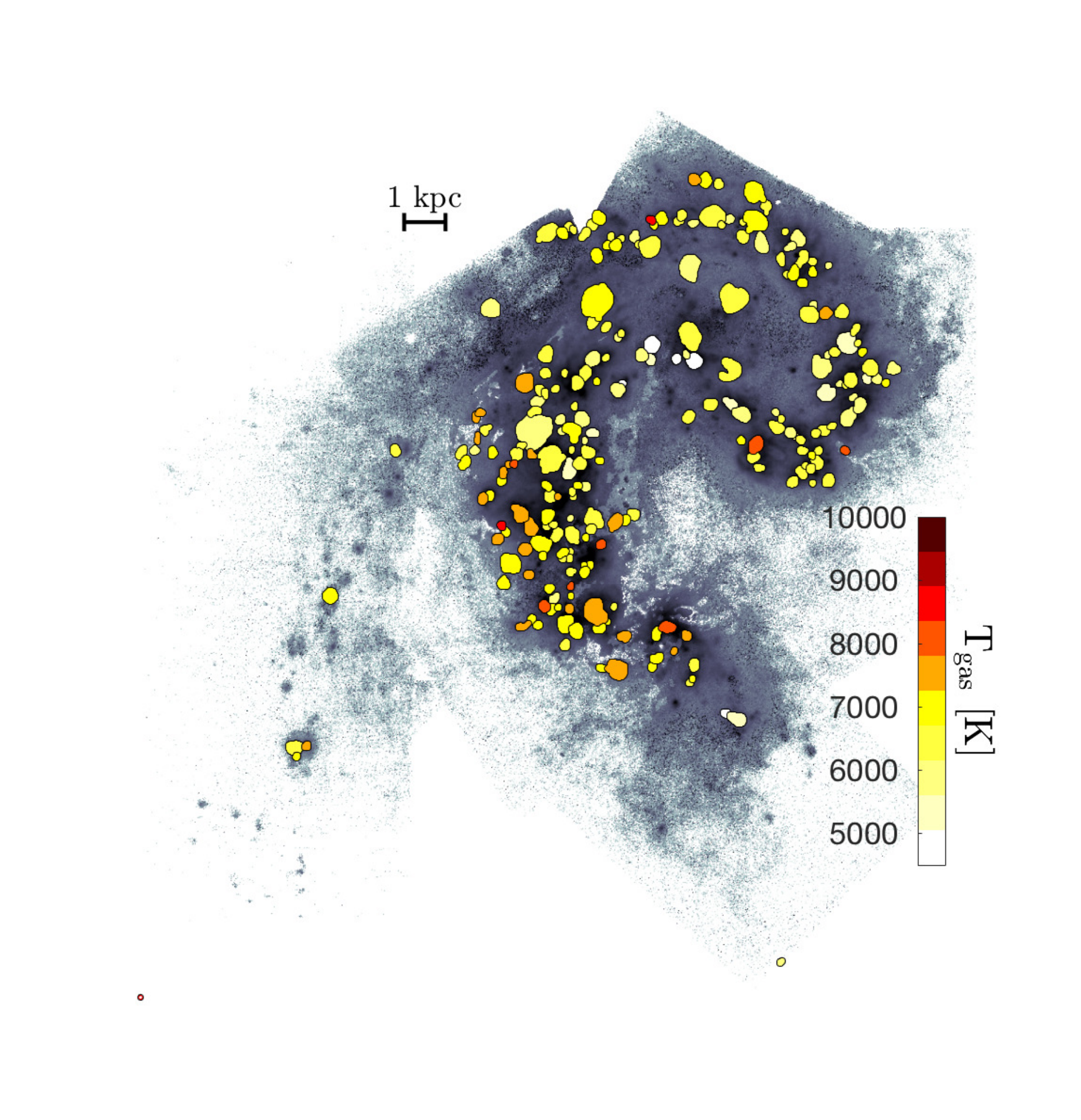}
\caption{The Antennae HII regions colour-coded by the physical properties derived from fitting the Geneva high mass loss models. From left-to-right, top-to-bottom: The light weighted age of the young (i.e.\,<10 Myr) stellar population, the age of the dominant old stellar population, the metallicity of the stars relative to solar metallicity, the metallicity of the star forming gas relative to solar metallicity, $n_{\rm{H}}$ in unit of cm$^{-3}$ and T$_{\rm{e}}$ in unit of K. }
\label{fig:prop_distributions_geneva}
\end{center}
\end{figure*}

\begin{figure*}
\begin{center}
\includegraphics[width=0.43\textwidth, trim={3.5cm 3.5cm 1.6cm 1.7cm},clip]{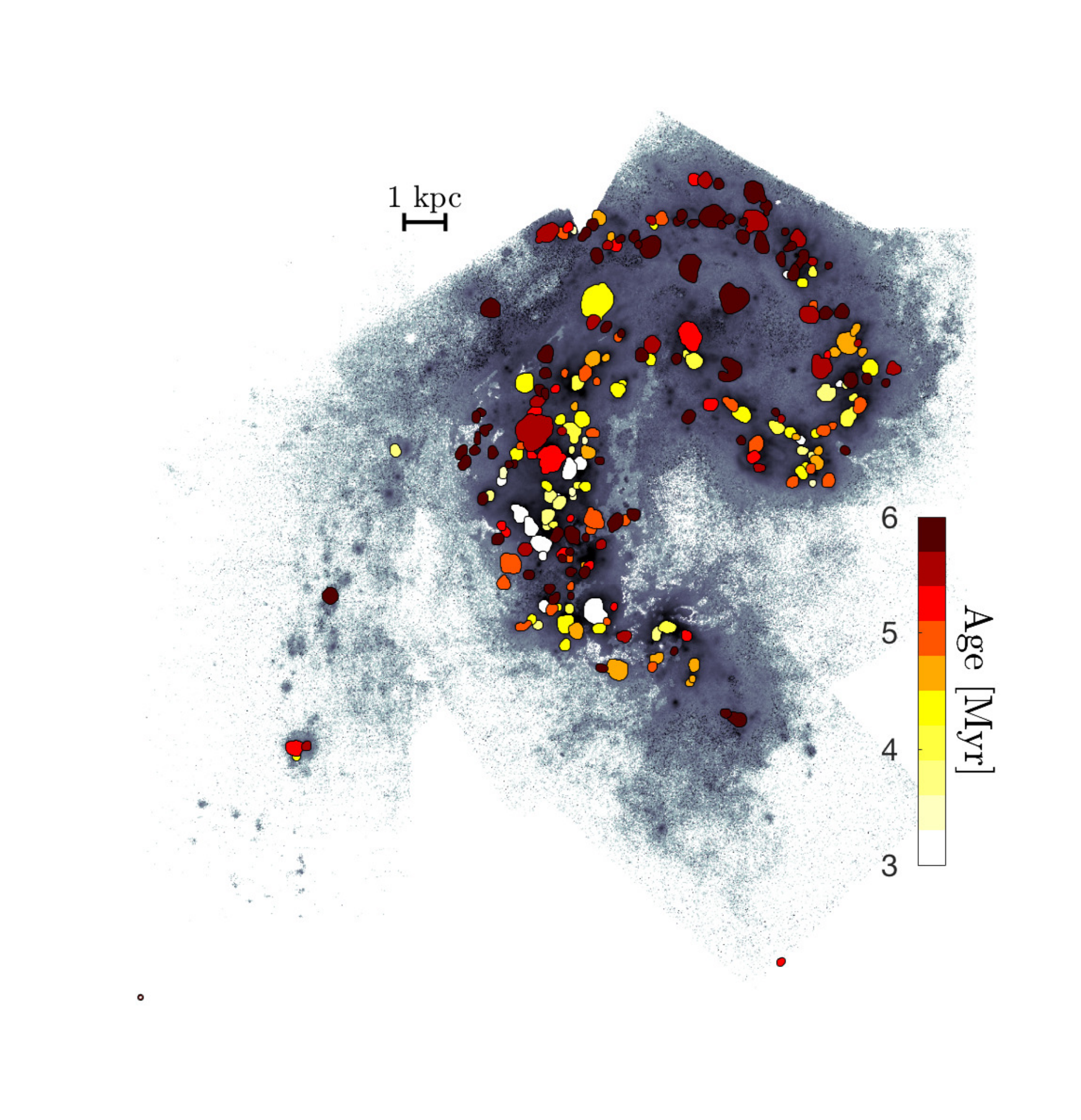} \vspace{-0.cm}
\includegraphics[width=0.43\textwidth, trim={3.5cm 3.5cm 1.6cm 1.7cm},clip]{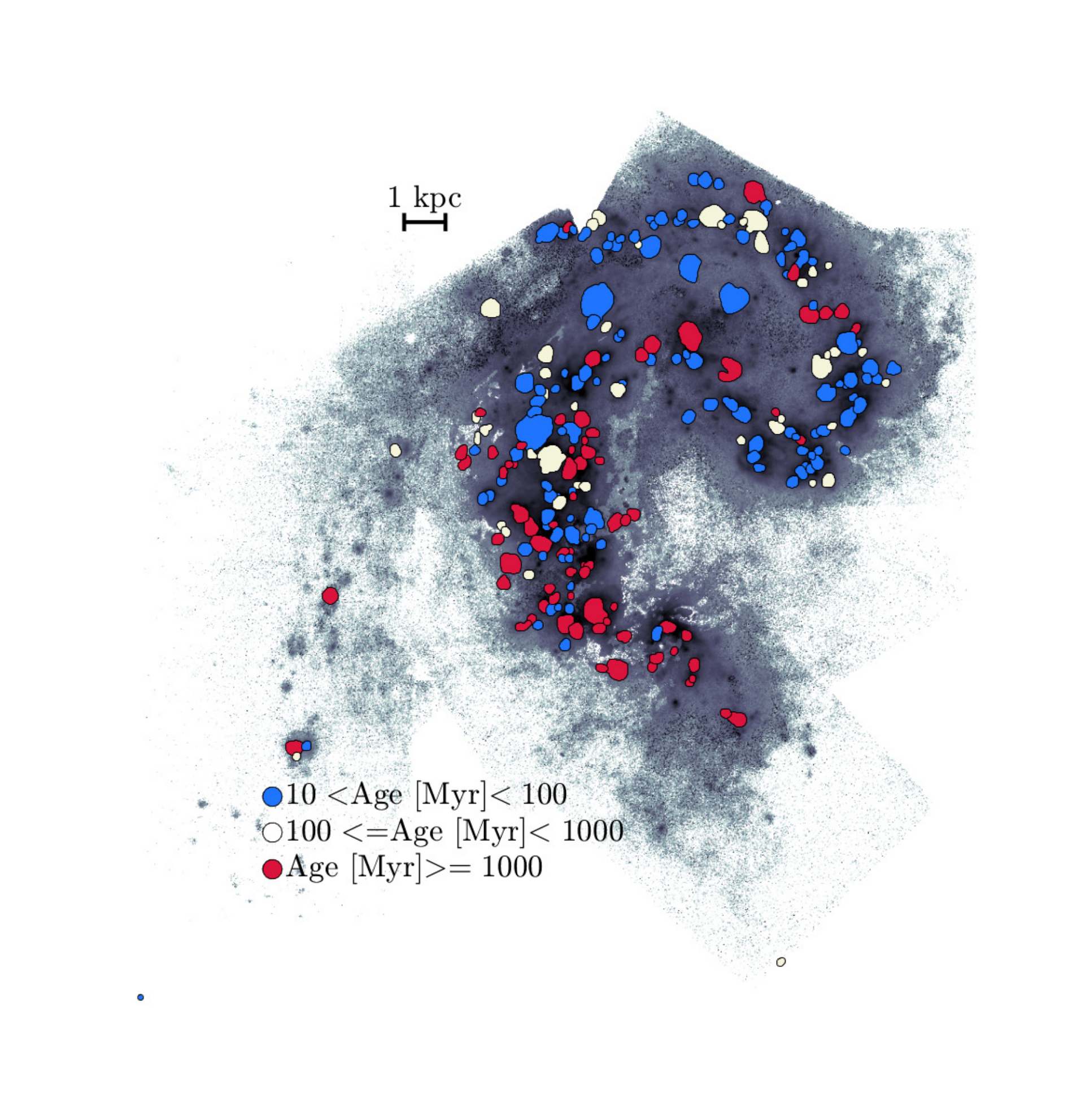} \vspace{-0.cm}
\includegraphics[width=0.43\textwidth, trim={3.5cm 3.5cm 1.6cm 1.7cm},clip]{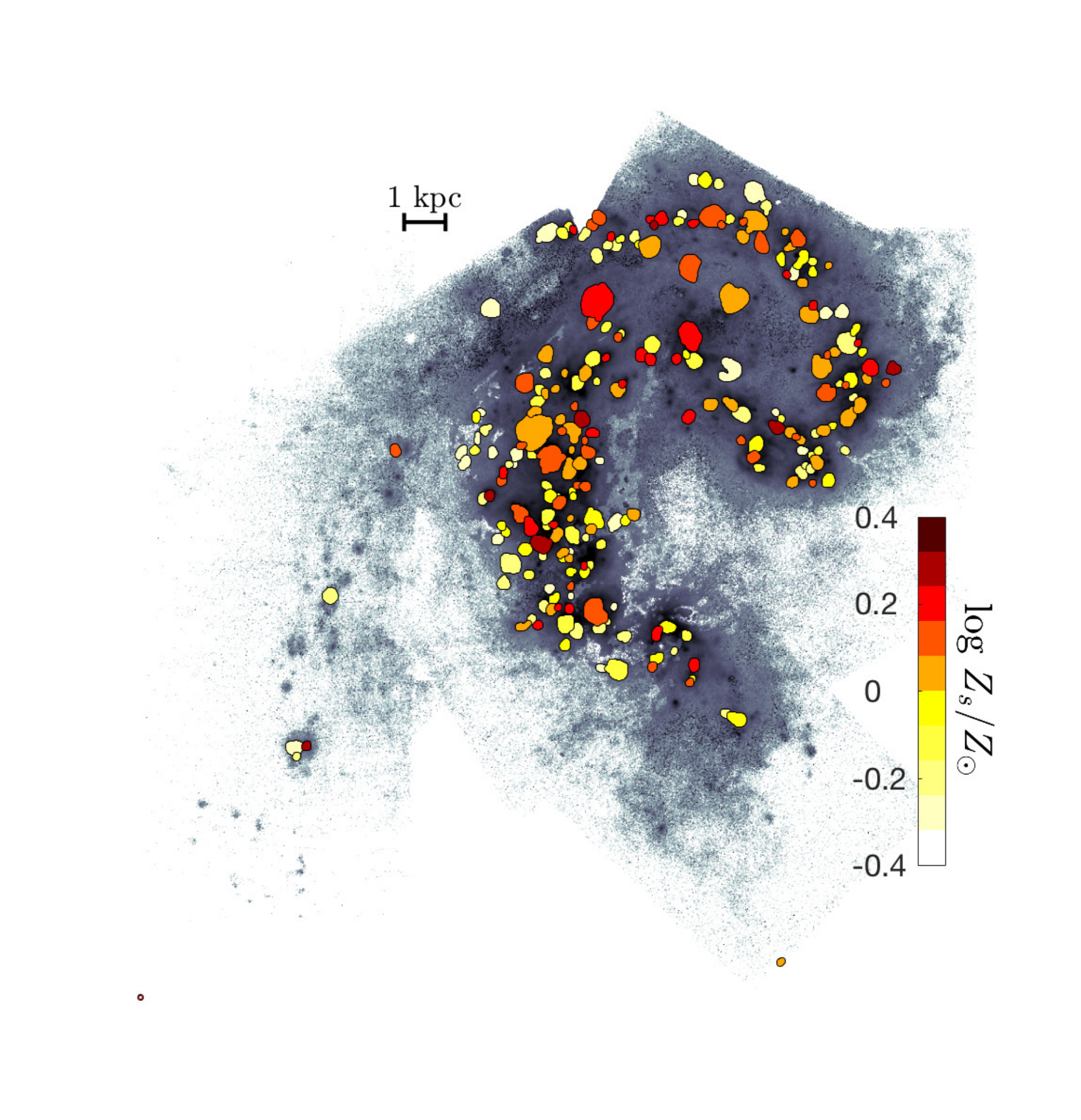} \vspace{-0.cm}
\includegraphics[width=0.43\textwidth, trim={3.5cm 3.5cm 1.6cm 1.7cm},clip]{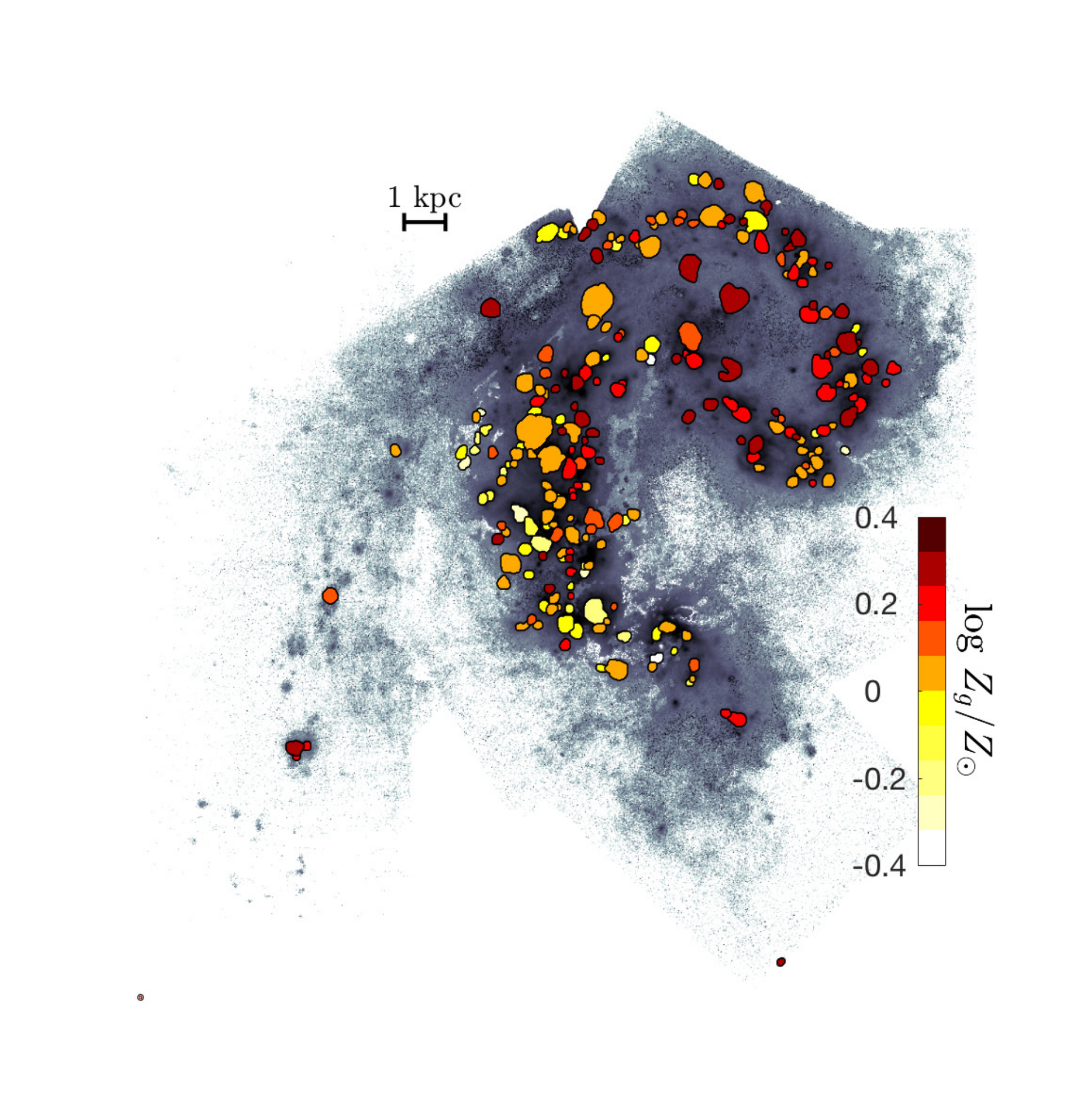} \vspace{-0.cm}
\includegraphics[width=0.43\textwidth, trim={3.5cm 3.5cm 1.6cm 1.7cm},clip]{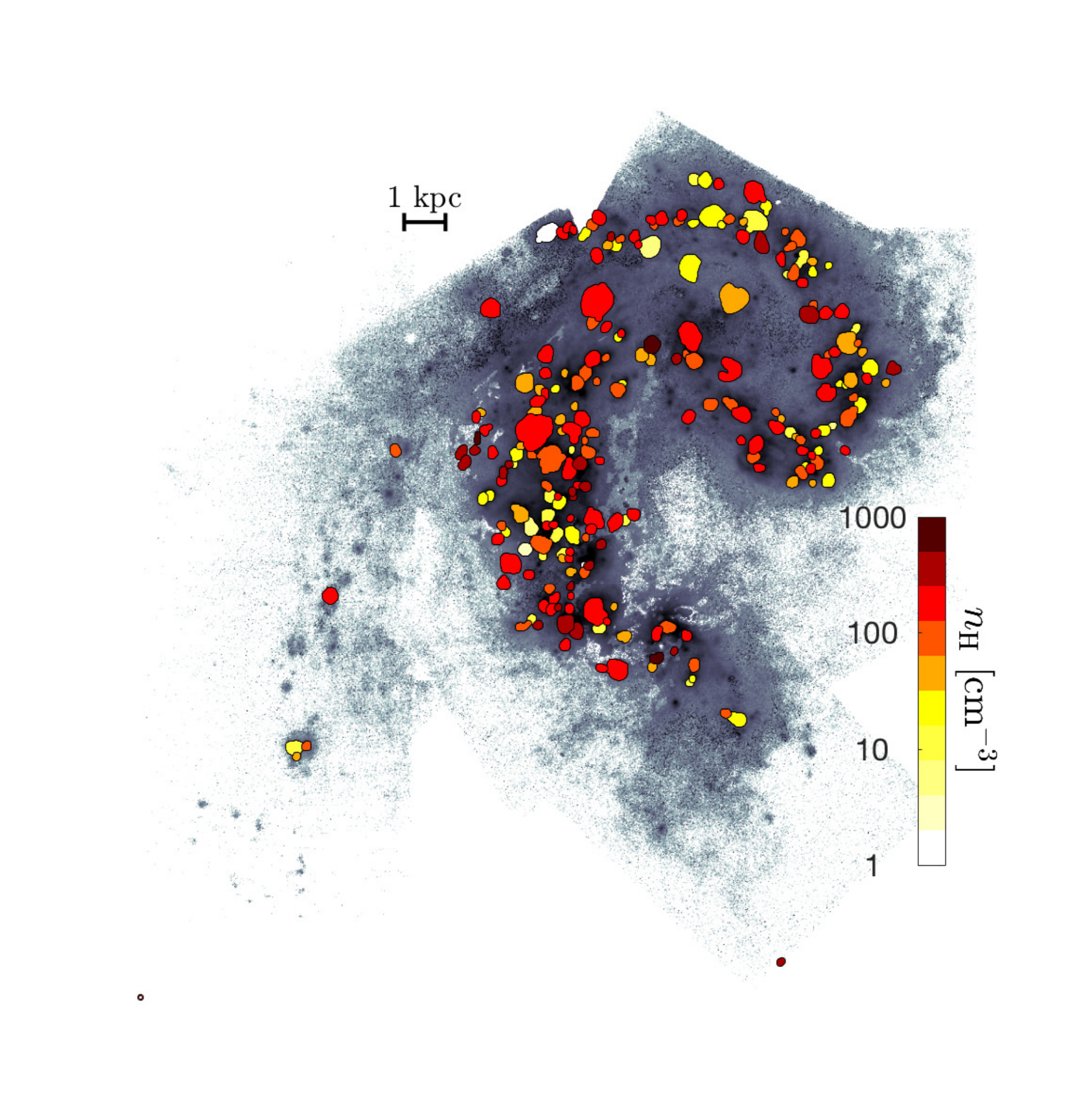} 
\includegraphics[width=0.43\textwidth, trim={3.5cm 3.5cm 1.6cm 1.7cm},clip]{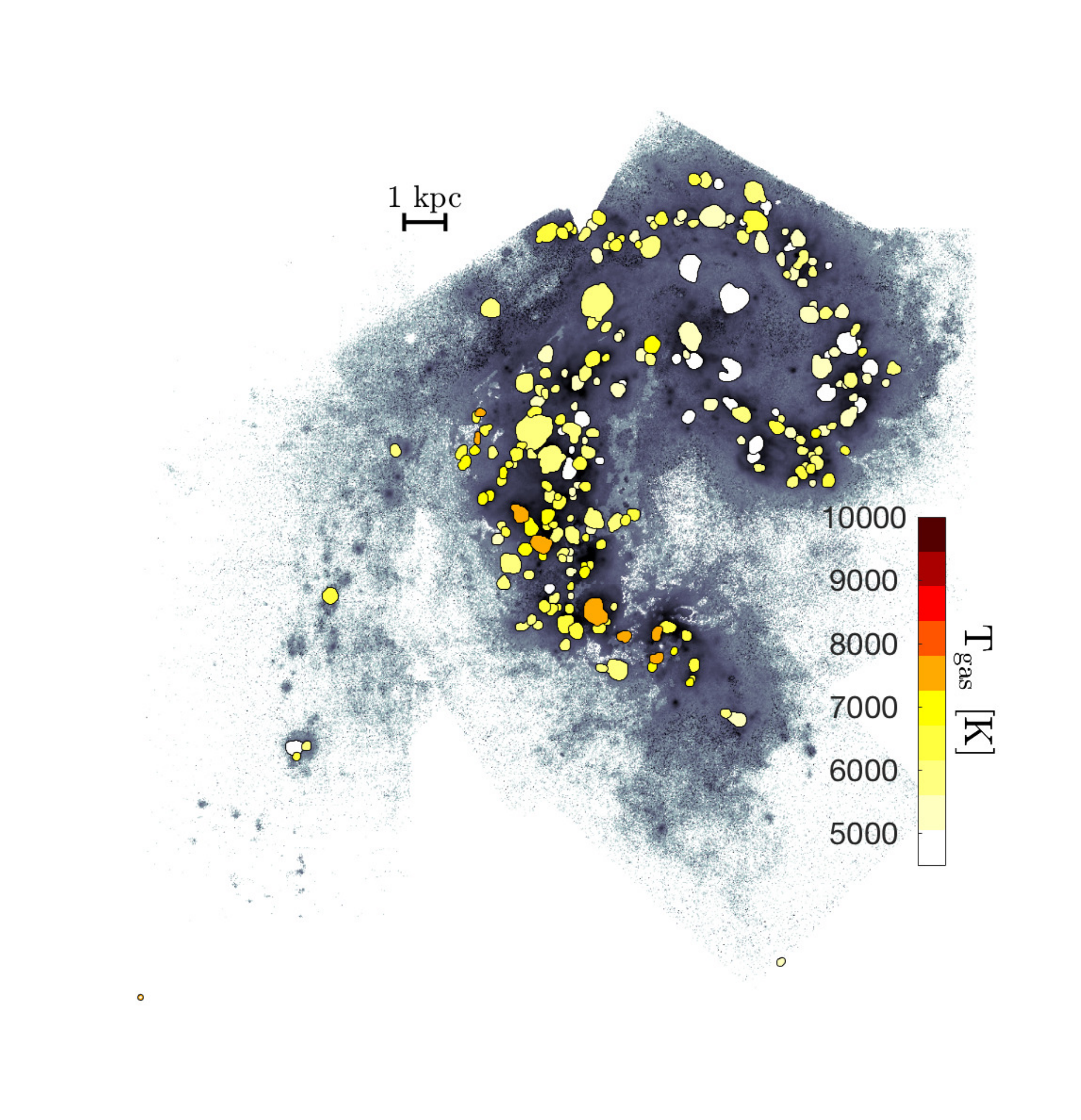}
\caption{Similar to Figure\,\ref{fig:prop_distributions_geneva}, but derived from fitting the Parsec model library. }
\label{fig:prop_distributions_parsec}
\end{center}
\end{figure*}

We apply the models described in \S\,\ref{high_res_models} and the method outlined in \S\,\ref{sec:method} to the MUSE spectra of starbursting (i.e.\,H$\alpha$ EW$>50$\AA) HII regions in the merging Antennae system. The spectra of HII regions are separately fitted with both the Geneva and Parsec model libraries to self-consistently derive stellar and gas metallicities, light-weighted age of the $<10$ Myr stars, the age of the dominant old ($>10$ Myr) stellar population, n$_{\rm{H}}$ and T$_{\rm{e}}$, which we present in Figures\,\ref{fig:prop_distributions_geneva} and \ref{fig:prop_distributions_parsec}, respectively. In each figure, from left-to-right, top-to-bottom, the HII regions are colour-coded by the light-weighted age of the young ($<10$ Myr) stellar population, the old stellar population that dominate in light-weight in the $10-100$ Myr, $100-1000$ Myr and $>1$ Gyr range, the stellar and gas metallicity in the unit solar metallicity, $n_{\rm{H}}$ and T$_{\rm{e}}$. 

In the subsequent sections, we discuss the distribution of each property shown in Figures\,\ref{fig:prop_distributions_geneva} and \ref{fig:prop_distributions_parsec} and compare our results with the previous observations of the Antennae star-forming regions. 

\subsection{The stellar populations in the Antennae galaxy}


\begin{figure*}
\begin{center}
\includegraphics[width=0.4\textwidth, trim={2.2cm 3.0cm 1.6cm 1.7cm},clip]{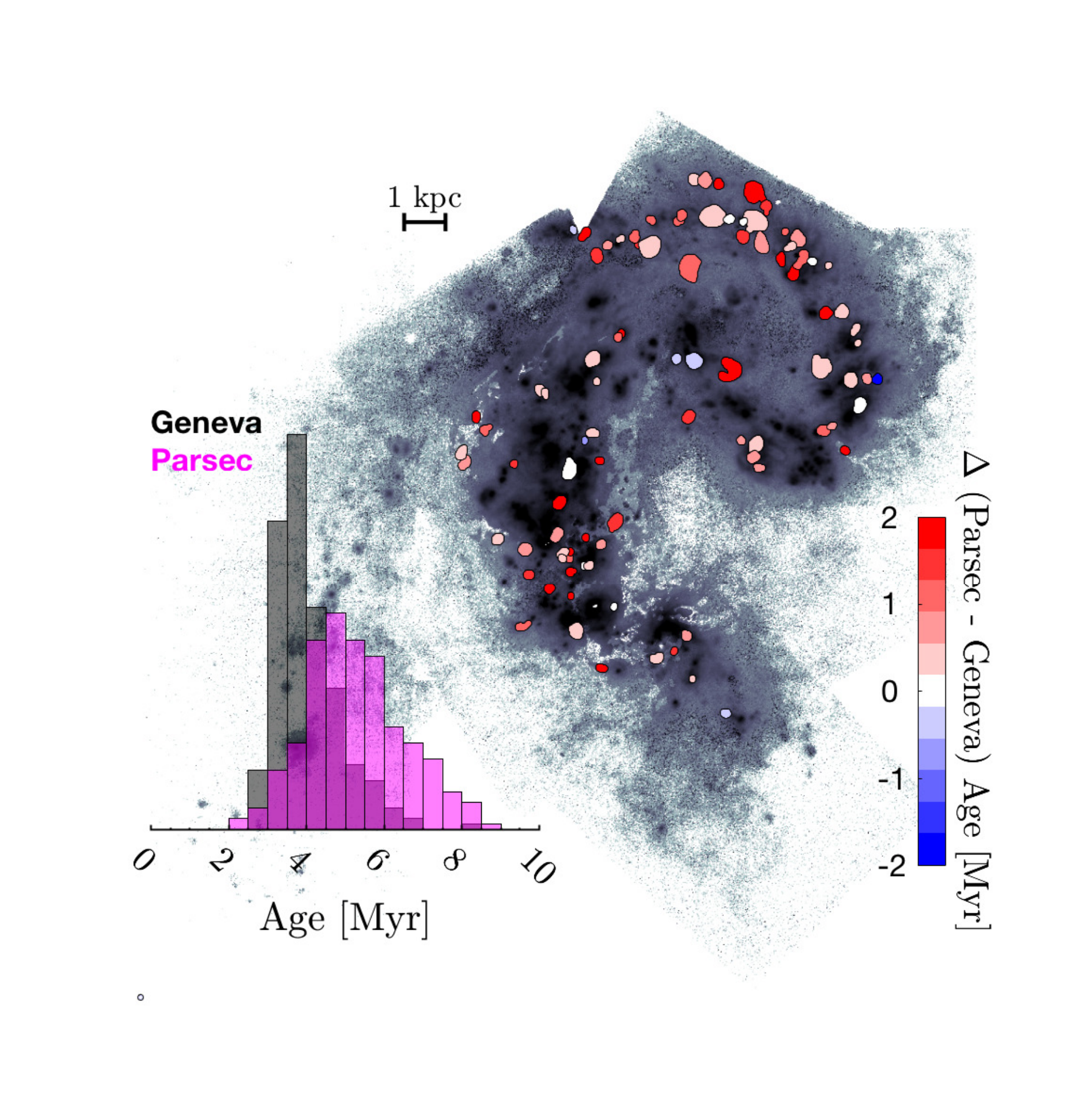} \vspace{-0.cm}
\includegraphics[width=0.4\textwidth, trim={2.2cm 3.0cm 1.6cm 1.7cm},clip]{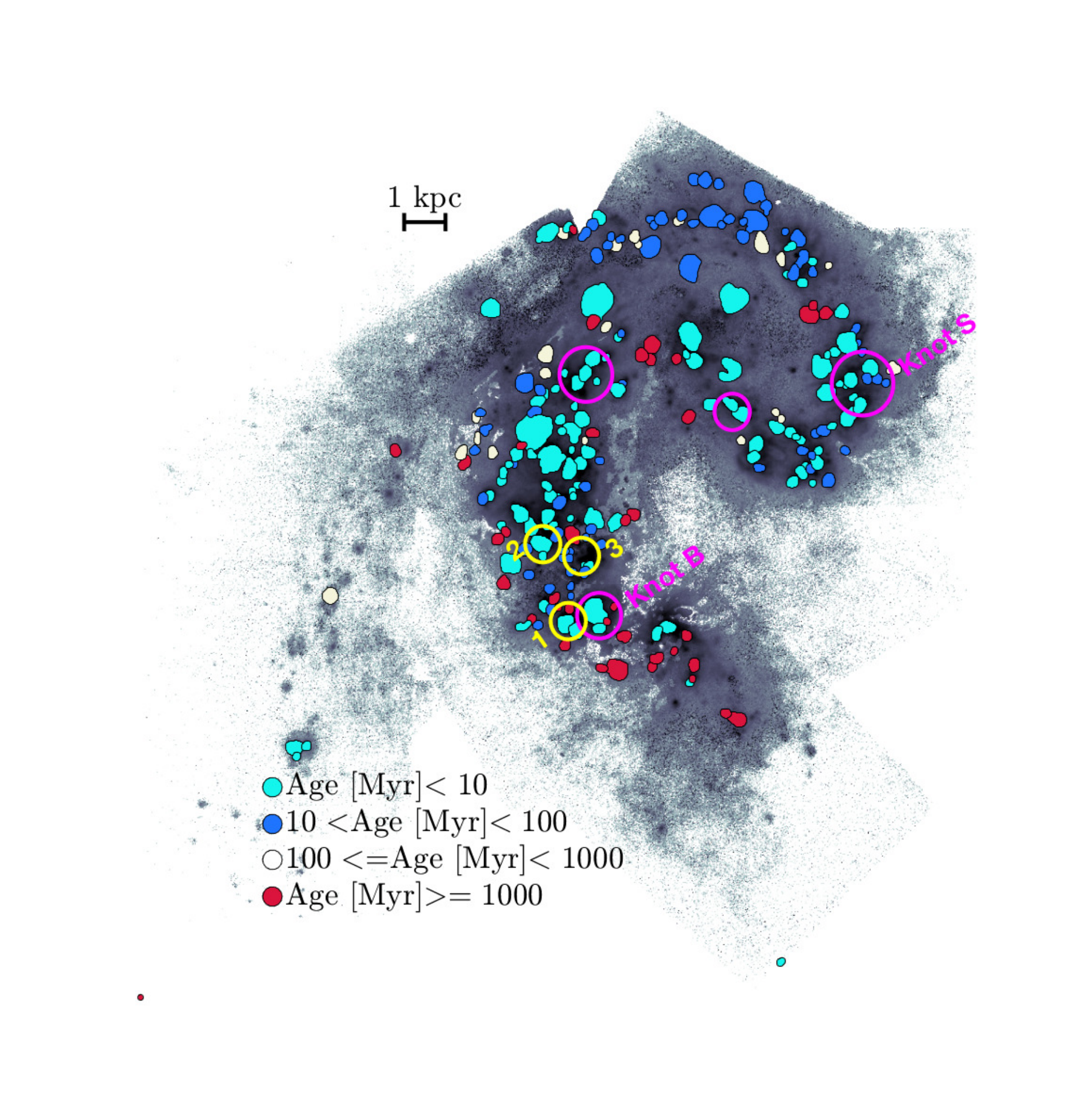}
\caption{Left panel: The residual map showing the differences in light-weighted young ($<10$ Myr) ages derived from fitting Geneva HML and Parsec models. The HII regions are colour-coded by their differences, and only those that show differences more significant than their respective one-sigma errors are shown. The inset compare the Geneva (black) and Parsec (magenta) distributions of light-weighted ages of all HII regions selected for this study. Right panel: The HII regions colour coded by the age of the dominant stellar population derived from the fitting Geneva HML models. The dominant age is now determined by considering the light-weights of individual spectral models constituting to a best-fitting model over four age bins, including the youngest ages (i.e.\,$<10$ Myr, $10-100$ Myr, $100-1000$ Myr, and $\gtrsim1$ Gyr). The HII regions indicated within circles are some of the regions that have been widely studied in the literature, which we discuss in detail in \S\,\ref{subsubsec:age_lit}. The distribution of dominant ages derived from fitting Parsec models (not shown here) is in agreement with the Geneva result shown.}
\label{fig:parsec_versus_geneva_AgeComparison}
\end{center}
\end{figure*}

For each starbursting region, we estimate the age of the "young"\footnote{Note that we define "young" stellar population to be a one that has a nebular model, which essentially means a $<10$ Myr population} component using the light-weights assigned to each $<10$ Myr spectral model contributing to the best-fitting solution. The dominant old\footnote{"old" is broadly defined to be any stellar population $>10$ Myr in age} stellar population, on the other hand, is determined by simply summing the light-weights of $>10$ Myr spectral models in $10-100$ Myr, $100-1000$ Myr, and $>1$ Gyr age bins. 

\subsubsection{The distribution of the $<10$ Myr old stellar population}
The light-weighted young stellar ages (Figure\,\ref{fig:prop_distributions_geneva}, top-left) shows a trend, albeit weak, across the merger, where the youngest ages are localised to the overlap region, and some of the oldest ages are found in more extended HII regions near the northeastern star-forming ridge (towards the centre of the northern galaxy NGC 4038). The western-loop of NGC 4038 also hosts young star-forming regions that, relative to the overlap region, appear to be slightly older. Compared to other regions, the HII regions in the overlap region are experiencing a more significant enhancement in star formation as evident from both their atypically high H$\alpha$ EWs (Figure\,\ref{fig:Antennae_HIIregions}) and the average $\sim3$ Myr age distribution. The lack of a sizeable dynamical range in young stellar ages (Figure\,\ref{fig:prop_distributions_geneva}, top-left) is likely a consequence of the fact that we are only showing the light-weighted ages of the young ($<10$ Myr) component with no regard to the light-weights assigned to any underlying $>10$ Myr stellar populations.  

We find that the light-weighted young ages derived from fitting the Parsec models (Figure\,\ref{fig:prop_distributions_parsec} top-left) show broadly the same qualitative trends as Geneva. Overall, there is a good agreement between the Geneva and Parsec results. The subset of HII regions where the respective Parsec and Geneva-based light-weighted young ages differ by more than their one-sigma errors are shown in the left panel of Figure\,\ref{fig:parsec_versus_geneva_AgeComparison}. As can be seen, the young ages derived from fitting Parsec and Geneva libraries are consistent within error for a significant fraction of the HII regions. Interestingly, however, Parsec models tend to yield young ages for star-forming regions that are mostly older than the Geneva models (Figure\,\ref{fig:parsec_versus_geneva_AgeComparison} left panel inset). As a consequence, the Parsec versus Geneva residual map of the ages shown in Figure\,\ref{fig:parsec_versus_geneva_AgeComparison} is biased somewhat towards positive differences. Moreover, the PDFs derived from fitting Parsec models, on average, show larger dispersions, which in turn yields larger uncertainties for Parsec derived physical properties than the Geneva-based properties.  We discuss this in detail in Appendix\,\ref{appB}. Briefly, a direct comparison between the  Geneva and Parsec PDFs derived for 20 HII regions across NGC 4038/39 is presented in Figure\,\ref{fig:parsec_vs_geneva_pdfs_20regions} in Appendix\,\ref{appB}, which shows that in most cases the $Z_{\rm{g}}$ derived from fitting Parsec models are biased towards slightly super-solar metallicities than Geneva. Also, the Parsec PDFs of $Z_{\rm{s}}$ are generally broader than that of Geneva.

\subsubsection{The distribution of the $>10$ Myr stellar population}
\begin{figure*}
\begin{center}
\includegraphics[width=0.9\textwidth, trim={0cm 0cm 0cm 0cm},clip]{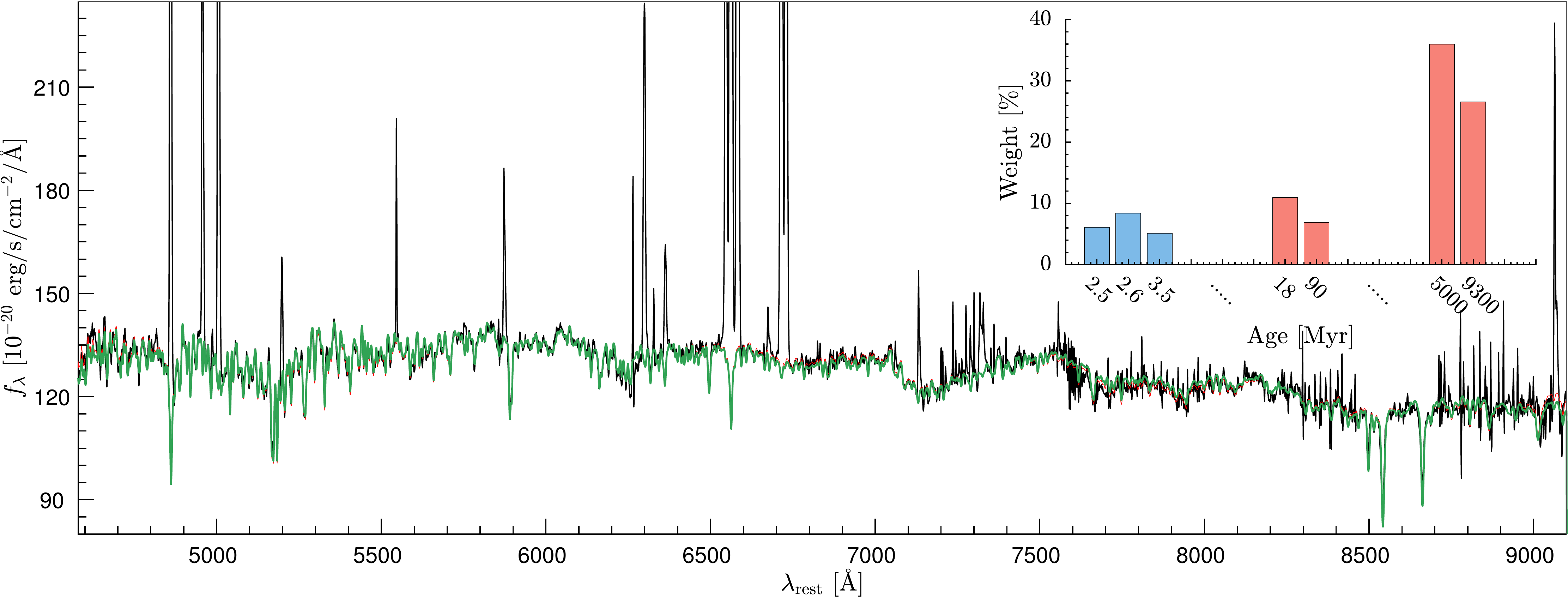}
\includegraphics[width=0.9\textwidth, trim={0cm 0cm 0cm 0cm},clip]{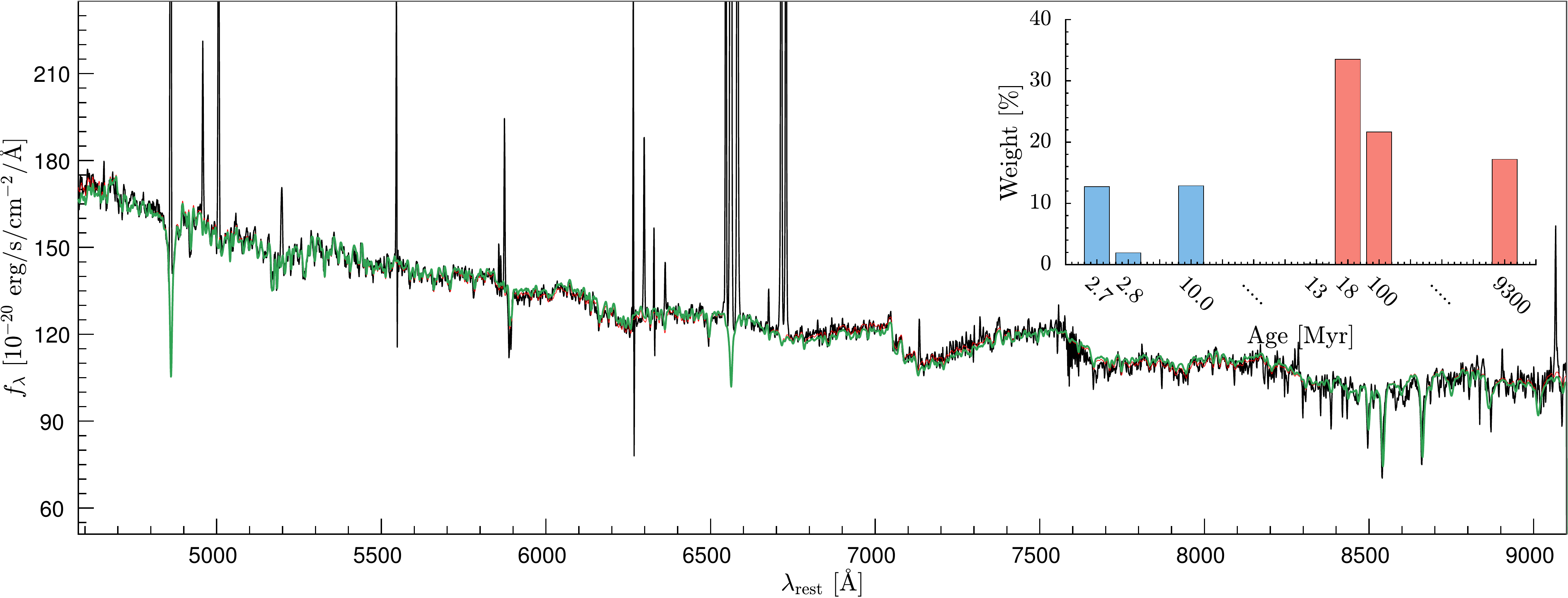}
\caption{The Geneva HML best-fitting spectral models (shown in green) obtained for the Antennae HII regions A884 (top) and A904 (bottom). The region A884 resides in the southern galaxy NGC 4039, and show distinct continuum spectral features common to $\sim1$ Gyr stellar population. The region A904 is part of the northern system NGC 4039, and show clear continuum features common to stellar populations of few tens of million years old. The Parsec best-fitting spectral models (not shown here) also clearly discriminate these distinct stellar populations and are in agreement with the Geneva result. The best-fitting models adjusted with a fit to the residuals are also shown in red to highlight that in most cases the best-fitting model alone provides a good description of the spectra of Antennae HII regions without needing any form of adjustment. The insets show age distributions corresponding to each best-fitting solution.}
\label{fig:spectra_geneva_oldStellarPops}
\end{center}
\end{figure*}

In Figures\,\ref{fig:prop_distributions_geneva} and \ref{fig:prop_distributions_parsec} top-right panels, we show the distributions of the dominant old ($>10$ Myr) stellar populations contributing to the Geneva and Parsec best-fitting spectral models. Both Parsec and Geneva ages show a clear dichotomy in the distribution of the old stellar population across the merger. The HII regions in the northern galaxy NGC 4038 appear to host an underlying old stellar population of between $10-100$ Myr in age, whereas the regions in the southern NGC 4039 show signs of the presence of a much older, a $\sim$Gyr-type, underlying stellar population. The spectral features of a $\sim$Gyr versus a few Myr stellar population are distinct. Therefore to demonstrate further that the starbursting regions in the two galaxies host two distinctive old stellar populations indeed, we show the best-fitting spectral models for two HII regions, one residing in NGC 4038 (HII region A884) and other in NGC 4039 (HII region A904), in Figure\,\ref{fig:spectra_geneva_oldStellarPops}. The spectrum of A884 shows more pronounced absorption features, particularly towards the bluer wavelengths, that rivals in strength with Balmer absorption features. In comparison, however, the same features in the A904 spectrum are less prominent, although its blue continuum is more enhanced than that of A884. 

For completeness, we also show the distribution of the ages of the dominant stellar populations in the Antennae, including the $<10$ Myr populations, in the right panel of Figure\,\ref{fig:parsec_versus_geneva_AgeComparison}. The dominant ages of the HII regions residing in the overlap region and the southern portion of the western loop appear to be $<10$ Myr, the northern portion of the western loop appears to be inhabited by HII regions with stellar populations in the $10-100$ Myr age range, while most HII regions in NGC 4039 show signatures of a $\sim$Gyr-old stellar population. Qualitatively, the trend in the distribution of the dominant stellar populations across the Antennae is similar to that observed for the light-weighted young stellar ages (Figures\,\ref{fig:prop_distributions_geneva} and \ref{fig:prop_distributions_parsec}).

It is interesting to note that the HII regions in the northern portion of the western loop appear to have an underlying dominant stellar population of $10-100$ Myr in age. The Parsec light-weighted young ages of these regions are notably older than those based on the Geneva (Figure\,\ref{fig:parsec_versus_geneva_AgeComparison}, left panel). 

\subsubsection{Comparison with previous age estimates of the HII regions in the Antennae} \label{subsubsec:age_lit}
\citet{Zhang2010} compared the broad-band spectral energy distributions for 34 24$\mu$m dust emission peaks across the Antennae to determine the SFH along the merging discs. Based on the large ratios of 24$\mu$m-to-8$\mu$m observed across the overlap region as well as the strong ultraviolet emission detected over the western loop, they conclude the overlap region and western loop to be the most intense star-forming sites in the Antennae. Between the HII regions in the overlap region and the western loop, \citet{Zhang2010} report that the regions in the western-loop are in a relatively later stage of star formation than those in the overlap region, in agreement with our findings above. 

Using the HST ACS (Advanced Camera for Surveys) observations of the Antennae in the $UBVIH_{\alpha}$ filters, \citet{Whitmore2010} predict the ages, extinction and masses for a large number of star clusters across the mergers using the population synthesis models of Charlot \& Bruzual \citep[2007;][]{BC2003}. For a part of the overlap region, called Knot B in their study (also indicated in the right panel of Figure\,\ref{fig:parsec_versus_geneva_AgeComparison}), they find that except for a single $10-50$ Myr cluster lying at the heart of the H$\alpha$ shell, the rest of Knot B is dominated by $<10$ Myr clusters with a majority $<5$ Myr in age. A direct comparison between our work and that of \citet{Whitmore2010} is difficult as our sample comprises of starbursting HII regions, which likely encompass many individual star clusters. Nonetheless, the ages reported by \citet{Whitmore2010} for a large fraction of the young star clusters embedded in the vast HII region within Knot B (right panel of Figure\,\ref{fig:parsec_versus_geneva_AgeComparison}) agree with the age estimates derived for that region in this study. Moreover, \citet{Mengel2005, Snijders2007, Brandl2009} have also studied the young HII region within Knot B. The ages they report are between $2.3-4$ Myr, again in agreement with the result of this study. 

The HII regions to the left of Knot B (the yellow circle labelled 1 in the right panel of Figure\,\ref{fig:parsec_versus_geneva_AgeComparison}) have been reported to have ages between $2.3-4$ Myr \citep[e.g.][]{Mengel2005, Gilbert2007, Snijders2007, Brandl2009}, for the regions within circle-2, ages between $1-4.9$ Myr \citep[][]{Mengel2005, Gilbert2007, Snijders2007, Brandl2009, Whitmore2010} and for circle-3, ages between $3-5.7$ Myr \citep[right yellow circle]{Mengel2005, Gilbert2007, Snijders2007}. Except for the last region, which is not bright enough in H$\alpha$ to be considered a starburst based on its H$\alpha$ EW, the ages we derive for the rest are within the age ranges reported in the literature. There is also a good agreement between the Geneva and Parsec predictions of these regions.  

For the older cluster within Knot B, to the right of the large HII region, \citet{Whitmore2010} estimate an age of $10-50$ Myr. We find that the light-weighted age of the young component of this region to be $\lesssim5$ Myr, however, its spectrum is similar to that shown in the top panel of Figure\,\ref{fig:spectra_geneva_oldStellarPops}, implying the presence of an underlying stellar population of around a Gyr or so in age. Interestingly, the spectra of most of the HII regions inhabiting NGC 4039 show signs of a $\sim$Gyr$-$type underlying population, which we discuss below. 

We have also indicated `Knot S' studied by \citet{Whitmore2010} in the right panel of Figure\,\ref{fig:parsec_versus_geneva_AgeComparison}. For this region, \citet{Whitmore2010} find a mixture of ages, with $<10$ Myr clusters dominating in number. The youngest clusters ($<5$ Myr) in the region appear to be mostly positioned to the left of the highest star-formation peaks, with the $5-10$ Myr clusters scattered across the region, which \citet{Whitmore2010} interpret as a sign of the progression of star formation towards the dust reservoir.  They also find a scattering of $10-100$ Myr clusters through much of the region. We find evidence for a similar variation in ages in the Geneva and Parsec distributions of light-weighted young ages (Figures\,\ref{fig:prop_distributions_geneva} and \ref{fig:prop_distributions_parsec}), largely in agreement with the results of \citet{Whitmore2010}. However, we remind the reader, again, that we cannot perform a direct comparison with the \citet{Whitmore2010} results as the HII regions shown may encompass several star clusters.  

Magenta circles (without labels) in Figure\,\ref{fig:parsec_versus_geneva_AgeComparison} indicate two other knots studied by \citet{Whitmore2010}. According to their study, many of the star clusters residing within these knots have ages $<5$ Myr, with $5-10$ Myr old clusters positioned approximately at the centres of the peaks in star formation. Also, for some of the star clusters within the two circles, \citet{Bastian2009} report ages between $3.16-6.3$ Myr. According to this study, some to the HII regions within these circles have Geneva-derived ages $<4$ Myr and Parsec-derived ages in the range $4-5$ Myr, which are broadly in agreement with \citet{Bastian2009} and \citet{Whitmore2010}, though, there are some discrepancies between the Geneva and Parsec estimates as can be seen in their residual map (Figure\,\ref{fig:parsec_versus_geneva_AgeComparison}).

On the old stellar populations in the Antennae, the numerical modelling studies \citep[e.g.][]{Barnes1988, Mihos1993} of the merger suggest that the first encounter between the two progenitors occurred $\sim$$200-400$ Myr ago. In support of this timeframe, several observational work \citep[e.g.][]{Whitmore1999, Zhang2001, Bastian2009, Whitmore2010} find evidence for the existence of intermediate-age clusters scattered throughout the merger, primarily across the northeastern star formation ridge of NGC 4038, with ages varying between $100-600$ Myr. 

In this analysis, we also see evidence for the presence of underlying old stellar populations in the starbursting regions. Our models predict both intermediate-age and older stellar populations, and intriguingly, there seems to be a dichotomy between the distribution of these two populations. The intermediate-age populations, comprising mainly of $10-100$ Myr regions as well as some between $100-1000$ Myr, are scattered across NGC 4038, including the overlap region, whereas the spectra of the HII regions in NGC 4039 show signs of a much older, around Gyr or so in age, population.  In comparison, the distribution of the intermediate-age population across regions belonging to NGC 4038 appears to be similar to that reported by \cite{Whitmore1999}. In terms of ages, \cite{Whitmore1999} report an age of $\sim$100 Myr for the HII regions residing mostly around the northeastern ridge, which is broadly consistent with the ages of $10-100$ Myr found for that region in this study. \cite{Whitmore1999} also find a second older, $\sim$500 Myr, population, mainly residing in the tidal tails, which they associate with the initial encounter between the progenitors. In our study, for the two large HII regions out of the four in the tidal tails that are currently in a starburst phase, we find dominant old ages of around $\sim$Gyr. We also find evidence for the existence of $\sim$Gyr$-$type populations in starbursting regions that appear to belong to the southern system NGC 4039.       

Finally, we note that it is challenging to constrain the ages of the old stellar population accurately from the current analysis alone. For the determination of young ages (i.e.\,$<10$ Myr), for example, we have utilised constraints from stellar and nebular continua as well as nebular lines, adding thorough constraints on young ages, whereas the old stellar populations are constrained only using continuum spectral features. Also, the age-selective dust attenuation can be necessary for the derivation of old ages. This is one of the main reasons as to why we report old stellar population ages in bins of $10-100$ Myr, $100-1000$ Myr and $\gtrsim1$ Gyr. Another being that, as discussed in \S\,\ref{subsec:props_wr} and demonstrated in Figures\,\ref{fig:prop_distributions_geneva} and \ref{fig:prop_distributions_parsec}, the ages derived are dependent on the isochrones used.

\subsection{The metallicities of the stars and the star-forming gas in the Antennae}
\begin{figure*}
\begin{center}
\includegraphics[width=.4\textwidth, trim={2.2cm 2.7cm 0.8cm 1.7cm},clip]{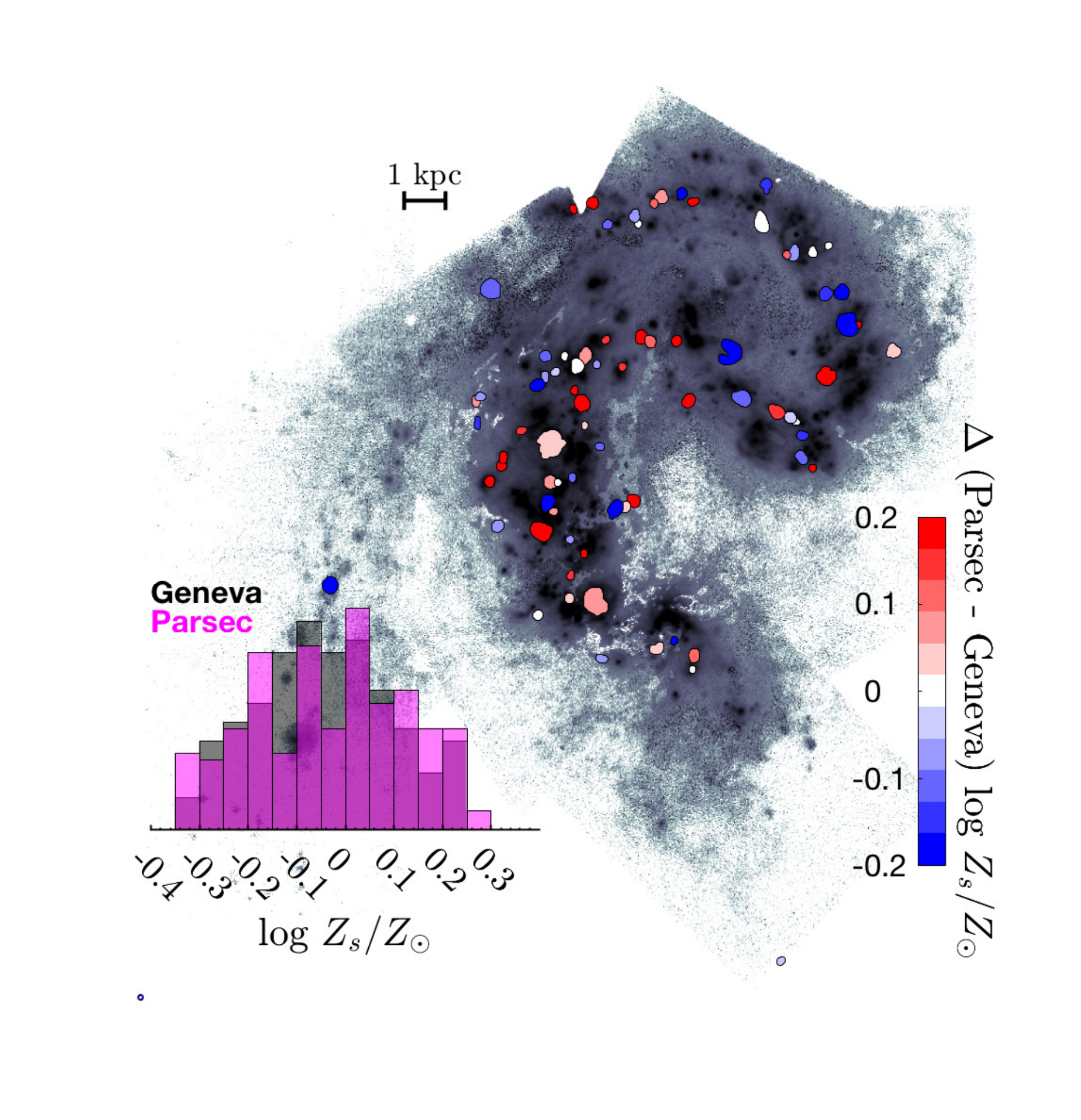} \vspace{-0.cm}
\includegraphics[width=.4\textwidth, trim={1.6cm 2.7cm 1.6cm 1.7cm},clip]{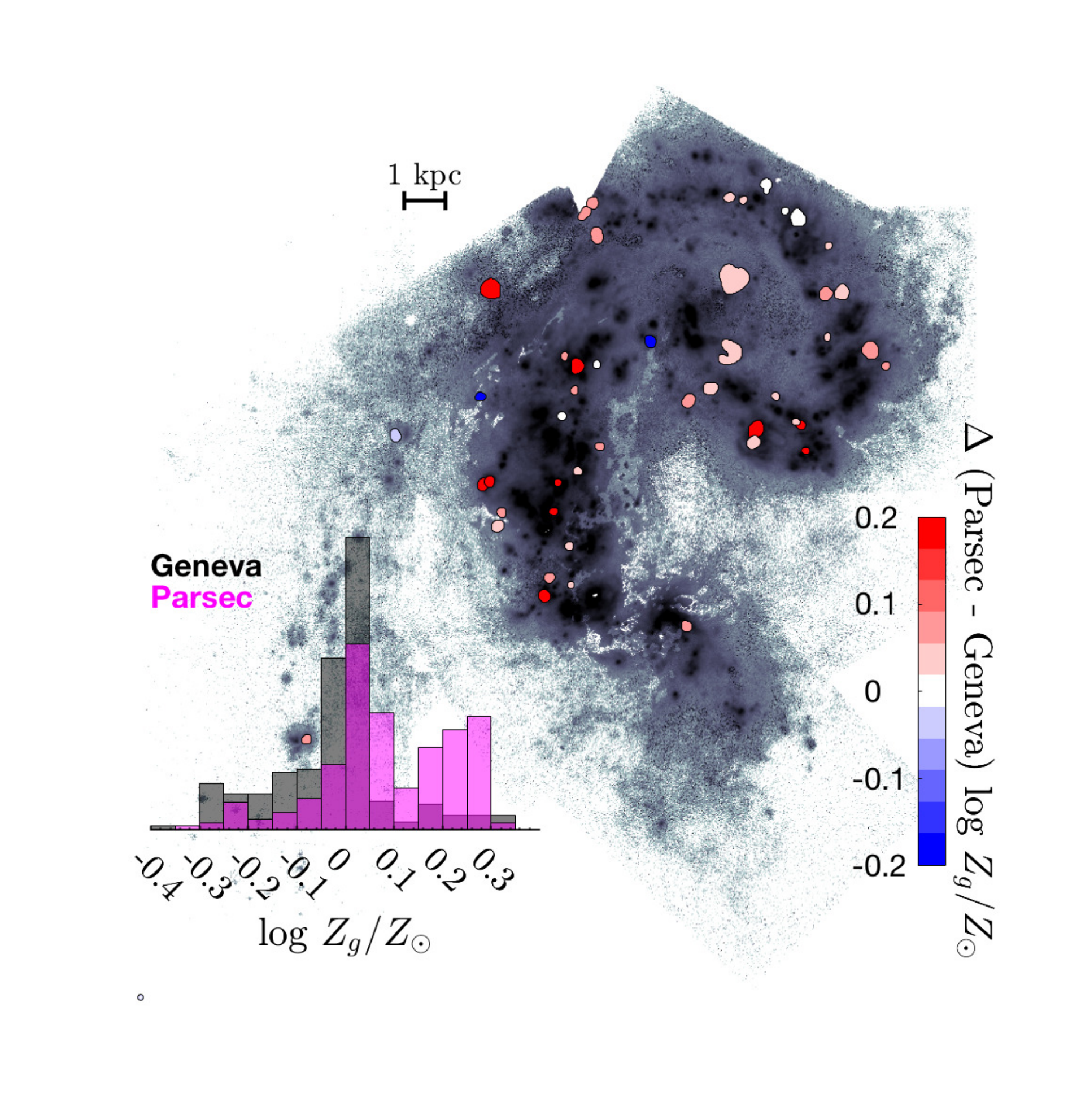} \vspace{-0.cm}
\caption{Residual maps showing the differences in stellar (left) and gas (right) metallicities derived from fitting Geneva HML and Parsec models. The HII regions are colour-coded by their differences, and only those regions that show differences more significant than their respective one-sigma uncertainties are shown. The insets compare the Geneva (black) and Parsec (magenta) distributions of $Z_{\rm{s}}$ and $Z_{\rm{g}}$ of all HII regions selected for this study.}
\label{fig:parsec_versus_geneva_MetalComparison}
\end{center}
\end{figure*}

The middle panels of Figures\,\ref{fig:prop_distributions_geneva} and \ref{fig:prop_distributions_parsec} present the HII regions colour-coded by their stellar and gas metallicities (in units of solar metallicity) derived from fitting Geneva and Parsec model libraries, respectively. 

\subsubsection{The stellar metallicity}

We find that, on average, both Geneva HML and Parsec best-fit stellar metallicities for the HII regions in the Antennae are around solar. There is a weak trend towards slightly super-solar stellar metallicities evident in the Geneva results (Figure\,\ref{fig:prop_distributions_geneva}, middle-left), which is, however, not apparent in the Parsec distribution (Figure\,\ref{fig:prop_distributions_parsec}, middle-left). 

In Figure\,\ref{fig:parsec_versus_geneva_MetalComparison} left-panel, we show the agreement between Geneva and Parsec derived stellar metallicities. As in Figure\,\ref{fig:parsec_versus_geneva_AgeComparison}, only the HII regions with differences in stellar metallicities more significant than their one-sigma error overlap are plotted. Overall, there is a good agreement between the predictions of the two model libraries. The inset in the left panel of Figure\,\ref{fig:parsec_versus_geneva_MetalComparison} compares the Geneva versus Parsec derived metallicities for all HII regions, and both distributions indicate similar broad dispersions. This is perhaps a result of the age $-$ stellar metallicity degeneracy yielding different potential stellar metallicity and age solutions, and thus increasing the dispersion of the stellar metallicity PDFs of the best-fit solutions (Figure\,\ref{fig:parsec_vs_geneva_pdfs_20regions}). As discussed in \S\,\ref{subsec:double_solutions}, the presence of prominent WR features and the Paschen jump in the nebular continuum can alleviate the effects of the age $-$ stellar metallicity degeneracy to an extent. 

\subsubsection{The gas metallicity}
In the middle-right panels of Figures\,\ref{fig:prop_distributions_geneva} and \ref{fig:prop_distributions_parsec}, we show the distributions of the gas metallicities derived from fitting the Geneva and Parsec model libraries. Both sets of models broadly predict solar-like metallicities for the HII regions in the Antennae, although, on average, the Parsec predictions appear to be slightly biased towards super-solar values. The Geneva versus Parsec residual map is shown in the right panel of Figure\,\ref{fig:parsec_versus_geneva_MetalComparison}, which demonstrate that within one-sigma error, the two model predictions for a significant fraction of the HII regions are indeed consistent.  

There is a trend in the distributions of gas metallicities across the merging discs. Both models predict the star-forming gas in the HII regions around the western loop to be more enriched than those inhabiting the overlap region. The HII regions in the overlap region and NGC 4039 show a mix of metallicities in the range between slightly sub- to super-solar. The inset in the right panel of Figure\,\ref{fig:parsec_versus_geneva_MetalComparison} present a comparison of the Geneva and Parsec derived gas metallicities for all HII regions, which shows that solar-like gas metallicities characterise a significant fraction of the HII regions in the Antennae, supporting our earlier assertion. Interestingly, the Parsec distribution shows a more distinct bimodality than the Geneva, reflecting the fact that the Parsec derived metallicities of HII regions across the western loop indicate, on average, a slightly greater enrichment that the respective Geneva metallicities (the right panels of Figures\,\ref{fig:prop_distributions_geneva} and \ref{fig:prop_distributions_parsec}). 

\subsubsection{Comparison with previous metallicity estimates for the HII regions in the Antennae }\label{subsubsec:metal_lit}
\citet{Mengel2002} report solar-like stellar metallicities for six HII regions in NGC 4038 based on an analysis of the metallicity-sensitive \ion{Mg}{i} line at 8806.8\AA. Similarly, \citet{Lardo2015} measure Fe, Mg, Si and Ti abundances for three super star clusters (two are part of the western loop, and the third resides in the northeastern ridge) and infer slightly super-solar metallicities of $\sim0.07\pm0.03$. \citet{Lardo2015} also calculate direct metallicities (i.e.\,gas-phase metallicities), which are consistent with being slightly super-solar. 

\citet{Bastian2009} derived metallicities for 16-star clusters in the Antennae from a comparison of observed Balmer and metal line strengths with stellar population synthesis models, as well as using strong emission lines where available. Out of the 16 star clusters, 15 are part of the central merger, with four residing in NGC 4039 and two of which are young, and the rest distributed mostly around the western loop of NGC 4038 and five of which are young. For all these star clusters, \citet{Bastian2009} infer metallicities that are, on average, solar to super-solar.  The metallicities of the HII regions that these star clusters are members of likely similar. 

Overall, the observations of solar to slightly super-solar stellar and gas metallicities for the star-forming regions are consistent with the results of this analysis. Finally, utilising the observations of \citet{Bastian2009}, \citet{Lardo2015} find a flat abundance gradient across the merger. In this analysis, however, we find a weak trend; the gas metallicities of HII regions in the western loop appear to be more metal-rich than those in the overlap and NGC 4039 regions. This trend is, however, not evident in the distribution of stellar metallicities. 

\subsection{On the distribution of the electron densities and temperatures in the Antennae}
The electron densities and temperatures derived from fitting Geneva and Parsec models are shown in the bottom left and right panels of Figures\,\ref{fig:prop_distributions_geneva} and \ref{fig:prop_distributions_parsec}. 

By large, the derived $n_{\rm{H}}$ values are between the $100-1000$ cm$^{-3}$ range, and we find no clear trend in their distribution across the central regions of the merger. As discussed in \S\,\ref{subsec:double_solutions}, however, most $n_{\rm{H}}$ PDFs show double peaks, which appear to be associated with the higher and lower ionisation nebular constraints used in the fitting.  The T$_{\rm{e}}$ of HII regions are, on average, centred around 7000 K, and as with $n_{\rm{H}}$, show no clear variation across the central merger.  

\subsubsection{Comparison with $n_{\rm{H}}$ and T$_{\rm{e}}$ estimated from PyNeb}\label{subsubsec:temp_density_comp}
\begin{figure*}
\begin{center}
\includegraphics[width=0.45\textwidth, trim={0.5cm 0.8cm 7.2cm 0.45cm},clip]{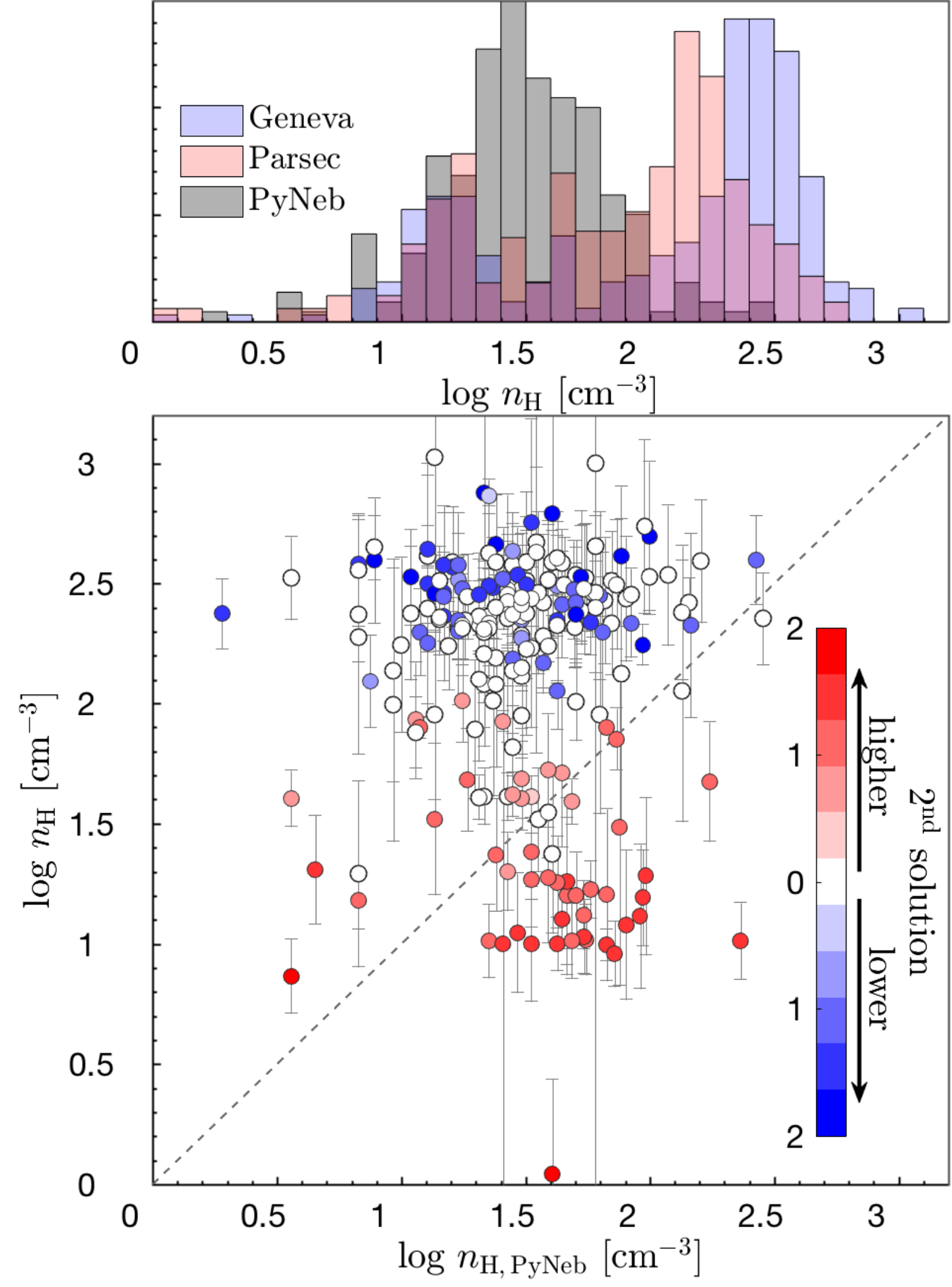}
\includegraphics[width=0.45\textwidth, trim={0.5cm 0.8cm 7.2cm 0.45cm},clip]{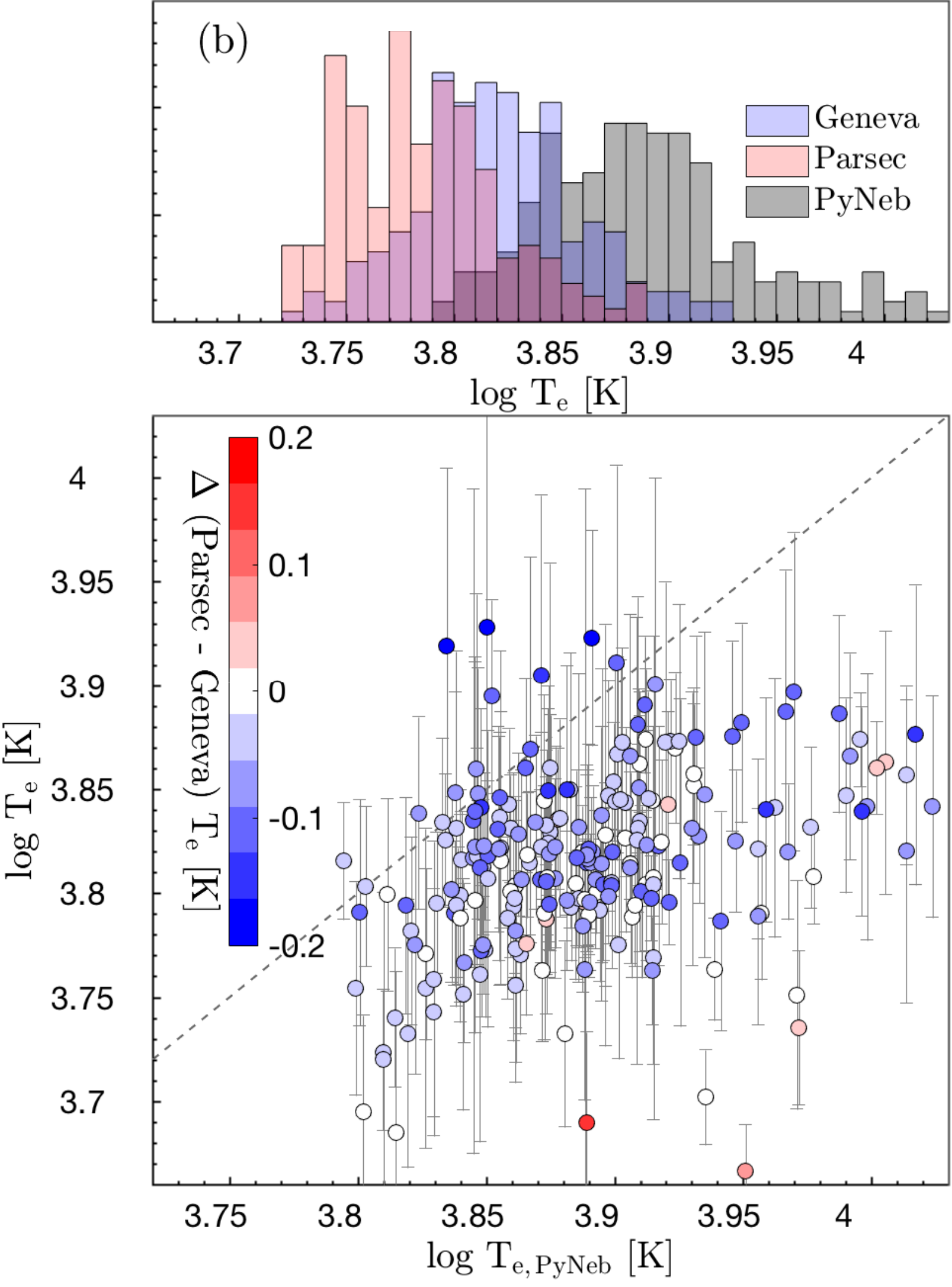}
\caption{Comparison between Geneva and Parsec model predictions, and \textsc{PyNeb} \citep{Luridiana2015}. (a) The main panel shows the distribution of $n_{\rm{H}}$ derived from fitting the Geneva models compared that estimated from \textsc{PyNeb} based on [\ion{S}{ii}]~$\lambda\lambda$6716,31\AA, with the dashed line denoting the one-to-one relation. As discussed in the text, there are cases where the best-fitting $n_{\rm{H}}$ PDF indicate possible double solutions. For these cases, the colour-code denote the difference between the most probable and the second most probable $n_{\rm{H}}$ values for a given spectrum.  The redder colours mean the second-most likely $n_{\rm{H}}$ is higher than the most probable value estimated from the respective PDF, and vice versa. The top panel of (a) compares the $n_{\rm{H}}$ distributions obtained from fitting Geneva (blue) and Parsec (red) models, including the double solutions, with the distribution of \textsc{PyNeb} (black) estimates. (b) The distribution of T$_{\rm{e}}$ derived from fitting the Geneva models compared that estimated from \textsc{PyNeb} based on [\ion{N}{ii}]~$\lambda\lambda$5755, 6548\AA\,is shown in the main panel. The dashed line, again, indicates the one-to-one relation, and the colour-code denotes the difference between the Parsec and Geneva predictions. Note that within one-sigma uncertainty, the T$_{\rm{e}}$ values obtained from Geneva and Parsec models are consistent for almost all the HII regions, therefore the colour-code simply denote the difference between the peaks of the Gaussian fits to the T$_{\rm{e}}$ PDFs. The top panel of (b) shows the T$_{\rm{e}}$ distributions obtained from fitting Geneva (blue) and Parsec (red) models, and \textsc{PyNeb} (black) estimates.}
\label{fig:comparisons_with_pyneb}
\end{center}
\end{figure*}
\begin{figure*}
\begin{center}
\includegraphics[width=1\textwidth, trim={0.cm 0.cm 0.cm 0.cm},clip]{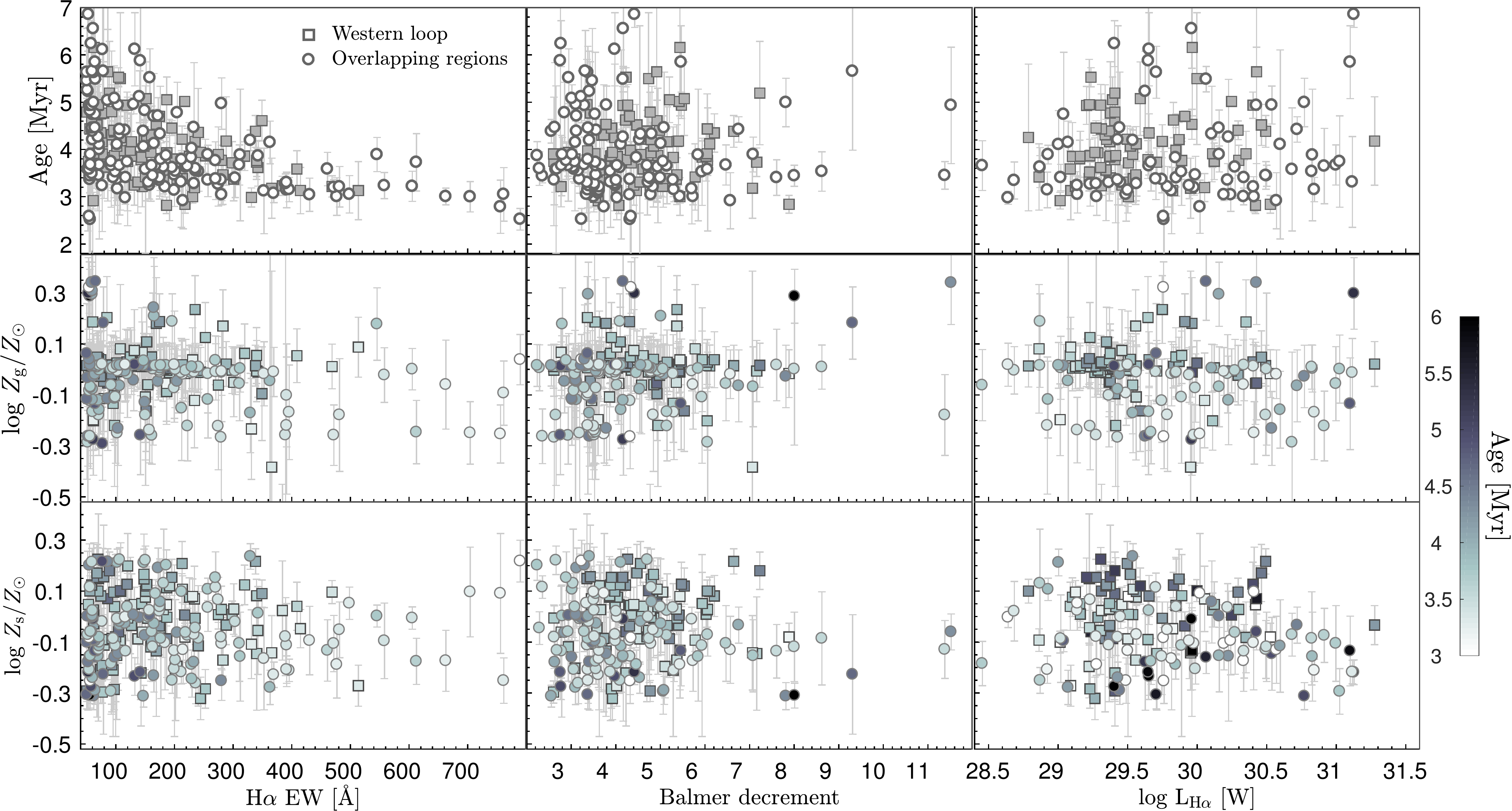} \vspace{-0.cm}
\includegraphics[width=1\textwidth, trim={0.cm 0.cm 0.cm 0.cm},clip]{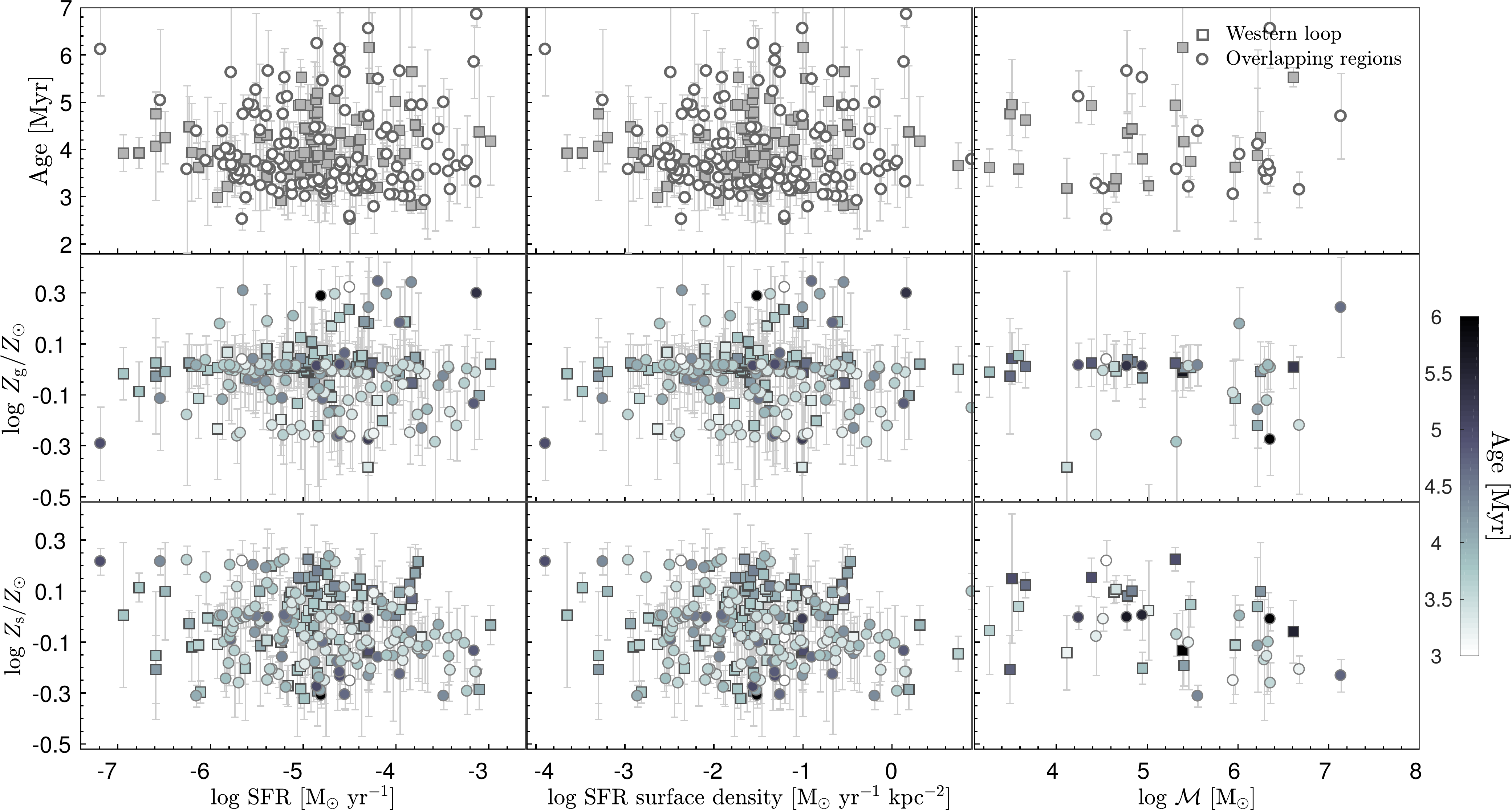} \vspace{-0.cm}
\caption{The model-based properties (e.g.\,light weighted young age, and gas and stellar metallicities) as a function of H$\alpha$ equivalent width in unit of \AA, Balmer decrement, H$\alpha$ luminosity in unit of W spaxel$^{-1}$ (top set of nine panels), SFR in unit of M$_{\odot}$ yr$^{-1}$, SFR surface density in unit of M$_{\odot}$ yr$^{-1}$ kpc$^{-2}$, and cluster mass in unit of solar masses (bottom set of nine panels). The circles and squares denote the HII regions residing in the overlap region, including the southern NGC4039, and the western loop, respectively, and the colour code indicate the light weighted young ages of the HII regions.}
\label{fig:correlation_between}
\end{center}
\end{figure*}

As mentioned earlier, the spectra used for this study are abundant in some of the most conspicuous emission lines in the optical, some which are widely utilised in the literature to probe the physical conditions of star-forming gas. For example, the optical diagnostics of $n_{\rm{H}}$ include the intensity ratios of [\ion{S}{ii}]~$\lambda$6717,31\AA, [\ion{Cl}{iii}]~$\lambda$5518,38\AA\,and [\ion{Ar}{iv}]~$\lambda$4711,40\AA, where each is only applicable over a specific range of densities. Likewise, the intensity ratios of [\ion{N}{ii}]~$\lambda$5755/$\lambda$6548 + 6584 and [\ion{S}{iii}]~$\lambda$6312/$\lambda$9532 + 9069 are highly sensitive to T$_{\rm{e}}$ \citep{Osterbrock, Peimbert2017}. As the density sensitive [\ion{S}{ii}]~$\lambda$6717, 6731\AA\,and the temperature sensitive [\ion{N}{ii}]~$\lambda$5755, 6584\AA\,are relatively stronger than the others lines mentioned above, we adopt them to calculate T$_{\rm{e}}$ and $n_{\rm{H}}$ with aid of the publicly available \textsc{PyNeb} software \citep{Luridiana2015}. 

The comparison between the \textsc{PyNeb}-based estimates of $n_{\rm{H}}$ and T$_{\rm{e}}$ versus those derived from fitting the Geneva and Parsec models is presented in Figure\,\ref{fig:comparisons_with_pyneb}. The left and right panels show the distributions of $n_{\rm{H}}$ and T$_{\rm{e}}$, respectively. The $n_{\rm{H}}$ and T$_{\rm{e}}$ values derived from fitting the Geneva models are for the comparisons shown in the main panels, however, note that the qualitative trends are unchanged if Parsec derived values have been used instead. 

As noted in \S\,\ref{subsec:double_solutions}, we infer more than one probable $n_{\rm{H}}$ solution for a number of HII regions, therefore, for the comparison with \textsc{PyNeb} shown in Figure\,\ref{fig:comparisons_with_pyneb} (left main panel), we use the most probable $n_{\rm{H}}$, but colour-code that to show the second $n_{\rm{H}}$ solution. The redder (bluer) shadings indicate HII regions with a second $n_{\rm{H}}$ that is higher (lower) than the best-fitting $n_{\rm{H}}$, and the intensity of the shading denote the magnitude of the difference between the two $n_{\rm{H}}$ values. A direct comparison between the $n_{\rm{H}}$ distributions derived from fitting Geneva (blue) and Parsec (red) models, including all second solutions, and that calculated from \textsc{PyNeb} is shown in the top panel of Figure\,\ref{fig:comparisons_with_pyneb}.      

According to the $n_{\rm{H}}$ comparison, the majority of the model-based predictions of $n_{\rm{H}}$ are, on average, higher than the \textsc{PyNeb} estimates, with a small subset of HII regions having  $n_{\rm{H}}$ values lower than the respective \textsc{PyNeb} values. Those regions with a probable second $n_{\rm{H}}$ solution also show a similar dichotomy. This is in the sense that the regions that lie above (below) the one-to-one relation have a second $n_{\rm{H}}$ lower than their best-fit $n_{\rm{H}}$. The preference for the best-fitting model-based $n_{\rm{H}}$ solutions to be biased towards higher $n_{\rm{H}}$ values than the \textsc{PyNeb} solutions is further demonstrated in the top-left panel of Figure\,\ref{fig:comparisons_with_pyneb}. The Geneva distribution peaks at a somewhat higher $n_{\rm{H}}$ than the Parsec, though there is a significant overlap between the two. In relation, the \textsc{PyNeb} distribution shows a peak at much lower $n_{\rm{H}}$. Furthermore, both Geneva and Parsec model-based $n_{\rm{H}}$ predictions exhibit a dichotomy in their distribution, which is absent in the \textsc{PyNeb} distribution.  

The lack of agreement between the model versus \textsc{PyNeb} predictions is expected given our choice of the $n_{\rm{H}}$ diagnostic. The [\ion{S}{ii}] doublet is largely sensitive to the low density, low ionisation zones within an HII region, whereas, an array of low and high ionisation emission line ratios are used as constraints in the model fitting. Alternatively, [\ion{Cl}{iii}]~$\lambda$5518,38\AA\,can be utilised as an indicator of the high density, central zones of an HII region. The [\ion{Cl}{iii}] doublet is, however, much weaker than [\ion{S}{ii}], and resides in a part of the spectrum where the model residuals can be large.

Similarly, in the right panels, we show the model- versus \textsc{PyNeb}-derived T$_{\rm{e}}$ comparison. On average, we find that the model-based  T$_{\rm{e}}$ is lower than the \textsc{PyNeb} estimates. This is probably due to \textsc{PyNeb} estimates been based on the [\ion{N}{ii}]~$\lambda$5577, 6584\AA\,diagnostic, which likely represent a single temperature zone within an HII region, while the model-based estimates provide an overall T$_{\rm{e}}$. To depict the differences between Parsec and Geneva derivations of T$_{\rm{e}}$, we colour-code the data points by their differences, where the bluer (redder) shadings portray the HII regions with Geneva-based T$_{\rm{e}}$ larger (smaller) than that of Parsec. For almost all HII regions, the Geneva predictions of T$_{\rm{e}}$ are larger than the Parsec values.    

The top-right panel of Figure\,\ref{fig:comparisons_with_pyneb} presents a T$_{\rm{e}}$ comparison between Geneva (blue) and Parsec (red) models, and \textsc{PyNeb} (black), which, again, illustrate that on average \textsc{PyNeb} estimates of T$_{\rm{e}}$ are larger than the model predictions. The Geneva and Parsec distributions show a substantial overlap, however, the Geneva distribution peaks at a slightly higher T$_{\rm{e}}$ than Parsec. This behaviour of T$_{\rm{e}}$ between the Geneva and Parsec models is expected given the tight correlation between gas metallicity and T$_{\rm{e}}$. Recall that Parsec, on average, yield gas metallicities that are slightly higher than Geneva, which in turn results in lower T$_{\rm{e}}$. 

\subsection{On the distribution of burst strengths in the Antennae}\label{subsec:burst_strengths}
The final parameter of the model HII regions is the strength of the starburst. This parameter is very sensitive to the ionisation structure of an HII region, and as the ionisation structure is not homogeneous throughout a region, it can be challenging to constrain the burst strengths. For example, Figures\,\ref{fig:A901_fits} and \ref{fig:double_solutions} illustrate the typical PDFs derived for burst strengths, which tend to show several distinct distinct peaks.  

The range in burst strengths probed by the current library of models is $10^2-10^4$ M$_{\odot}$, assuming a fixed radius for the model HII regions (Table\,\ref{tab:model_libraries}). The main reason for choosing this range in burst strengths is that the strength of a burst is directly proportional to $U$ (Eq.\,\ref{eq:ionisation_relation}). As such, increasing the strength significantly tend to yield high $U$ values (i.e.\,log $U>$-0.5) for ages around $\sim$1 Myr, which have not been explored much in the literature. Given the radius dependence of the burst strength parameter, to extract the most likely burst that would have formed a given HII region in the Antennae, the above range in burst strengths needs to be compensated for the difference in radii following Eq.\,\ref{eq:ionisation_relation}.  

Overall, almost all HII regions studied show a preference for starbursts of  $>10^4$ M$_{\odot}$ in strength. \citet{Whitmore2010} find masses in the $10^4-10^7$ range for the star clusters in the Antennae, similar to the range in burst strengths derived in this study. It is, however, difficult to perform a direct comparison with \citet{Whitmore2010} work as their work is focussed on star clusters and largely use photometry in the determination of masses. In contrast, our work focus of HII regions that likely consist of several star clusters and principally utilise the emission line ratios in constraining the strength of the starburst that likely formed a given HII region. 
 
\subsection{Correlations between properties}
To explore the correlations between different physical properties, we plot the stellar and gas metallicites, and the light-weighted young ages derived from fitting the model libraries as a function of various properties, e.g.\,H$\alpha$ EW, Balmer decrement, H$\alpha$ luminosity per unit spaxel, H$\alpha$ SFR, H$\alpha$ SFR surface density and stellar mass of HII region estimated by summing the masses of star clusters within that region, in Figure\,\ref{fig:correlation_between}. The star cluster masses are drawn from the catalogue of \cite{Whitmore2010}. 

Different marker types are used to distinguish between the HII regions in the overlap region (including NGC 4039), and those residing in the western loop. The data points are colour coded by the light-weighted young ages.  Except for weak trends apparent in the age versus H$\alpha$ EW and the age versus H$\alpha$ luminosity per unit spaxel plots, there appear to be no strong correlations between these properties. In the case of age versus H$\alpha$ EW, the HII regions show a trend where most regions with high H$\alpha$ EWs are characterised by younger ages, and as expected, this trend is more notable in the HII regions residing in the overlap region than those in the western loop. Similarly, the HII regions in the overlap region show preference for high H$\alpha$ luminosities than those residing in the western loop.  

\section{Summary}

In this paper, we discuss a methodology to exploit the full spectrum (continuum and emission lines) of an HII region. Our goal is to be able to model all the spectral features contained in a spectrum, and thereby extract information on stellar and gas metallicity, age of the stellar population, T$_{\rm{e}}$ and $n_{\rm{H}}$, and the strength of the zero-age starburst that would have produced the region, simultaneously.    

To achieve this goal, first, we develop a suite of comprehensive self-consistent and time-dependent models for characterising starbursting HII regions. The model HII regions are defined by five physical parameters; the effective temperature and age of the stellar population, chemical abundance, ionisation parameter, and density. One of the most important features in spectra of starbursts is the WR feature, and any model-based methodology that attempts to provide a physical description for a starbursting region must be able to reproduce these features. Therefore, in our construction of the stellar models, we use the latest information on the stellar and Wolf-Rayet libraries, and Wolf-Rayet classifications. Moreover, we employ two different stellar isochrones; the Geneva isochrones that are generally preferred for the modelling of massive stars, and the latest Parsec isochrones, in order to explore possible systematics arising from the different implementations of stellar evolution. 

Second, we discuss a potential model-fitting strategy that allows us to exploit the full spectrum, and thereby extract a number of physical properties simultaneously. The fitting procedure is two-fold. In the first instance, we fit the observed (stellar and nebular) continuum to extract the models for the unique underlying stellar populations contributing to an SFH of a region. Then, we use the age distribution of the extracted models and their light-weights to construct a nebular model, and all of this is done as a function of the model library parameters. 

By applying the methodology summarised above to the MUSE spectra of the highly star-forming regions  (i.e.\,H$\alpha$ EW>50\AA)  in the Antennae galaxy, we were able to extract the following physical properties; the light-weighted young ($<10$ Myr) age of the stellar populations, the ages of the dominant underlying old (i.e.\,$>10$ Myr)  stellar populations, stellar and gas metallicities, T$_{\rm{e}}$ and $n_{\rm{H}}$. The distributions of each of these properties corresponding to the Geneva- and Parsec-based model libraries are shown in Figures\,\ref{fig:prop_distributions_geneva} and \ref{fig:prop_distributions_parsec}.     

Thanks to its proximity, the Antennae galaxy is well-studied, and there is abundant literature on the properties of its HII regions. In \S\,\ref{subsubsec:age_lit}, we present a detailed comparison between the ages of the stellar populations derived in this study and the literature to-date. Likewise, a thorough discussion of the stellar and gas metallicities obtained from full-spectrum fitting compared to other methods employed in the literature is presented in \S\,\ref{subsubsec:metal_lit}. In summary, \textit{we find that the HII regions with the youngest ages lie in the overlap region between the two merging discs. The HII regions in the western-loop of NGC 4038 also appear to host young stellar populations, although, their ages are older than those in the overlap region.} Our results are consistent with a number of studies have also made similar observations\citep[e.g.][]{Mengel2005, Snijders2007, Zhang2010, Whitmore2010}. For the stellar and gas metallicities, we find around solar-like values for the starbursts in the Antennae, which is again consistent with the literature \citep[e.g.][]{Bastian2009, Lardo2015}.

\textit{For the Antennae starbursting regions, we derive T$_{\rm{e}}$, on average, around $\sim$7000 K, and average $n_{\rm{H}}$ of slightly $>100$ cm$^{-3}$, and again, there appears to be no significant trend in their distributions.} A detailed discussion on the distribution of these parameters as well as on the comparisons between the T$_{\rm{e}}$ and $n_{\rm{H}}$ derived from the full-spectrum fitting method and that estimated from widely used \textsc{PyNeb} software \citep{Luridiana2015} is presented in \S\,\ref{subsubsec:temp_density_comp}. 

\textit{The zero-age burst strengths derived for almost all HII regions studied are in the range $10^4-10^7$ M$_{\odot}$.} The burst strength is highly sensitive to the ionisation structure of an HII region, as such it is somewhat more difficult to constrain than the other parameters. While the range in burst strength derived is similar to that found by \citet{Whitmore2010} for the star clusters, a one-to-one comparison cannot be performed given the differences between the methodologies used. Refer to \S\,\ref{subsec:burst_strengths} for a discussion on the burst strength parameter. 

Finally, there are several caveats that we have not considered in this study. 

One of the caveats is the Lyman Continuum (LyC; $\lambda<$912\AA) photons leakage from HII regions. \citet{Weilbacher2018} find that a significant fraction of HII regions in the Antennae galaxy is leaking LyC photons at various levels.  In this analysis, however, we construct model HII regions assuming no leakage of LyC photons, which likely increase the dispersion of the reported properties. 

The second caveat being our assumption of bursty-like SFHs for starbursts. While episodic SFHs are considered to be a good approximation, some of these high SFR regions may be characterised by a different type of SFH. 

The third is that the current library of models for starbursts does not consider the stellar evolutionary scenarios that take the effects of binary evolution and stellar rotation into account. These scenarios are critical for the modelling of massive star formation. While stellar evolutionary models, like BPASS \citep{Eldridge2009} and MESA \citep{Choi2016}, allow these scenarios to be incorporated into the models of HII regions, both evolutionary predictions for massive stars can be vastly different from those of single stellar evolutionary models. BPASS, for example, predicts $\sim$100 Myr duration for the WR stars in contrast to a $\sim$4 Myr durations predicted by the single stellar models \citep{Eldridge2017}. It is also challenging to find high-resolution stellar libraries that adequately describe binary interactions. Therefore, before using these evolutionary scenarios for constructing models of HII regions, further exploration is needed to understand not only the best ways of constraining models, but also the adjustments needed in fitting routines to minimise computational costs.

\section*{Acknowledgements}
{\small
M.L.P.G. acknowledges the support from European Union's Horizon 2020 research and innovation programme under the Marie Sklodowska-Curie grant agreement No 707693. P.M.W. was supported by BMBF Verbundforschung (project MUSE-NFM, grant 05A17BAA). A.M.I. acknowledges support from the Spanish MINECO through project AYA2015-68217-P. 

Data used in this paper is based on observations collected at the European Organisation for Astronomical Research in the Southern Hemisphere under ESO programs 095.B-0042, 096.B-0017, and 097.B-0346

This work used the DiRAC Data Centric system at Durham University, operated by the Institute for Computational Cosmology on behalf of the STFC DiRAC HPC Facility (www.dirac.ac.uk). This equipment was funded by a BIS National E-infrastructure capital grant ST/K00042X/1, STFC capital grant ST/K00087X/1, DiRAC Operations grant ST/K003267/1 and Durham University. DiRACis part of the National E-Infrastructure.}

\appendix
\section{On the modelling of Wolf-Rayet features}\label{appA}
\begin{figure*}
\begin{center}
\includegraphics[scale=0.4, trim={0.1cm 0.1cm 0.1cm 0.1cm},clip]{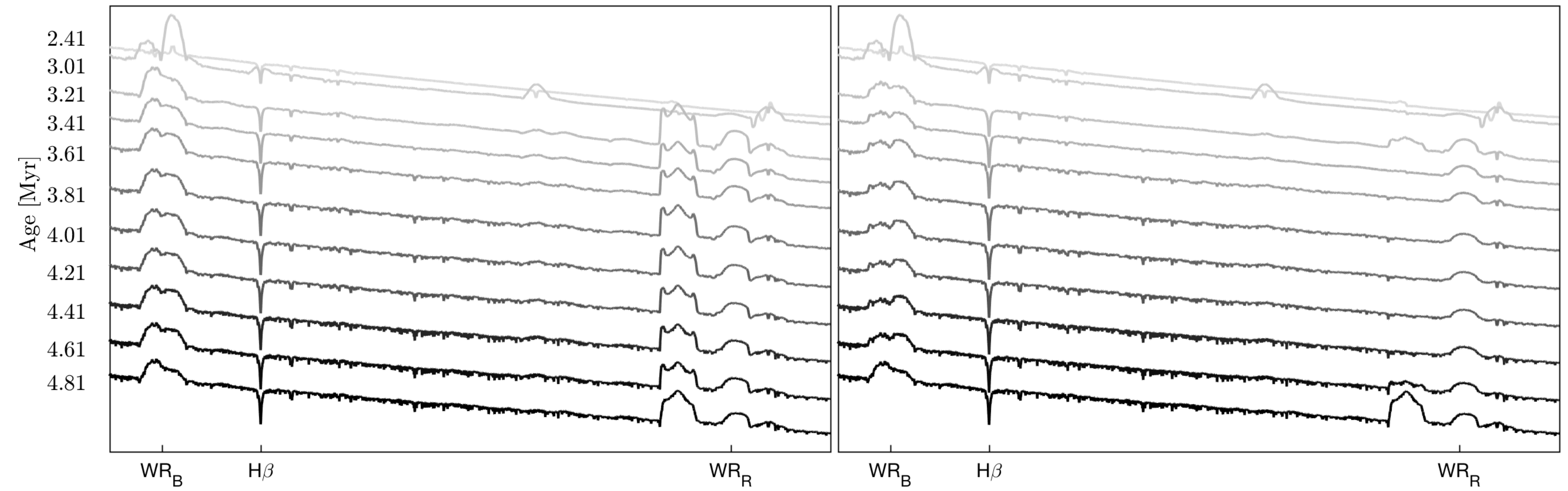}
\caption{With (left) and without (right) the inclusion of the PoWR Galactic WC spectra 12-18 and 12-19 in the \textsc{starburst99} synthesis of stellar evolution guided by the Geneva isochrones.The inclusion of these spectra produce a pronounced blue emission peak of the red WR feature, which is a feature the WC9 WR stars. }
\label{fig:wr_removal}
\end{center}
\end{figure*}
In \S\,\ref{subsubsec:wr_lib} and \ref{subsec:props_wr}, we discuss the selection of the spectra from the PoWR library to input into \textsc{starburst99}, and the properties of the generated SSPs, respectively. Out of a total of 1385 spectra provided in the PoWR WC, WNE and WNL libraries, we remove four spectra from the PoWR WC library (two from the Galactic and two from the LMC) in order to generate stellar templates in cooperation with the Geneva HML isochrones with WR features similar to those that have been observed in the spectra of starbursting HII regions. How the SSPs of solar metallicity change as a result of this small change to the WC library is shown in Figure\,\ref{fig:wr_removal}. 

In the left panel of Figure\,\ref{fig:wr_removal}, we show the stellar templates produced by incorporating the full Galactic PoWR library, and on the right panel, the stellar templates generated by removing the WC spectra 12-18 and 12-19 from the Galactic library. By removing the spectra, the significantly enhanced blue component of the red WR bump, a strong signature of WC9 WR stars, evident across a wide range of young ages is suppressed.  Likewise, we remove the 12-21 and 12-22 spectra from the PoWR LMC WC library, which, again, act to reduce the enhancement of the blue emission peak of the red WR bump. 

These modifications to the PoWR WC libraries are motivated by the fact that the observations of HII regions and galaxies in the nearby Universe do not show such heightened WC9 WR features. Also, the stellar metallicity effects can be crucial in the formation of different subtypes of WR stars. For instance, \citet{Sander2012} illustrate that the WC4 stars in the LMC are significantly brighter than their Galactic counterparts for the same mass-loss rates, and conclude that the WC4 stars in the LMC likely have higher current mass than the Galactic WC4 WR stars, as well as higher luminosities in order to drive the same mass-loss.  

It is worth re-iterating that, as discussed in \S\,\ref{subsubsec:wr_lib}, the mass-loss rates from the stellar isochrones play a significant role in the selection of the WR spectra. The enhancement of the blue peak of the red WR bump appears to be primarily related to the use of the Geneva HML isochrones. The mass-loss rates, among to other parameters, derived from the PARSEC isochrones, on the other hand, do not lead to a substantial enhancement of the features of the red WR bump. However, for consistency, we have used the modified PoWR library throughout.    

Finally, as evident in Figure\,\ref{fig:wr_removal}, the removal of the spectra also affect other WR features. In particular, the blue component of the blue WR feature is somewhat suppressed at certain ages. 

\section{PARSEC versus Geneva}\label{appB}
\begin{figure*}
\begin{center}
\includegraphics[width=0.45\textwidth, trim={1.5cm 0.5cm 0.6cm 0.cm},clip]{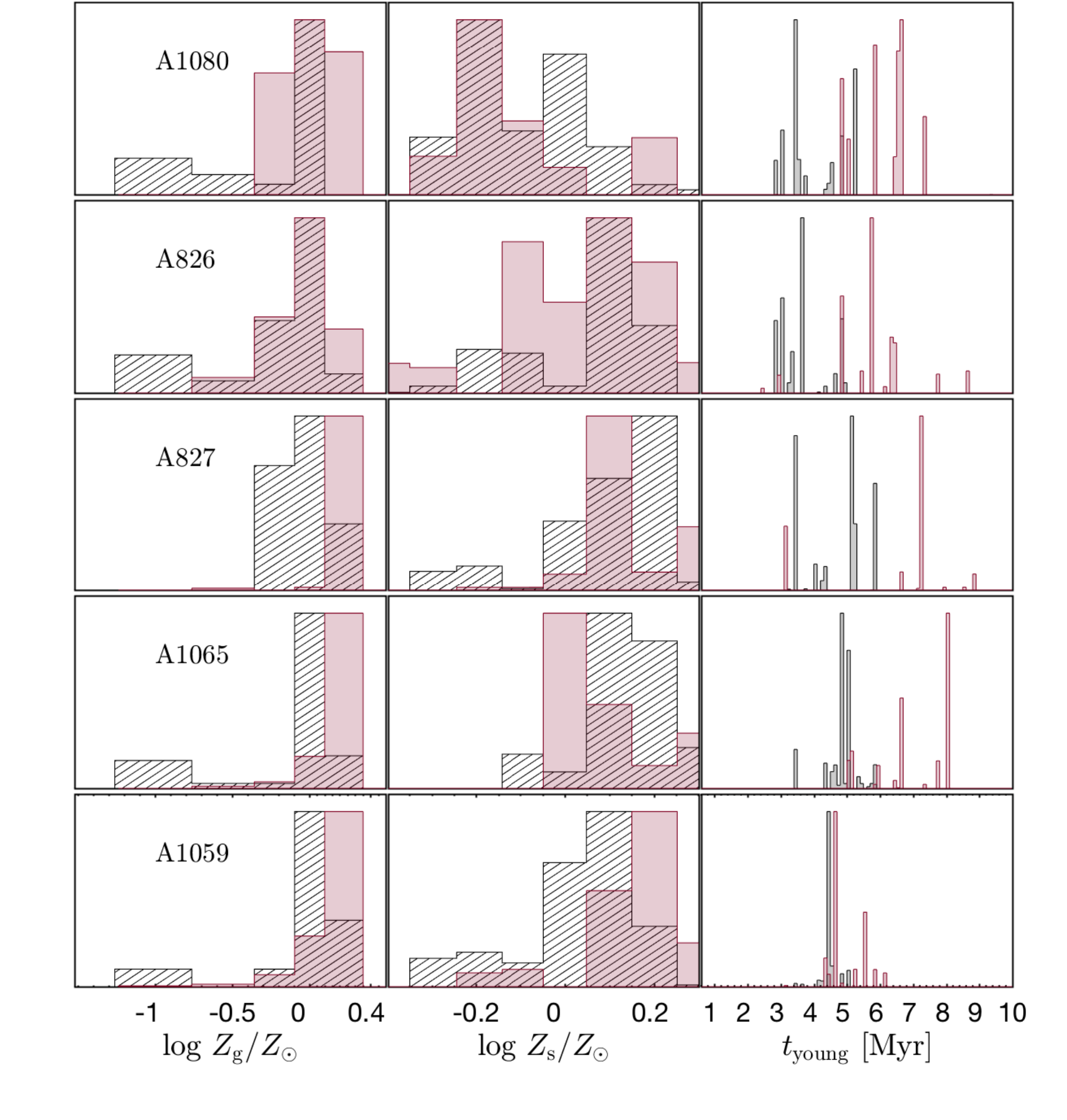} \vspace{-0.cm}
\includegraphics[width=0.45\textwidth, trim={1.5cm 0.5cm 0.6cm 0.cm},clip]{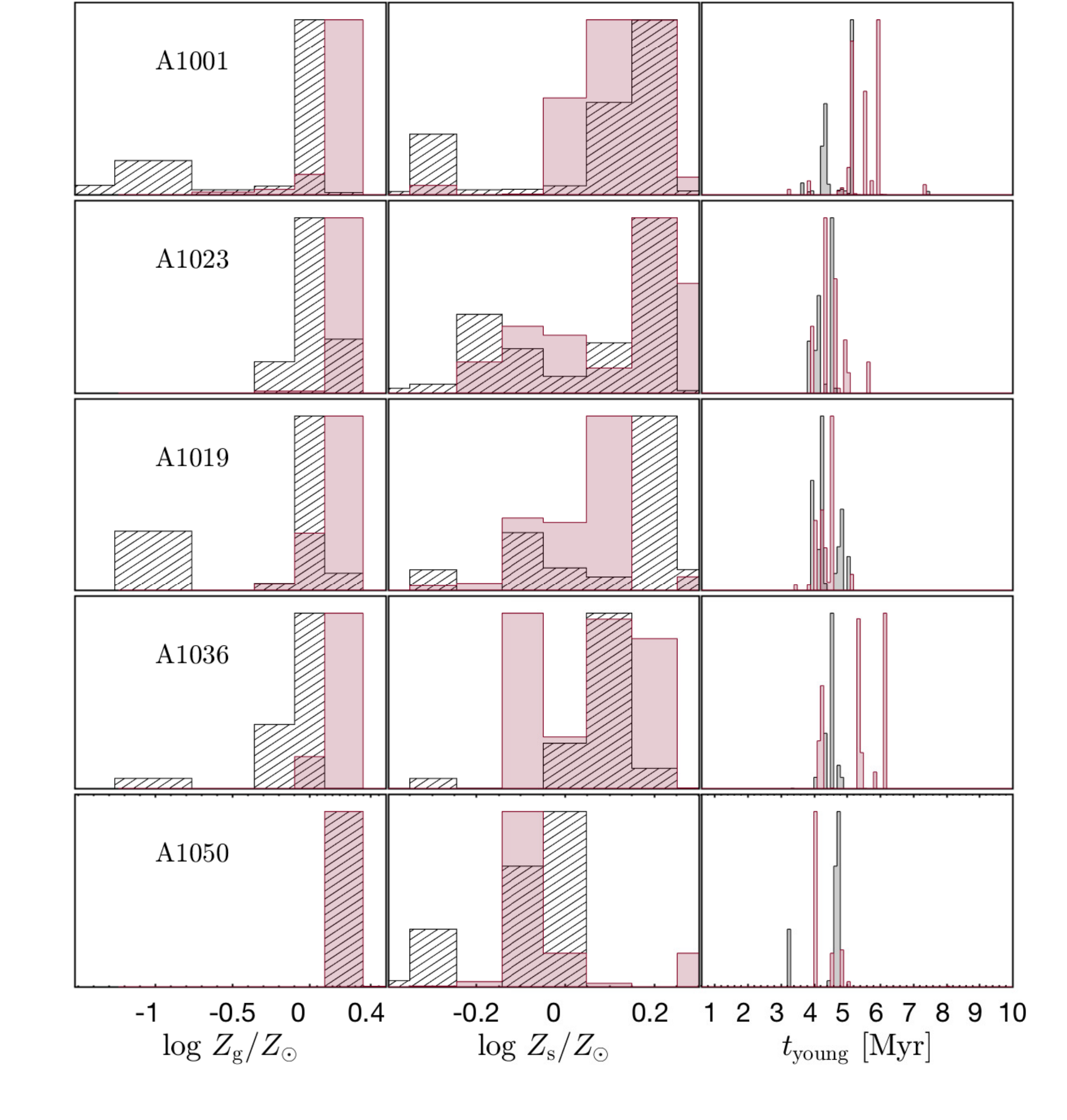} \vspace{-0.cm}
\includegraphics[width=0.45\textwidth, trim={1.5cm 0.5cm 0.6cm 0.cm},clip]{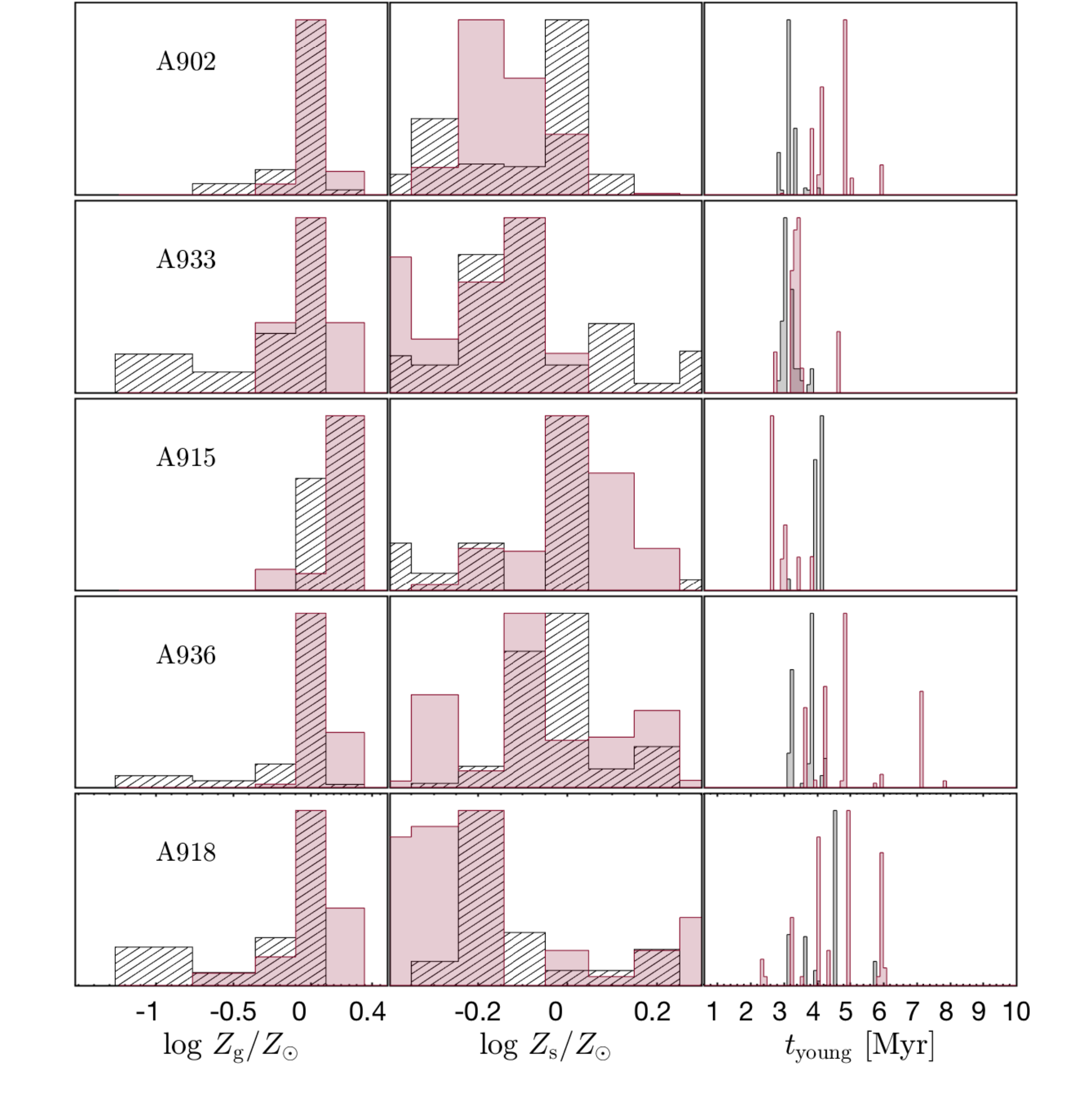} \vspace{-0.cm}
\includegraphics[width=0.45\textwidth, trim={1.5cm 0.5cm 0.6cm 0.cm},clip]{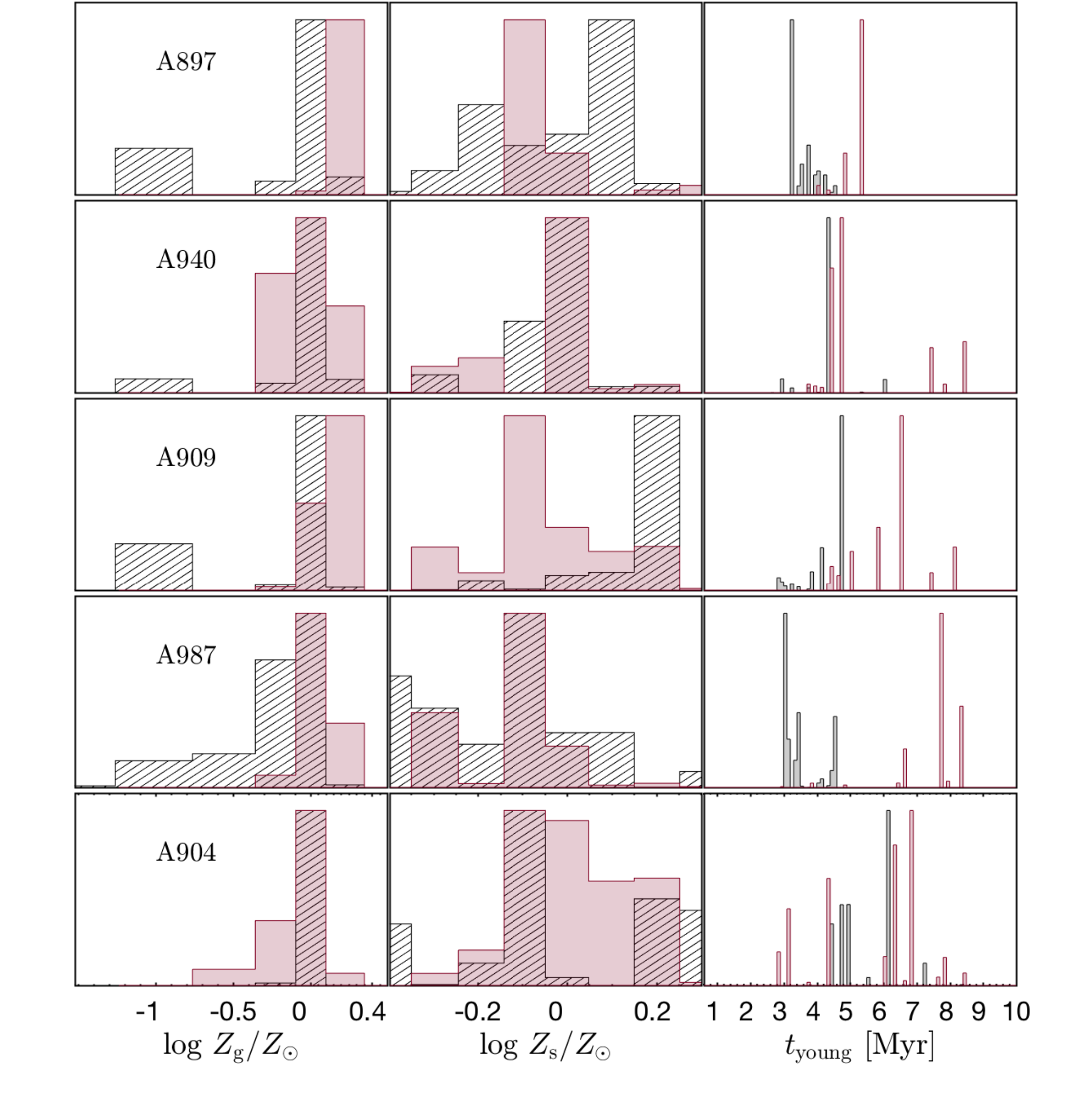} \vspace{-0.cm}
\caption{The Geneva (black dashed) versus Parsec (brown filled) fits for 20 HII regions. The top set of panels show the PDFs of gas and stellar metallicities, and light-weighted young ages (from left-to-right) for ten HII regions scattered across the western-loop. Similarly, the bottom set of panels show the PDFs for ten regions residing in the overlap region and NGC 4039.}
\label{fig:parsec_vs_geneva_pdfs_20regions}
\end{center}
\end{figure*}
In Figure\,\ref{fig:parsec_vs_geneva_pdfs_20regions}, we present a comparison between Geneva (black dashed) and Parsec (brown filled) fits for twenty HII regions in total; ten (top panels) distributed across the western-loop and the rest across the overlap region and NGC 4039.  On average, the Parsec-based light-weighted young ages are relatively older, and the dispersion is larger than the respective Geneva-based values. We discuss some of the potential reason for these discrepancies below. 

Certain cases, like A915 and A1050 in Figure\,\ref{fig:parsec_vs_geneva_pdfs_20regions} and A927 in Figure\,\ref{fig:double_solutions}, indicate Parsec-based light-weighted young ages predictions to be somewhat younger than the Geneva predictions. The spectra of these regions also show the blue WR bump very prominently, and in comparison, a less prominent red WR bump. In the best-fitting models shown in Figure\,\ref{fig:double_solutions} for A927, for example, both Parsec and Geneva assign highest light-weights to a stellar population of $\sim$2.7 Myr in age. While Parsec predicts that a $\sim$2.7 Myr population solely dominates the young component of A927, Geneva also assigns some weights to a $\sim$3.3 Myr and $\sim$4.5 Myr populations, hence biasing the estimated light-weighted young age towards a slightly older age than Parsec.   

In contrast, the Parsec and Geneva fits of, for example, A1080 and A987 in Figure\,\ref{fig:parsec_vs_geneva_pdfs_20regions}, point to distinctly different light-weighted young ages, where the Parsec-based age is clearly older than the Geneva-based value. In the case of A987, both Parsec and Geneva models predict the presence of a $\sim$3 Myr stellar population. Additionally, both Parsec and Geneva models also predict the presence of a second young population, however, at slightly discrepant ages; Parsec models predict the age to be around $\sim$9 Myr, and Geneva, around $\sim$12-13 Myr. While the $\sim$9 Myr versus $\sim$12-13 Myr is essentially a small difference, given the $<$10 Myr age requirement imposed in the calculation of light-weighted young ages, means that in this case, the Parsec-based light-weighted young age is biased towards older age than Geneva. Likewise, with A1080, both Parsec and Geneva models, again, predict the presence of two young stellar populations; the first around $\sim$3 Myr in age, and second with slightly discrepant ages with Parsec predicting a $\sim$7 Myr population and Geneva preferring  $8-13$ Myr population. This behaviour is likely primarily driven by the differences in the stellar isochrones. 

The age--stellar metallicity degeneracy is also a likely reason for the differences as can be seen for A897 and A909 in Figure\,\ref{fig:parsec_vs_geneva_pdfs_20regions}. 

Finally, there could also be some uncertainty arising from the differences in the assumed definition of solar. The solar definition of Parsec is $\sim$0.0169, whereas the stellar library combined with the Parsec models assume a solar value of $\sim$0.02. Throughout this study, we have also assumed a solar metallicity of $\sim$0.02. 

\break
\footnotesize
{
\bibliographystyle{apsrmp}
\bibliography{references}
}
\bsp

\label{lastpage}

\end{document}